\documentclass[usenatbib]{mnras}
\pdfoutput=1
\usepackage{mathptmx}
\usepackage{txfonts}

\usepackage[T1]{fontenc}
\usepackage{ae,aecompl}

\usepackage{graphicx}
\usepackage{longtable}
\usepackage[utf8]{inputenc}
\usepackage[english]{babel}

\newcommand{\Rs}{\ensuremath{R_{\odot}}}
\newcommand{\Ms}{\ensuremath{M_{\odot}}}
\newcommand{\Zs}{\ensuremath{Z_{\odot}}}
\newcommand{\eg}{{\it e.g.}}
\newcommand{\cf}{{\it c.f.~}}
\newcommand{\ie}{{\it i.e.}}

\newcommand{\beq}{\begin{equation}}
\newcommand{\eeq}{\end{equation}}
\newcommand{\mtot}{\ensuremath{M_{\rm tot}}}

\newcommand{\mchirp}{\ensuremath{M_{\rm chirp}}}

\newcommand{\kmps}{\ensuremath{{\rm~km~s}^{-1}}}

\newcommand{\mcl}{\ensuremath{M_{cl}}}
\newcommand{\rh}{\ensuremath{r_h}}
\newcommand{\rhstar}{\ensuremath{R_{h,\ast}}}

\newcommand{\tkl}{\ensuremath{T_{\rm KL}}}

\newcommand{\nmrgin}{\ensuremath{N_{\rm mrg,in}}}
\newcommand{\nmrgout}{\ensuremath{N_{\rm mrg,out}}}
\newcommand{\nsling}{\ensuremath{N_{\rm sling}}}
\newcommand{\nbhbound}{\ensuremath{N_{\rm BH,bound}}}
\newcommand{\nbbhbound}{\ensuremath{N_{\rm BBH,bound}}}
\newcommand{\ndynbbhbound}{\ensuremath{N_{\rm dynBBH,bound}}}
\newcommand{\nbhmsbound}{\ensuremath{N_{\rm BH-MS,bound}}}
\newcommand{\ndynbhmsbound}{\ensuremath{N_{\rm dynBH-MS,bound}}}

\newcommand{\nbseven}{{\tt NBODY7~}}

\newcommand{\bse}{{\tt BSE }}

\newcommand{\chain}{{\tt CHAIN }}
\newcommand{\tevol}{\ensuremath{T_{\rm evol}}}
\newcommand{\fbin}{\ensuremath{f_{\rm bin}}}
\newcommand{\nbin}{\ensuremath{N_{\rm bin}}}
\newcommand{\fobin}{\ensuremath{f_{\rm Obin}}}
\newcommand{\fgwp}{\ensuremath{f_{\rm GWp}}}
\newcommand{\opt}{{\tt option }}

\title[Stellar-mass black holes in open clusters III]
{Stellar-mass black holes in young massive and open stellar
clusters and their role in gravitational-wave generation III:
dissecting black hole dynamics}

\author[S. Banerjee]{
Sambaran Banerjee$^{1,2}$\thanks{E-mail: sambaran@astro.uni-bonn.de (SB)}
\\
$^{1}$Helmholtz-Instituts f\"ur Strahlen- und Kernphysik (HISKP),
Nussallee 14-16, D-53115 Bonn, Germany\\
$^{2}$Argelander-Institut f\"ur Astronomie (AIfA),
Auf dem H\"ugel 71, D-53121, Bonn, Germany
}

\pubyear{2018}

\begin{document}
\label{firstpage}
\pagerange{\pageref{firstpage}--\pageref{lastpage}} 
\maketitle

\begin{abstract}
Stellar-remnant black holes (BH) in dense stellar clusters
comprise a natural setup to trigger
general-relativistic (GR) inspiral and merger of
binary black holes (BBH), detectable by the LISA and
the LIGO-Virgo, through dynamical encounters inside such environments. In this work,
the intricacies of such dynamical interactions are probed utilizing realistic,
self-consistent, post-Newtonian, direct N-body evolutionary models
of young massive and open stellar clusters. Particularly, the configurations
of the compact subsystems, that drive the in-cluster GR BBH coalescences,
are tracked on the fly.
Such an approach reveals that the GR coalescences
within the open clusters take place primarily via chaotic interactions involving
triple BH systems. Although less frequently, such mergers are found to happen also in higher-order
subsystems such as quadruples and in subsystems involving non-BH members;
the mergers can themselves be BH---non-BH, which events would leave electromagnetic signatures.
Close, fly-by encounters inside the clusters can also make BBHs and other types of
double-compact binaries temporarily post-Newtonian;
such binaries would potentially contribute to the GW background for the LISA and the PTA.
These calculations,
furthermore, suggest that open clusters are potential hosts for not only detached BH---main-sequence
binaries, as recently identified in the globular cluster NGC 3201, but also
a wide variety of other types of remnant---non-remnant binaries, which are assembled via dynamical
interactions inside the clusters and which have the prospects of being discovered
in radial-velocity surveys. 
\end{abstract}

\begin{keywords}
open clusters and associations: general -- globular clusters: general --
stars: kinematics and dynamics -- stars: black holes -- methods: numerical -- 
gravitational waves
\end{keywords}

\section{Introduction}\label{intro}

Recent studies show that the retention of stellar-mass black holes (hereafter BH) in
dense stellar clusters of wide mass range, beginning from low-/medium-mass young and open clusters
\citep[\eg,][]{2010MNRAS.402..371B,Mapelli_2011,Mapelli_2013,Ziosi_2014,Mapelli_2016,Park_2017,Banerjee_2017,Mapelli_2017,Banerjee_2017b}
through globular clusters
\citep[\eg,][]{2013MNRAS.430L..30S,Morscher_2013,Rodriguez_2016,2016arXiv160300884C,Wang_2016,Askar_2016}
to galactic nuclear clusters
\citep[\eg,][]{Antonini_2016,ArcaSedda_2017,Hoang_2017},
make up environments where the BHs pair up
through dynamical interactions which, furthermore, potentially lead to general-relativistic (hereafter GR)
coalescences of these binary black holes (hereafter BBH).
Being much more massive compared to the rest of the stellar members in a cluster,
the BHs sink and remain highly concentrated in the cluster's
central region \citep[\eg,][]{2010MNRAS.402..371B,Morscher_2015}
due to the dynamical friction \citep{1987degc.book.....S} from the stellar background, where
they undergo frequent close and energetic encounters, giving rise to such events. 
That way, all the BBH merger events,
as inferred so far through the detection of inspiral gravitational waves (hereafter GW)
by the ground-based Advanced
Laser Interferometer Gravitational-wave Observatory (hereafter LIGO) and the Virgo interferometric
GW observatory
\citep{2016PhRvL.116f1102A,2016ApJ...818L..22A,2016PhRvL.116x1103A,Abbott_GW170104,Abbott_GW170814,Abbott_GW170608},
could be \emph{naturally} 
reproduced \citep[\eg,][see also \citealt{Mandel_2017,Mapelli_2018}]{Banerjee_2017b}.
The ``dynamical channel'' would easily sum up to a BBH merger detection
rate of $\sim100{\rm~yr}^{-1}$ by the LIGO at its proposed full sensitivity
\citep[see][and references therein]{Banerjee_2017,Banerjee_2017b}.

Alternatively, BBH mergers, like those detected,
can be derived through the evolution of isolated massive-stellar
binaries in the field. Such a scenario typically involves a
common-envelope (CE) phase, that tightens an initially-wide
(semi-major-axis $a\sim1000$ AU) massive binary, combined with  
the formation of direct-collapse BHs (see Sec.~\ref{stellrem}) from the members
\citep{Belczynski_2016,Mandel_2017a,2017arXiv170607053B}. The alternative to this
scenario involves chemically-homogeneous evolution of the
members in a massive ``over-contact'' binary (MOB): here tidally-induced
spins in the members of a close ($a\sim100\Rs$) massive binary allow them to
become chemically mixed and helium rich throughout
so as to avoid coalescence until they collapse into BHs \citep{DeMink_2009,DeMink_2016,Marchant_2016}.
In such studies, a low to a very high BBH inspiral detection rate of $\sim10-\sim1000{\rm~yr}^{-1}$
has been estimated for the full-sensitivity LIGO.

Recent studies by \citet[][hereafter Paper I]{Banerjee_2017} and
\citet[][hereafter Paper II]{Banerjee_2017b} have shown, for the
first time through detailed and completely self-consistent relativistic (post-Newtonian; hereafter PN) direct
N-body simulations, the
importance of triple-/higher-order-dynamical
interactions in
triggering in-cluster GR BBH coalescences, particularly inside low-velocity-dispersion systems  
such as open clusters and lower-mass globular clusters (hereafter GC)
\footnote{See also the related blog article from the Oxford University Press:
\url{https://blog.oup.com/2017/03/dance-of-black-holes/}.}.
A lower velocity dispersion makes gravitational focusing \citep{1987degc.book.....S}
generally more efficient, especially when the most massive cluster members
such as the BHs are involved, facilitating the dynamical formation of triple and
higher-order multiple systems comprising (but not limited to) them. Such
dynamically-formed subsystems are also generally wider in a lower velocity-dispersion
environment, due to the lower hard-soft energy boundary \citep{1987degc.book.....S},
making the subsystems' geometrical cross sections larger.
The efficient gravitational focussing and larger geometric cross
section make the subsystems both highly efficient in undergoing dynamical
interactions, potentially competing in frequency
with those involving only singles and binaries \citep{Leigh_2013}, and highly
vulnerable to close, dynamical encounters leading to their significant alteration
or destruction \citep[``dynamical interruption'';][]{Geller_2015a}.
Hence, the success of a GR coalescence within
a triple or a higher-order multiple system often relies on the large
eccentricity growth in chaotic encounters, during either 
a resonant binary-single interaction \citep{Hut_1983,Samsing_2014}
or a chaotic phase of the evolution of a
hierarchical triple (higher-order multiple; \eg, \citealt{Antonini_2016b}), rather than
the secular evolution of a compact subsystem, \eg, the (eccentric) Kozai-Lidov (hereafter KL) oscillation
of a triple \citep{Kozai_1962,Katz_2011,Lithwick_2011}.
The timescales of eccentricity boost during the chaotic processes
are, typically, much shorter compared to the secular timescales of the compact subsystem,
fostering the opportunity for rapid GR inspiral and merger before
the subsystem can be significantly altered or destroyed by the
next close encounter.
Note that such triples can form as a result of both binary-single and
binary-binary geometrically-overlapping or close encounters
\citep[see, \eg,][]{2003gmbp.book.....H,SigurdPhin_1993,Leigh_2011}.
As estimated in Paper II based on conservative assumptions,
the possibility of triple-mediated,
in-cluster BBH coalescences combined with their abundance would make
the contribution of open clusters, to the dynamical BBH merger rate,
comparable to those from the GCs and nuclear clusters.

In fact, a substantial increase in such in-cluster BBH mergers has been
demonstrated even for GC-like systems in subsequent semi-analytic and
Monte Carlo studies when PN effects are
incorporated \citep{Samsing_2018,Rodriguez_2018}.
Given that any BBH (as such, any binary) that remains bound to the
cluster for a while is necessarily hard \citep{2003gmbp.book.....H},
close, \emph{fly-by} binary-single encounters would typically \emph{not}
cause sufficient eccentricity growth of the BBH so that
it can promptly undergo GR inspiral and merger before the next close encounter;
however, the dynamically-induced eccentricity can still be sufficient
to make the BBH ``temporarily relativistic'' (see Secs.~\ref{nbprog} \& \ref{sling}).
Of course, if a BBH (or even a BH-neutron star; hereafter NS or NS-NS binary) ends up
getting ejected from the host cluster as the outcome of 
a resonant or a fly-by binary-single interaction or a chaotic
breakup of a triple, then, given sufficient tightness and eccentricity,
it will have time until the current epoch for merging
\citep{2010MNRAS.402..371B,Askar_2016,Rodriguez_2016,2016arXiv160906689C}. 
Massive GCs, with velocity dispersion typically $\sim10\kmps$ and
moderate escape velocities, which host fairly tight BBHs that are also
sufficiently dynamically active, are the most efficient machines for
ejecting such ``merge-able'' BBHs. 

Such intricacies of the dynamical interactions in stellar clusters that
lead to BBH and possibly also
other types of GR binary coalescences make it tempting to follow
them \emph{live} (\ie, on the fly)
in evolutionary cluster models comprising
realistic ingredients in a fully self-consistent manner,
which motivates the current Paper III. Perhaps more importantly,
such a study, which is for the first time of its kind, will be directly useful
for interpreting the existing and the forthcoming inspiral GW detections
by the LIGO-Virgo and as well the GW signals from BBHs detected by
the forthcoming space-based interferometric GW observatory
Laser Interferometer Space Antenna \citep[hereafter LISA][]{eLISA}.
It will as well aid the anticipations and the interpretations of
combined LISA-LIGO BBH inspiral detections and BBH eccentricity
detections in the LISA band \citep{Sesana_2016,Nishi_2017,Banerjee_2017b}.
In particular, for the first time from ab initio, long-term, relativistic, direct N-body simulations,
\citet[][see their Fig.~7]{Banerjee_2017b} has shown that all on-the-fly-detected, in-cluster
BBH in-spirals and
at least half of the ejected BBH in-spirals pass through high eccentricities
($e>0.1$), but not too high to be inaudible \citep[$e\lesssim0.7$;][]{Chen_2017},
in the LISA's characteristic detection band. Similar conclusions
have subsequently been arrived at in Monte Carlo and semi-analytical
studies such as \citet{Kremer_2018a,Samsing_2018a}.
Note that until now the configuration and evolution of relativistic
triple systems have been probed in the
field \citep[\eg][]{Silsbee_2017,Antonini_2017},
``isolated'' from their parent globular or open clusters to
study them separately \citep[\eg][]{Kimpson_2016,Antonini_2016b}, or 
such triples have been formed in three-body scattering
experiments \citep[\eg,][]{Samsing_2014,Samsing_2017}.
From a pedagogical point of view, a live investigation of relativistic subsystems, that
continue to form and destroy in a dynamically-active environment such as
inside a dense star cluster, will lead to a better understanding of
the dynamical interactions that cause extreme GR effects in an otherwise
Newtonian system.

The current study, which is a preliminary step towards such ``hands-on'' probe
into the internal mechanics of BH subsystems in stellar
clusters, involves ab initio, long-term, relativistic,
direct N-body computations of young massive and
open clusters along the lines of Paper II (Sec.~\ref{runs}). The enhancement and diversity of
dynamical interactions and their consequences, due to the inclusion of even a
small fraction of primordial binaries, has been conceived in Paper II: in
this work, a wider range of primordial-binary fraction is considered (5\%-50\%; Sec.~\ref{runs}).
Also, a wide initial-mass range of the primordial binary-containing models
are considered ($7.5\times10^3\Ms-5.0\times10^4\Ms$; Sec.~\ref{runs}).

Furthermore, in some
of the runs here, specialized output arrangements are made
to record the configuration of the triple/higher-order-multiple system
hosting a GR coalescence (Sec.~\ref{getsys}):
this is one step forward towards on-the-fly probing of GR-merger-hosting subsystems (Sec.~\ref{bbhmrg})
and also GR---fly-by interactions (Sec.~\ref{sling}) within the clusters.
Also, the ``ambient'' population of BBHs,
as a function of the host cluster's properties, is investigated here in detail (Sec.~\ref{bbhpop}).

Apart from the BBHs and their mergers, the formation of BH--non-BH binaries, and, in particular, 
of detached BH-main sequence (hereafter MS) binaries are looked at in this work.
The recent discovery of a detached ($a\gtrsim1$ AU) BH-MS binary in the 
GC NGC 3201 by \citet{Giesers_2018}, through radial-velocity measurements,
has generated high interest in theoretically investigating
formation channels of such binaries in GCs \citep{Askar_2018,Kremer_2018}.
Other GCs, where stellar-mass BH candidates accreting from stellar companions
have been identified through radio and X-ray observations, are M22 \citep{Strader_2012},
M62 \citep{Chomiuk_2013}, 47 Tuc \citep{Miller_Jones_2015},
and M10 \citep{Shishkovsky_2018}; the potential
doner companion has also been optically identified for the M10 object which is
a ``red straggler'' star.
It is natural
to expect that the dynamically-active environment of young massive and open
clusters would as well give rise to such binaries, which is also investigated
in this work (Sec.~\ref{bhmspop}). The results are recapitulated in Sec.~\ref{summary}.

\section{Model computations}\label{runs}

Table~\ref{tab1} summarizes the model computations: some of the computations
from Paper II are recycled here, which are indicated in Table~\ref{tab1}. 
The computed models \emph{initiate} with \citet{Plummer_1911} profiles with masses
$7.5\times10^3\Ms\lesssim\mcl(0)\lesssim5.0\times10^4\Ms$, half-mass radii
$1.0{\rm~pc}\lesssim\rh(0)\lesssim2.0{\rm~pc}$, \emph{overall} (see Sec.~\ref{primbin})
primordial-binary fractions (the fraction of the
total number of stars in binaries initially) $0.05\lesssim\fbin(0)\lesssim0.5$, 
and have metallicities $0.001\leq Z\leq0.02$. 
All the computed models are taken to be initially
unsegregated and are subjected to a solar-neighbourhood-like
external galactic field.
In this work, some stress is provided
to the models in the lower side of the covered mass range, which
initiate with a high fraction of primordial binaries ($\fbin(0)\gtrsim0.3$; see Sec.~\ref{primbin},
Table~\ref{tab1}). Since these models are time consuming (and challenging) to run
for long term, only $Z=0.02$ and $0.001$ are considered for them; the BH mass distribution,
that is retained in the cluster at birth (see Sec.~\ref{stellrem}), does
not depend strongly on $Z$ at its sub-solar values
\citep[see, \eg, Fig.~1 of][]{Banerjee_2017} so that a $Z=0.001$ model would
serve to mimic typical $Z<\Zs$ clusters, \eg, as in
dwarf galaxies.

Such model clusters represent young massive and open clusters that
continue to form and dissolve throughout a gas-rich galaxy, as in, \eg,
the Milky Way and the other Local Group galaxies. Therefore, considering
only a solar-neighbourhood-like external field ignores the fact that
such a cluster may be subjected to a much stronger galactic field that
would accelerate its dissolution. However, recent test computations
(Riaz et al., in preparation) suggest that such clusters continue to
produce relativistic subsystems and mergers (see Secs.~\ref{bbhmrg} \& \ref{sling})
even when they orbit at 1 kpc Galactocentric radius as long as
they survive long enough to allow the BHs in them to become centrally segregated and
hence dynamically active \citep{2010MNRAS.402..371B}.
A larger set of models covering a range of external field
will be undertaken in a future study.

\subsection{Introduction of primordial binaries}\label{primbin}

Although having $\fbin(0)$
as high as 0.5 makes the long-term direct N-body computations, as
done here (see below), quite challenging, it has the potential
to generate BBH coalescences even in the least massive models, due
to the enhanced BBH formation through exchange encounters (see Paper II).
More importantly, as observations imply, intermediate- and old-aged open clusters
typically have significantly higher 
primordial-binary fractions than those in GCs \citep{Leigh_2015}
--- often several 10s of percent
\citep[][]{Khalaj_2013,Milliman_2014,Geller_2015}. Hence, assuming such high
$\fbin(0)$ is as well more realistic. For computational ease, only the
$\mcl(0)\approx7.5\times10^3\Ms$ and $1.5\times10^4\Ms$ models are computed
with $\fbin(0)>0.1$, in this study.

Note that the values of $\fbin(0)$ as quoted above and in Table~\ref{tab1}
represent the \emph{overall} initial binary fraction in the cluster
for the entire stellar mass distribution, the latter being
taken to be canonical \citep{Kroupa_2001}.
However, as in Paper II (see their Sec.~2), the initial binary fraction of the
O-type stars (stellar mass, $m_s\gtrsim16\Ms$), taken separately,
is $\fobin(0)\approx100$\%, to be consistent with the observed high binary fraction
among O-type stars in young clusters and associations \citep[see, e.g.,][]{Sana_2011,Sana_2013}.
As elaborated there, the O-type-stellar binaries are taken to initially obey
the orbital-period distribution of \citet{Sana_2011} and a uniform mass-ratio distribution
(an O-star is paired only with another O-star, as typically observed, and the pairing among the
lower-mass stars is obtained separately; see below).
The orbital periods of the non-O-star primordial binaries are taken to follow a
\citet{Duq_1991} distribution that represents a dynamically-processed binary population
\citep{Kroupa_1995a}, and their mass-ratio distribution is also taken to be uniform.
The initial binary fraction among the non-O-type stars is taken to be
$\fbin(0)$; since for the adopted canonical IMF the non-O-type stars comprise by far
the majority of the stellar population, the overall initial binary fraction is
also $\approx\fbin(0)$. As in Paper II, the initial eccentricities of the O-type stellar
binaries follow the \citet{Sana_2011} eccentricity distribution and those
for the rest of the binaries are drawn from the thermal eccentricity distribution
\citep{1987degc.book.....S}. As explained in Paper II, such a scheme of including
primordial binaries provides a reasonable compromise between the economy of computing
and consistencies with observations.

\subsection{Stellar remnant formation}\label{stellrem}

In this work, stellar-remnant NSs and BHs are formed as in
Papers I \& II, \ie, with a modified version of the semi-analytical
binary-stellar-evolutionary program \bse \citep{Hurley_2000,Hurley_2002}
which is coupled to the N-body program (see Sec.~\ref{nbprog}) that
evolves the stars and stellar binaries from their zero age until
remnant formation in tandem with the dynamical evolution; see
Sec.~2.2 of Paper I and references therein for a detailed description. Here, direct-collapse
BHs (hereafter DCBH) and electron-capture-supernova (hereafter ECS; \citealt{Podsiadlowski_2004})
NSs, whose natal kicks are taken to be zero, retain in the
clusters at birth and participate in dynamical interactions for long term.
In the presence of massive primordial binaries, as in here, the mass distribution of the BHs
are moderately influenced by massive-binary coalescences, the occurrences of
which can be dynamically-influenced (see Sec.~2.1 of Paper II).
As elaborated in Paper I (see their Sec.~2.2), the
DCBHs are formed based on the supernova-fallback results of
\citet{Fryer_1999,Fryer_2001} which, as well as the formation
of ECS-NSs, are implemented in \bse
in similar ways as in \citet{Belczynski_2008,Belczynski_2010}.

Note that the evolution of a binary containing at least one non-compact
star, be it detached or Roche-lobe overflown (hereafter RLO), is affected both by
stellar-evolutionary mass loss and dynamical encounters; also by mass
transfer, for the RLO binaries. However, in the majority of the model
computations in Table~\ref{tab1}, the optional tidal interactions
among single stars and within a binary, as
implemented in \nbseven according to the prescriptions described in \citet{Mardling_2001},
is disabled. Such tidal energy dissipation schemes, in addition to being
generally poorly understood, apply only to low-mass stars.
The tidal interaction would mainly affect a handful of very tight primordially-derived binaries
(orbital period $\lesssim10$ days)
that anyway project negligible cross sections for dynamical interactions
and would typically not play a role for the formation and/or
modification of the much wider dynamically-formed/dynamically-active binaries.
However, the members of such a dynamically-formed/dynamically-active binary can also
interact tidally at the periastron
provided the binary is sufficiently eccentric; that way, it can circularize
and undergo orbital decay even up to the point of RLO \citep{Kremer_2017}. 
Therefore, to see whether tidal dissipation would influence the dynamical formation
and the properties of BH-MS binaries, a set of test, shorter-term runs
are performed for three of the $Z=0.02$ initial models with tidal
dissipation activated, as indicated in Table~\ref{tab1}.
Tidal dissipation may also likewise tighten a fraction of the star-star binaries,
resulting in tighter BBH and BH-NSs through exchange encounters
\footnote{In \nbseven, \opt $27=1/2$ and 3 applies tidal dissipation
and GW-radiation dissipation mutually exclusively on a
KS-regularized pair (Sec.~\ref{nbprog}). However, since in the present
runs, the compact subsystems as well as the PN binaries are always treated
with the ARC (\opt $11=-1$ used; Sec.~\ref{nbprog}), no compromise is made
with the PN treatment of a binary or a higher-order subsystem even if
the tidal-dissipation \opt is chosen for the KS pairs (\opt $27=1$).}.

Note that according to the adopted primordial-binary population (Sec.~\ref{primbin}),
all of the BH progenitor stars are initially in massive binaries. However,
the members' mass loss until BH formation widens the binaries significantly,
which are then dissociated in a dynamically-active environment
\citep{Morscher_2015,2016arXiv160300884C}, resulting in predominantly
single BHs. Furthermore, strong dynamical encounters would often impart
high eccentricities onto a massive binary (see Sec.~\ref{intro}),
potentially resulting its members' merger into a single star; if the
merger happens after the members have already evolved until close to the
remnant formation, the single BH resulting from the merger product can be
substantially more massive than what a dynamically-inert population of
single stars with the same zero-age mass range would yield (see
Fig.~1 or Paper II). Nevertheless, a few of the massive primordial
pairs may survive even after both members have become BHs - the
properties of such primordially-derived BBHs are discussed in Sec.~\ref{bbhpop}.
Apart from these, any BBH existing inside a cluster must
have been dynamically paired, as discussed in Sec.~\ref{bbhpop}.

Finally, note that although in the present study the BH population
within a cluster, that participates in dynamical encounters, is derived
via a specific recipe (see above) the qualitative aspects of the outcomes,
especially those of the GR mergers and close encounters (Secs.~\ref{bbhmrg} \& \ref{sling}),
are likely to hold true
if the BHs' natal mass distribution is altered; in particular, if it
is widened. This is because the present BH mass distribution is
already wide by several factors, especially at
low metallicities (see Figs.~1 of Papers I \& II). However,
if one or two of the retaining BHs are distinctly more massive than the
rest, say, be intermediate-mass black holes \citep[IMBH; \eg][]{Lutzge_2016}, 
then the BH triples/multiples can be additionally dissociated
by the IMBH(s) \citep{Leigh_2014}, and (intermediate-mass-ratio) inspirals
(IMRI) onto them may be favoured. A detailed study of the latter case
is beyond the scope of this paper.

\subsection{The post-Newtonian, direct N-body evolution program \nbseven:
live tracking of relativistic interactions and subsystems}\label{nbprog}

As in Paper I \& II, the state-of-the-art, relativistic, direct N-body evolution
program \nbseven \citep{2003gnbs.book.....A,Nitadori_2012,2012MNRAS.422..841A}
is utilized to evolve these initial models. The key aspects of the
code are summarized in Sec.~2.1 of Paper II. In brief, \nbseven is a fourth-order
Hermite integrator for member-by-member tracking of trajectories in an arbitrary, self-gravitating,
many-body system where the close encounters and the bound subsystems (binaries and multiples) are treated
with \emph{regularization techniques} and \emph{no force softening} is
applied. The integration
is facilitated by applying a neighbour-based scheme \citep{Nitadori_2012} for the force contributions
at the shortest time intervals (the ``irregular'' force/steps).
At longer intervals (the ``regular'' force/steps), all bound members in the
system are included for the force evaluation. The irregular forces
are computed by parallel processing in CPUs, while the much more
expensive regular forces are evaluated on 
GPUs\footnote{All simulations in this work are done on
server-class workstations equipped with quad-core {\tt AMD} processors and {\tt NVIDIA}'s
{\tt Fermi} and {\tt Kepler} series GPUs.}. The near-diverging gravitational forces
during two-body close passages and in binaries are dealt with two-body or
Kustaanheimo-Stiefel (KS) regularization
\citep{2003gnbs.book.....A} and higher-membership interactions/multiples are treated with the
Algorithmic Regularization Chain
\citep[ARC;][]{Mikkola_1999,Mikkola_2002,Mikkola_2008}. As opposed to
the traditional or KS-Chain Regularization \citep{Mikkola_1993},
the adoption of the ARC in \nbseven allows the inclusion of members in arbitrary mass ratios,
in the Chain. This is particularly necessary for the
present calculations where the BHs span a wide mass range and predominantly
take part in binary-single and binary-binary interactions (see Sec.~\ref{intro},
Papers I \& II).

In \nbseven, the Chain members and the Chain perturbers are, nevertheless,
selected in the same well-proven way as in the standard KS-Chain implementation
\citep{2003gnbs.book.....A,2012MNRAS.422..841A}, except that the internal integration
of the Chain
is done in the ARC way (\citealt{Mikkola_2008} and references therein). Here,
the Chain members, that comprise a compact subsystem
(often a triple or a quadruple), are initially chosen based on the
strategies for identifying hierarchical systems. This comprises
identifying, for each member in the system, the dominant and the next-to-dominant
perturber and testing whether to form a
triple or a quadruple; see Chapter 9 and Algorithm 11.3 of \citet{2003gnbs.book.....A}
for the details. The
Chain is then constructed (and possibly switched during the Chain integration
due to either change in the subsystem's configuration or
the close approach of a perturber) in the usual way
by sequentially connecting the closest subsystem members with vectors as described in
\citet{Mikkola_1993}. The perturbers of the Chain are then selected
either from within its close-encounter radius or through a full perturber
search.
The key algorithms are discussed in Chapter 12 of \citet{2003gnbs.book.....A}.

An important aspect of \nbseven is its GR treatment of binaries and multiples
(and also of close two-body encounters) involving a BH or an NS through the ARC
\citep{Mikkola_2008}. This enables
on-the-fly GR orbital modifications and coalescences of relativistic subsystems (typically,
but not limited to, a binary or a triple containing one or more BH/NS)
that are bound to the system.
In principle, PN-1, PN-2 (GR periastron precession), PN-2.5 (orbital shrinking due to
GW radiation), PN-3, and PN-3.5 order terms can be included in the ARC
procedure, including their spin contributions.
However, for computational ease, the BHs' spins are taken to be zero in the present computations.
The spin terms would have moderately affected the times of the GR coalescences occurring within the cluster
(see Sec.~\ref{intro}), however, this is
not critical here due to the statistical nature of the dynamically-induced
BBH coalescences.

Furthermore, in practice when the stellar-remnant BHs have large spins,
a BBH would typically receive a large GW merger kick during its final
inspiral and plunge ($\sim100-1000\kmps$; \citealt{Campanelli_2007,Hughes_2009}).
This would cause almost all the newly-formed, merged BHs to escape from open and globular
clusters almost immediately,
without having a chance to participate in dynamical encounters further.
This effect is mimicked in \nbseven by applying a velocity kick onto the merged BH
immediately after a GR BBH
coalescence occurs within the cluster (Sec.~\ref{bbhmrg}).
In the current implementation, the applied kick
is kept only marginally above the escape speed to avoid large energy errors,
$\approx5$ times the central RMS speed,
which is enough to eject the merged BH out of the cluster;
in reality a BBH merger product would typically escape at a much higher
speed\footnote{Alternatively, the LIGO-observed BBH mergers to date may
suggest zero or small spins of the BHs participating in the mergers \citep{Farr_2017}.
In that case, the merger product would likely be retained in open clusters
and GCs, potentially giving rise to a second generation of merged BHs
\citep{Rodriguez_2018}. Multiple-generation BBH mergers
are likely within galactic nuclear clusters irrespective of
the magnitudes of the BHs' spins, where the escape speeds are
large ($\gtrsim100\kmps$). These cases are beyond the scope of
this paper.}.

To accommodate the possibility of a pure KS compact binary becoming relativistic,
say, due to a strong fly-by encounter that imparts a high eccentricity onto it
(see also \citealt{Samsing_2014} in this context),
GR orbital modifications are included in the KS treatment as well. Such encounters,
which are termed here as ``GR slingshot''s, do take place in the model computations
presented here, although relatively rarely (see Sec.~\ref{sling}). However, in these \nbseven calculations,
the two-body GR systems are also treated via the ARC (\opt 11$=-1$ used).
As pointed out in Sec.~\ref{intro}, GR slingshots would mostly give rise to
temporarily-relativistic binaries but are unlikely to lead to GR coalescences,
as found in the model computations presented in Table~\ref{tab1}.

As in Papers I \& II, the physical speed of light has been used in all the newer computations
in this work, except for one particular test computation with
$\mcl(0)\approx1.5\times10^4\Ms$, $\rh(0)\approx1.5$ pc,
$Z=0.001$, and $\fbin(0)\approx0.05$, where $c/100$ is assumed (see Table~\ref{tab1}). 

\subsubsection{Extracting (relativistic) Chain compact subsystems in \nbseven}\label{getsys}

As discussed in Sec.~\ref{intro}, it would be interesting and useful
to probe the compact subsystems in which the GR BBH and, possibly,
other types of compact-binary mergers occur. More generally,
it would be useful to extract the spectrum of GR subsystems that continue to appear (and disappear)
within a model cluster over its evolution; they would collectively
contribute to the total GW output from the cluster and a subset of them
may potentially serve as semi-persistent sources for the LISA and the
Pulsar Timing Array \citep[hereafter PTA;][]{Hobbs_PTA}.
Furthermore, at least for the Galactic and Local Group clusters,
the population of relativistic
subsystems in them may, potentially, contribute to the
GW background noise for the PTA and the LISA. In the present
work, among all the relativistic interactions that occur inside
the model clusters, only the GR coalescences and the GR-slingshot
interactions (see above) will be considered and a broader study
will be presented in a future paper.

The straightest way to determine the innermost configuration of a hierarchical,
compact subsystem, hosting a BBH or another type of compact-binary merger, is to
print the Chain members directly from the \nbseven subroutine
{\tt ARchain/chain.f} with name \chain that contains the
ARC algorithm (called as a separate subroutine
from \chain). Currently, this is done when a Chain, that is passed   
through the \chain routine, contains more than two members (see below),
at least one bound pair, and requires PN modification (as indicated by the PN-order indicator
{\tt IPN}$\geq1$) of the lowest order or above (see above). Distinction is
made between the cases when there is at least one Chain member that is bound
w.r.t. the innermost binary and when all other Chain members are unbound w.r.t.  
it, which are recorded against the N-body time in separate (machine-readable, text) files.
It is the latter record that identifies the GR-slingshot events.
In both cases, the member masses, semi-major-axis, and eccentricity ($m_1$, $m_2$, $a$, $e$)
of the innermost binary are recorded; in the bound case, the nearest bound orbit
w.r.t. the binary ($m_0$, $a_0$, $e_0$) and as well the corresponding KL period, $\tkl$,
are entered into the same record. When all the Chain members are unbound
w.r.t. the binary (see above), information regarding the binary's
closest fly-by member (its mass, stellar type, parameters for the relative
hyperbolic trajectory) is stored instead.

However, this alone is not sufficient since  
the Chain membership of a hierarchical configuration depends on the degree of
its hierarchy. Especially, from the point of view of the difference in dynamical timescales of
the innermost binary and of the outer members gravitationally
bound to it, additional hierarchy is
introduced when the innermost binary becomes relativistic (say, by gaining a high
eccentricity due to its interactions with the outer, bound members), shortening the binary's
orbital-evolutionary timescales due to the associated PN modifications (\eg, GR precession and
GW-radiation shrinking of its orbit). In such cases, even for a nearly
democratic triple (outer-periastron/inner-apoastron $\sim1$),
often, only the two members of the innermost (relativistic) binary
comprise the Chain, and the third member becomes a part of
the binary-Chain's perturbers (there can, of course, be additional
bound and unbound perturbers of the Chain).
As can be seen from such a PN inner binary's perturber outputs (see below),  
the number of its perturbers oscillate between zero and finite (typically, between 0 and 1
close to a coalescence)
over a very small fraction of an N-body time ($\sim$ the cluster's dynamical time),
as the outer, bound member(s) exit and enter the binary's radius of
gravitational influence. In fact, typically, when such a PN binary is marked as
``COALESCENCE'', following one of the coalescence conditions in the \chain routine,
the number of its perturbers is zero despite having bound outer members,
meaning its GR-inspiral timescale (according to orbit-averaged PN-2.5; \citealt{Peters_1964})
has become short enough that it can be
traced unperturbed up to coalescence \citep[see, \eg, Fig.~7 of][]{Banerjee_2017b} before the binary
can again be perturbed significantly by an outer orbiter, justifying the ``COALESCENCE'' tag.
However, although found more rarely in
the present computations, GR coalescence also
happens within a Chain with $>2$ members, in which case the innermost triple/multiple
configuration, hosting the merger, can be obtained directly from the \chain outputs arranged
as described above.

Therefore, in addition to recoding the Chain members as
they are passed through the ARC routine, it is important
to simultaneously scan and record members from the Chain's perturber list.
A new subroutine, currently named {\tt BBHNB}\footnote{As opposed to what the subroutine's name,
which is due to historical reasons, may indicate,
this routine deals with \emph{any} Chain that
is passed through \chain and not just those which contain a BBH
and/or are relativistic.},
which is also
called from within \chain, executes
this task by utilizing the handy {\tt CHAINC} common block that
contains information on the currently-executed Chain's center of mass (hereafter COM) 
and the list of the identities of its perturbers. By following
a similar procedure as in the default routine {\tt GPU2/bhplot.f}\footnote{This
procedure has been partly sketched in the ``Discussion session on
perspectives of NBODY6/7 with Sverre Aarseth'' during the MODEST17
conference in Prague; see \url{ftp://ftp.ast.cam.ac.uk/pub/sverre/discuss.pdf}},
{\tt BBHNB} records (against the N-body and physical times;
in machine-readable, text format) the masses, semi-major-axes,
and eccentricities of up to five nearest (in the order of increasing semi-major-axis;
the nearest one being denoted by $m_0$, $a_0$, $e_0$) bound
perturbers of the Chain. The room for five bound neighbours is kept for
just in case: hierarchies beyond triple are rare throughout the present model computations. 
In the case of a binary-Chain (see above), the mass and
the orbital parameters of the innermost (Chain) binary ($m_1+m_2$, $a$, $e$)
and as well the $\tkl$ w.r.t. the nearest bound perturber (if
at least one bound perturber exists) are entered into the same record.
Since, as found in the models in Table~\ref{tab1}, such computed models can produce
long-lasing, hierarchical triples, care should be taken \emph{not} to
print, particularly the {\tt BBHNB} records, on every call of \chain, to avoid
producing excessively large output files\footnote{This would affect the
print-output cadence from \chain and {\tt BBHNB}, so that a high-time-resolution
recording of the subsystem evolution is currently prevented. Another
parameter affecting the cadence of the subsystem output is the
regularized time step parameter which is taken to be {\tt ETAU}$=0.1$
in the present computations. At present, the output cadence from \chain
and {\tt BBHNB} is increased with the PN order of the innermost Chain binary.}.

\subsubsection{Extracting model snapshots and its binary population}\label{snap}

Apart from tracking compact subsystems, as described above, star-by-star snapshots
of the ``cluster'' part of the stellar system (\ie, of the stellar population within
the instantaneous tidal boundary that is gravitationally self bound but for the on-the-way
escapers) are obtained from the computed evolutionary models (Table~\ref{tab1}) 
in intervals of 10 or 20 Myr. It becomes much more convenient to track
specific stellar or binary type if the cluster data bank, as stored in the \nbseven's
default runtime output file {\tt OUT3} in the binary data format (double precision
for floating variables), is enhanced:
apart from the N-body time, identities, masses, and the Cartesian components
of positions and velocities, the
\bse stellar-evolutionary stage \citep{Hurley_2000} of each star is also entered
in every record (two N-body times apart).
Each such snapshot is then searched for the closest gravitationally-bound
pairs to retrieve the binaries (the members' identities, masses,
stellar types, derived orbital parameters, COM coordinates
and velocities) at the corresponding evolutionary time.
That way, the recovery of the binary population does not depend on the chosen KS
regularization parameters ({\tt RMIN} and {\tt DTMIN}) and, in principle,
all binaries, including those at the hard-soft limit \citep{2003gmbp.book.....H},
are obtained.
Utilizing the stellar types, the binaries in
a given snapshot can then be sorted for specific types such as
star-star, BBH, BH/NS-star, BH/NS-MS, etc and utilizing the identity values the evolution
of a given pair can be followed over the snapshots, through machine reading.

\begin{figure*}
\centering
\includegraphics[width=14.0cm,angle=0]{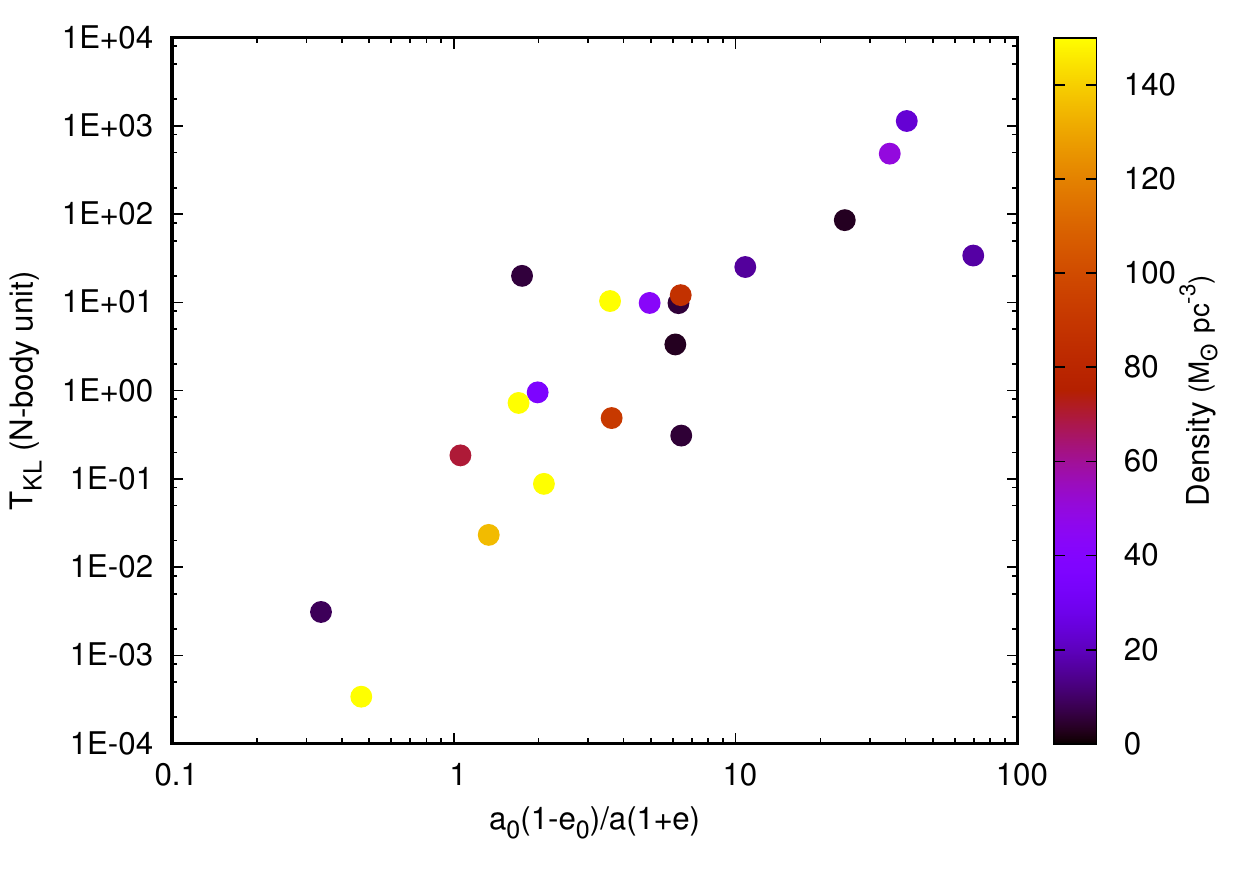}
	\caption{The configurations of the (innermost) triple subsystems hosting (a subset of) the GR BBH 
	(and other types of compact-binary; see Sec.~\ref{bbhmrg}) coalescences,
	that occurred inside the computed cluster models in Table~\ref{tab1}. The abscissa represents the
	ratio of the outer periastron to the inner apoastron of the triple, which measures the extent of hierarchy
	in the triple configuration. The ordinate represents the Kozai-Lidov timescales (N-body
	unit), $\tkl$, of these triples. These triple configurations correspond to
	times very close to the commencement of the GW inspirals of the inner binary
	(see Secs.~\ref{getsys} \& \ref{bbhmrg}).
	The colour coding (the colour bar) represents the
	mean density within the innermost 10\% Lagrangian radius of the host cluster at the
	time of the coalescence, representing the environment in which the merger event has taken place.}
\label{fig:stabplot}
\end{figure*}

\begin{figure*}
\centering
\includegraphics[width=12.0cm,angle=0]{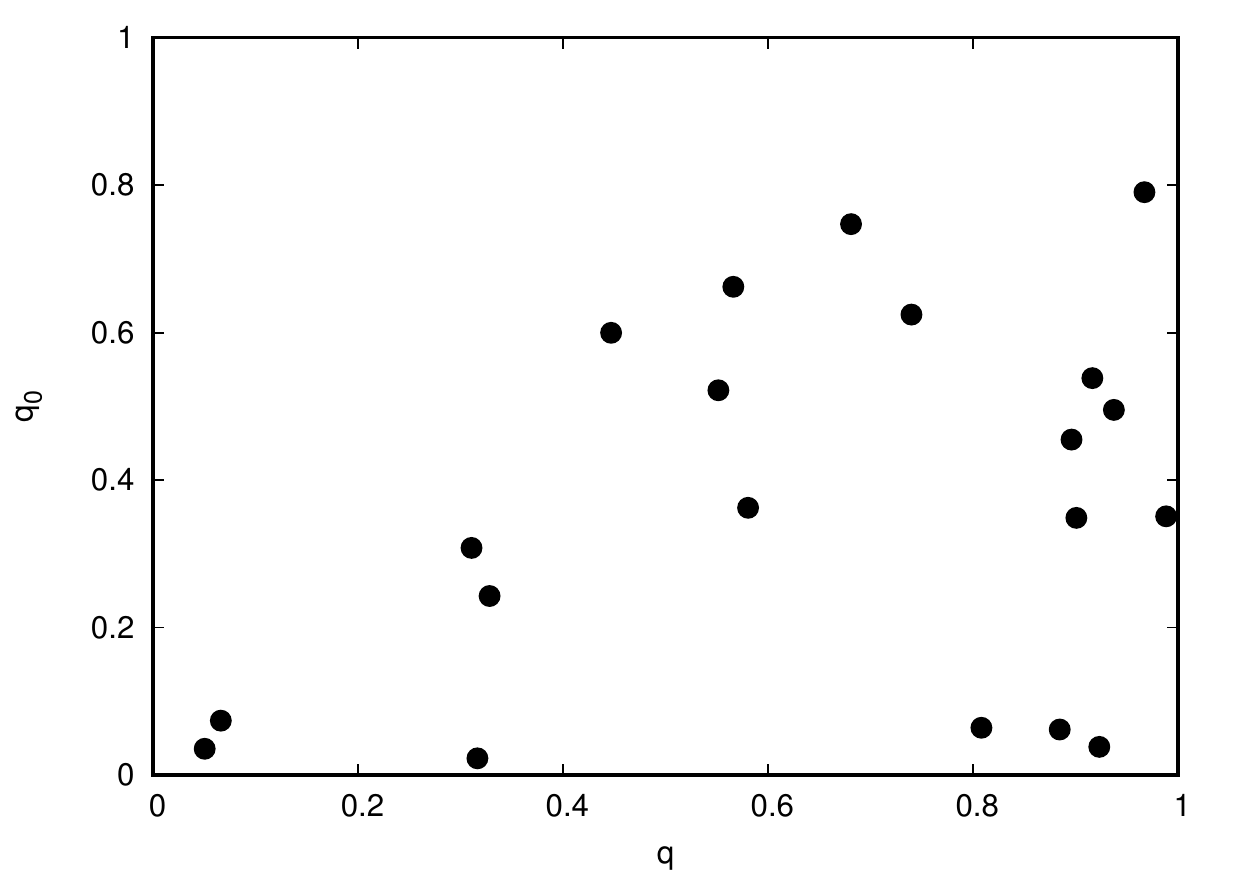}
\caption{The mass ratios corresponding to the triple-hosted GR coalescences that are
shown in Fig.~\ref{fig:stabplot}.
The mass ratio, $q$, of the inner, merged binary is plotted along the abscissa and that of the
outer binary, $q_0$, between the inner binary's total mass and the outer member's mass,
is plotted along the ordinate.}
\label{fig:stabplot_ratio}
\end{figure*}

\section{Results}\label{res}

In this section, various aspects of the model computations in Table~\ref{tab1}
are described.

\subsection{In-cluster, general-relativistic coalescences driven by compact subsystems}\label{bbhmrg}

As already shown in Papers I \& II and found true as well
for the broader set of models evolved here, the GR coalescences in young massive- and
open-range clusters take place primarily inside the clusters, driven
by the short-timescale, few-$N$ dynamics of dynamically-formed compact subsystems (see Sec.~\ref{intro})
that are composed of one or more of the retained stellar remnants, typically of BHs. 
The newly arranged live tracking system of Chain subsystems, as described in Sec.~\ref{getsys}, facilitates
the recovery of the subsystems hosting the GR coalescences.

Fig.~\ref{fig:stabplot} depicts the configurations of (a subset of
\footnote{The developments in the \chain routine, as described in Sec.~\ref{getsys},
have been implemented over time, in parallel to ongoing computations. The
latest developments were available only to the most recent runs which explored the models with
relatively low $\mcl(0)$ and high $\fbin(0)$. In the runs where both the direct \chain and the
{\tt BBHNB} outputs are available, making detailed extraction of the Chain-subsystem
configurations possible, the corresponding entry in the $\nsling$ column in Table~\ref{tab1}
shows a value, otherwise a `-' is shown. During the $\mcl(0)\approx5.0\times10^4\Ms$ models'
computations, only a part of the direct \chain output arrangement was available, allowing the
recovery of 4 of the triple configurations hosting GR mergers, from these models. One of the
triple configuration, very close to the merger, could be obtained from the
default Kozai output data. These 5 points are also included in
Figs.~\ref{fig:stabplot} \& \ref{fig:stabplot_ratio}. The rest of the runs in Table~\ref{tab1},
although couldn't be utilized for the purposes of the Secs.~\ref{bbhmrg} \& \ref{sling},
are still utilized for the study of BH---non-BH binaries presented here (Sec.~\ref{bhmspop}).})
the triple subsystems,
formed dynamically in the computed models, in which the
inner compact binary has undergone a GW-radiation inspiral and merger. The horizontal
axis of Fig.~\ref{fig:stabplot} represents the hierarchy of the triple, as measured
by the ratio, ${\mathcal R}\equiv a_0(1-e_0)/a(1+e)$, of the periastron of the outer orbit to the apoastron of
the inner orbit and the vertical axis represents the triple's secular KL time period, $\tkl$ (N-body units),
given by \citep{Kiseleva_1998} 
\begin{equation}
\tkl=\frac{2P_0^2}{3\pi P}(1-e_0^2)^{3/2}\frac{m_1+m_2+m_0}{m_0},
\label{eq:kozai}
\end{equation}
where $P$ and $P_0$ are the Keplerian periods of the triple's inner and outer orbits respectively.
Note that the configurations in Fig.~\ref{fig:stabplot} correspond to a time very close, $\lesssim10^{-4}$ N-body
times, to the commencement of the GW inspiral of the inner binary (see Sec.~\ref{getsys}).
Therefore, the inner eccentricity, $e$, used in the definition of ${\mathcal R}$ is the
maximum inner eccentricity reached by the triple before the commencement of the inspiral. 

Given that the majority of the configurations in Fig.~\ref{fig:stabplot} have
$1\lesssim{\mathcal R}\lesssim10$ (full range $10^{-1}\lesssim{\mathcal R}\lesssim10^2$),
the triples, that ultimately drive the inner binary
into coalescence, typically range from being marginally stable to stable. The
typical range of their secular time period is $10^{-2}\lesssim\tkl\lesssim10$ N-body unit
(full range $10^{-4}\lesssim\tkl\lesssim10^3$ N-body unit) and there is an overall
positive trend between ${\mathcal R}$ and $\tkl$. 
However, except for a few of those with the shortest $\tkl$s,
these triples typically appear and remain in the clusters for time spans
much shorter than their respective $\tkl$s, before the merger of the inner
binary. Therefore, essentially all of these inspirals are triggered
via chaotic (non-secular) interactions between the innermost (PN) binary and its
(nearest) outer companion, that results in a rapid
eccentricity growth of the binary (Sec.~\ref{intro}),
irrespective of the extent of the hierarchy (${\mathcal R}$);
see also \citet{Antonini_2016b} in this context.
Based on the \chain and {\tt BBHNB} outputs (Sec.~\ref{getsys}),
the typical duration of the
``final'' triple that has led to a merger is a fraction of an N-body time (the dynamical
time of the cluster). As can be expected, given the moderate to low velocity dispersion
of the cluster environment and the fact that a triple, often composed
of BHs (see below), would be the most massive compact-subsystem member in the cluster,
the triple would be susceptible to significant alteration or destruction due
to its dominant gravitational focusing. Therefore, it is unlikely
to have the chance of undergoing many cycles of secular oscillations \citep{Katz_2011,Lithwick_2011},
and any dramatic outcome (\eg, GR merger, high-velocity breakup), as a result of the triple evolution, would
happen only due to its chaotic part (as obtained via a full 3-body integration such as the ARC;
see Sec.~\ref{nbprog}). Like the secular Kozai-driven mergers \citep[\eg,][]{Liu_2017,Antonini_2017a},
one may plausibly expect that the mergers driven by such chaotic triple interactions
would also exhibit spin-orbit misalignment, as hinted in at least one of the
LIGO detections \citep{Abbott_GW170104,Farr_2017}, although it deserves a dedicated study.

In fact, in the few cases of relatively long-term (at least for a few N-body
times) Chains leading to a merger, it is also ambiguous to
identify a single triple leading to the event. In these cases, a given inner
binary have had its outer companion exchanged multiple times before its merger.
Well before the merger, such binaries would also have more than one perturbers bound
to it; although \emph{rarely}, this is true for the short-term ($<1$ N-body time) 
merger Chains too. However, except for one (which turns out to be a hierarchical quadruple),
the final configuration of all the
mergers shown in Fig.~\ref{fig:stabplot} is a triple, \ie, the merging binary
has only one orbiter. In a few cases (see below), this final outer member is actually a binary, \ie,
the final configuration hosting the merger is actually a binary-binary quadruple.
Only one of the mergers (not shown in Fig.~\ref{fig:stabplot})
happened following a GR slingshot (see Sec.~\ref{nbprog}, below). 
Clearly, a wide variety of dynamical interactions inside open clusters, involving a wide variety of
compact subsystems (also, see below), lead to GR coalescences in them, which, perhaps, deserves a classification.
This will be undertaken in a future study with a larger collection of merger events.

The colour coding in Fig.~\ref{fig:stabplot} (colour bar), which is according to the mean density
within the innermost 10\% Lagrangian radius of the host cluster at the time of the coalescence,
represents the stellar density of the environment in which the merger has taken place.
The qualitative trend in colour indicates that in higher-density environments,
the GR mergers tend to take place in tighter (less hierarchical) triples: such tighter
triples may exist in lower-density environments too but the reverse is unlikely.
This is expected since a more crowded environment is more likely to clobber a triple
with a looser outer member.

Fig.~\ref{fig:stabplot_ratio} shows the inner-binary vs. outer-binary mass ratio ($q$ and
$q_0$ respectively) of the mergers shown in Fig.~\ref{fig:stabplot}. Most of the
mergers are of (inner) BBHs having $0.2\lesssim q\lesssim1.0$ with the majority
of them having $q\gtrsim0.5$, as can be expected for mergers of dynamically-assembled
BBHs (\cf Fig.~6 of Paper II). The two mergers with $q<0.1$ in Fig.~\ref{fig:stabplot_ratio} 
are BH-NS and BH-white dwarf (hereafter WD); although rare, such BH---non-BH GW-inspiral mergers do occur in these model
computations which events would be accompanied by electromagnetic signatures \citep{Abbott_GW170817a}
\footnote{Among the merger events shown in Figs.~\ref{fig:stabplot} \& \ref{fig:stabplot_ratio},
the BH-NS one has happened in the computation with the speed of light $c/100$ (Table~\ref{tab1}).
However, the BH-WD merger here and also the BH-NS merger identified in one of
the models in Paper II (not shown here due to the unavailability of the subsystem data;
see above, Table~\ref{tab1})
have occurred in models where the physical speed of light has been used.}. The
outer members for these two mergers are also WDs (also, see below), as indicated by their small ($q_0<0.1$)
values.

The outer members corresponding to most of the mergers in Fig.~\ref{fig:stabplot_ratio}
are also BHs, making their $q_0$s spread around 0.5. The ones with $q_0\approx0.8$
have a BBH, instead, as the outer member, \ie, the subsystem configuration, just before
the merger, is a BBH-BBH quadruple. The few mergers with $q_0<0.1$ have a non-BH
outer member.

\begin{figure*}
\centering
\includegraphics[width=14.0cm,angle=0]{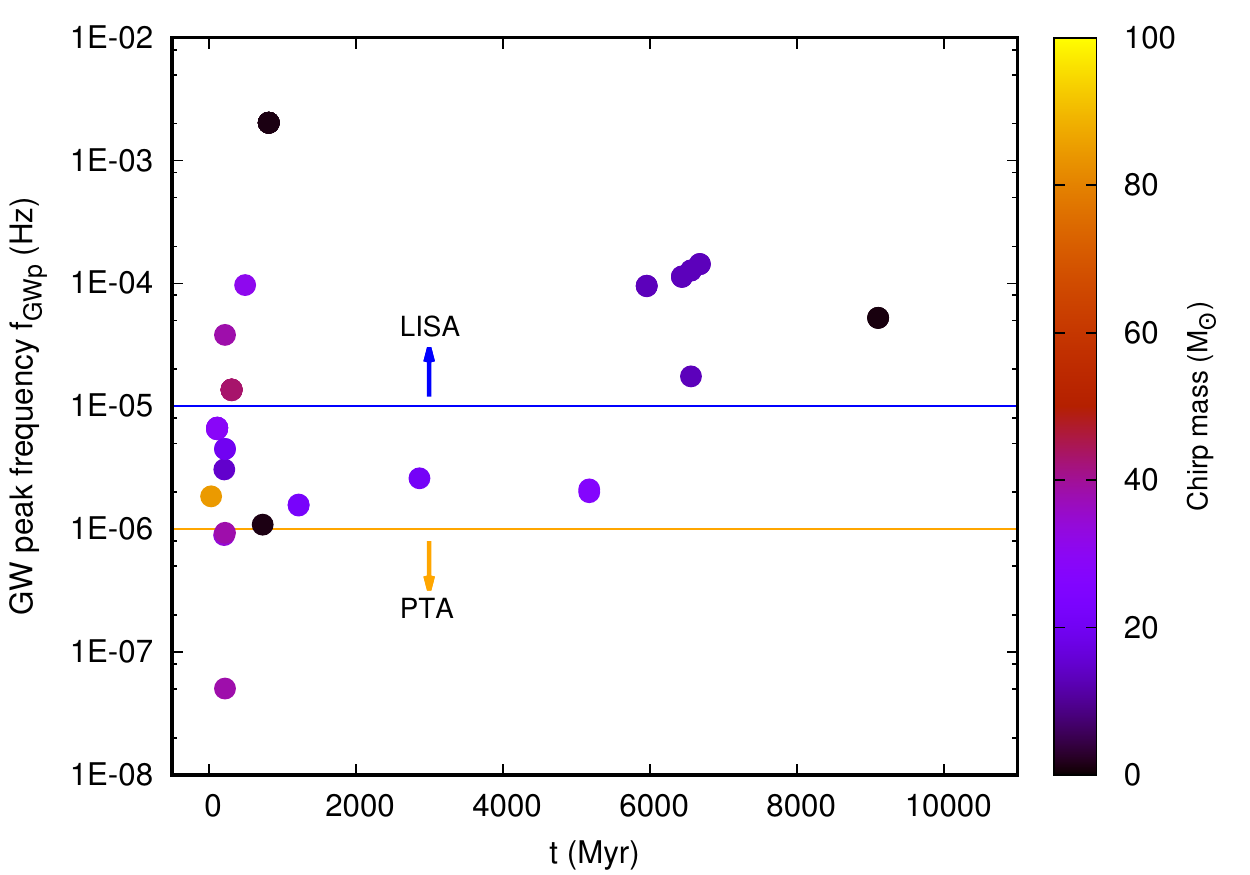}
	\caption{The peak-power GW frequency, $\fgwp$ (ordinate), during the ``GR slingshot''
	interactions recorded by (a subset of) the computed models (Table~\ref{tab1}). The
	abscissa represents the host-cluster-evolutionary time, w.r.t. a common
	$t=0$, of the occurrence of these interactions and the colour coding represents the
	chirp mass, $\mchirp$, of these ``temporarily-relativistic'' binaries. Although
	the majority of such binaries are BBHs, some of them are also BH-non-BH and
	non-BH-non-BH. The characteristic frequency bands for the LISA and the PTA are
	indicated; although such interactions did not result in complete GR inspirals,
	they would still potentially give rise to temporary sources for the LISA.}
\label{fig:sling}
\end{figure*}

\subsection{General-relativistic ``slingshot'' interactions inside clusters}\label{sling}

As already pointed out in Sec.~\ref{nbprog}, temporarily PN binaries are found to appear
\emph{inside} the computed models due to close fly-by interactions between a compact binary and
a single object. These binaries, although unlikely
to undergo GW inspiral (see Sec.~\ref{intro}), would nevertheless be interesting
as they, among others (\eg, temporarily-relativistic triples),
would potentially contribute to the low-frequency GW noise flux
from the cluster. Fig.~\ref{fig:sling} shows such GR slingshots which
occurred inside a subset of the computed model clusters in Table~\ref{tab1}.
As recovered from the direct \chain output (Sec.~\ref{getsys}),
these are Chains containing only one PN binary and the rest of the Chain's members (typically one)
are unbound w.r.t. it, indicating a close, fly-by encounter with the binary that has imparted a high
eccentricity onto it, making it PN. In Fig.~\ref{fig:sling}, the peak-power frequency of
the broadband GW emission of such eccentric binaries, as given by the expression \citep{Wen_2003}
\begin{equation}
\fgwp=\frac{\sqrt{G(m_1+m_2)}}{\pi}
	\frac{(1+e)^{1.1954}}{\left[a(1-e^2)\right]^{1.5}},
\label{eq:gwfreq}
\end{equation}
is plotted against the host cluster's evolutionary time. This reasonably assumes
that the binary passes through its periastron at least a few times before it is
altered again by the next significant encounter.

As the corresponding chirp masses (color coding in Fig.~\ref{fig:sling}) defined as
\begin{equation}
\mchirp\equiv\frac{(m_1m_2)^{3/5}}{(m_1+m_2)^{1/5}}
\label{eq:chirp}
\end{equation}
indicate, the majority of these GR-slingshot binaries are BBHs although a few
of them are WD-WD, BH-WD, and BH-MS. Note that such close, fly-by, binary-single
interactions would show up in \chain around their closest separations, essentially
extracting the configuration of the participants of the interaction over the small time interval during which
the slingshot happens, which is what are
plotted in Fig.~\ref{fig:sling}. The binary can thereafter be followed from the direct \chain
outputs: a more elaborate study of these GR-slingshot binaries plus other instances
of temporarily-relativistic systems that appear within the clusters
will be undertaken in a forthcoming work.

As seen in Fig.~\ref{fig:sling}, around half of the detected slingshot binaries
fall in the LISA's characteristic detection-frequency band, which would then, in principle,
serve as persistent sources for the LISA.
The rest of them, here, primarily fall in the gap between the characteristic
detection bands of the LISA and the PTA with a few but one of them marginally
entering the PTA band. Note that the GR-slingshot binaries are, generally, highly eccentric
($1-e<0.1$), that would make their ``proper'' detection including their orbital parameters
questionable. However, they would still contribute to the GW background
for these instruments.

The GR slingshots tend to occur more frequently over
$t\lesssim1$ Gyr cluster-evolutionary time
due to the generally larger number of BHs present in the clusters
over their initial evolution. 

\subsection{The ambient binary black hole population inside clusters}\label{bbhpop}

BBHs continue to form, dissociate, and get ejected via dynamical interactions from within the clusters'
innermost regions where the density of the BH population is the highest due to their segregation via
dynamical friction \citep{1993Natur.364..421K,1993Natur.364..423S}.
The two main dynamical channels of BBH formation are (a) binary formation via 3-body gravitational
interactions \citep{2003gmbp.book.....H} among single BHs and (b) multiple exchange interactions of single BHs
with stellar binaries. In the presence of a substantial population of primordial binaries, 
a population BBHs derived from the (dynamically-modified) evolution of massive primordial binaries
can also be expected inside the clusters. 

\begin{figure*}
\centering
\includegraphics[width=5.94cm,angle=0]{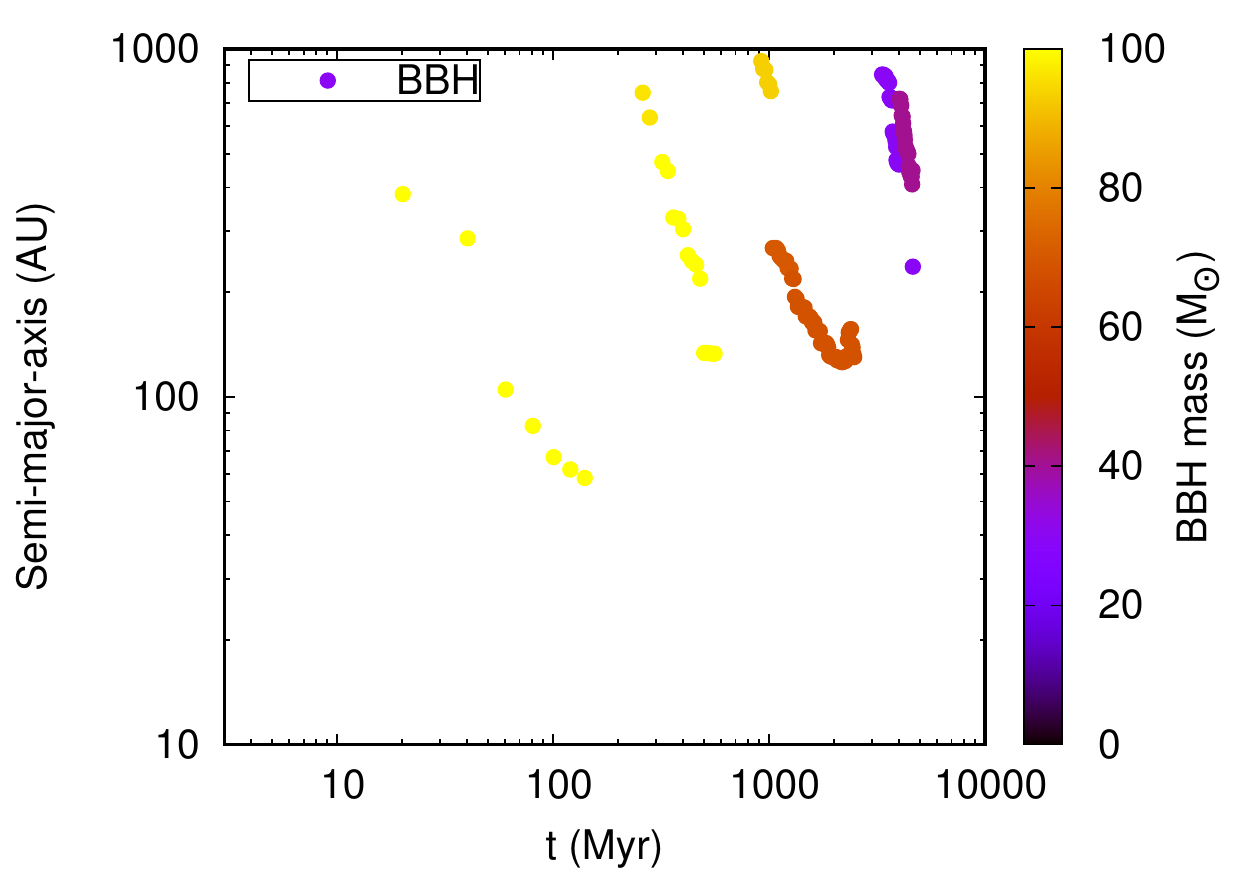}
\hspace{-0.25cm}
\includegraphics[width=5.94cm,angle=0]{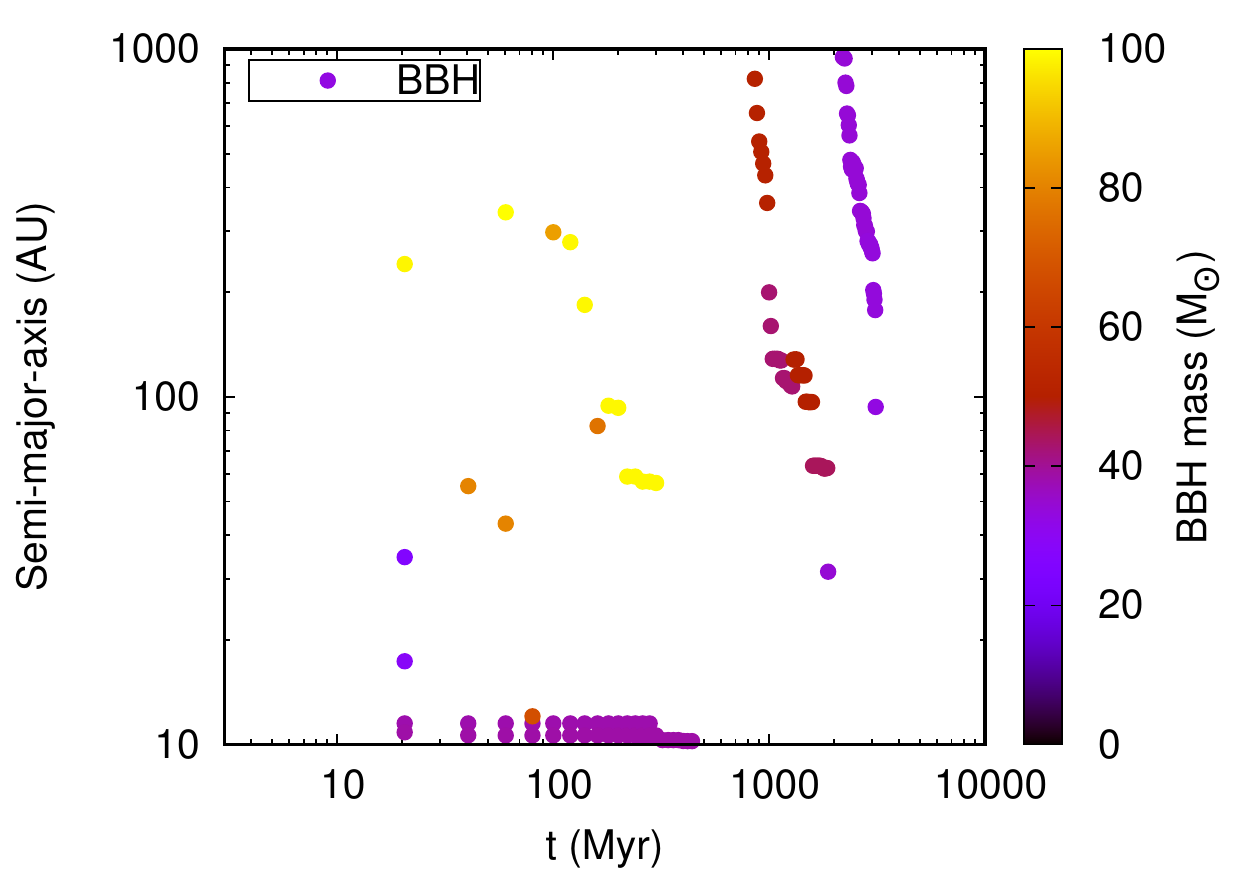}
\hspace{-0.25cm}
\includegraphics[width=5.94cm,angle=0]{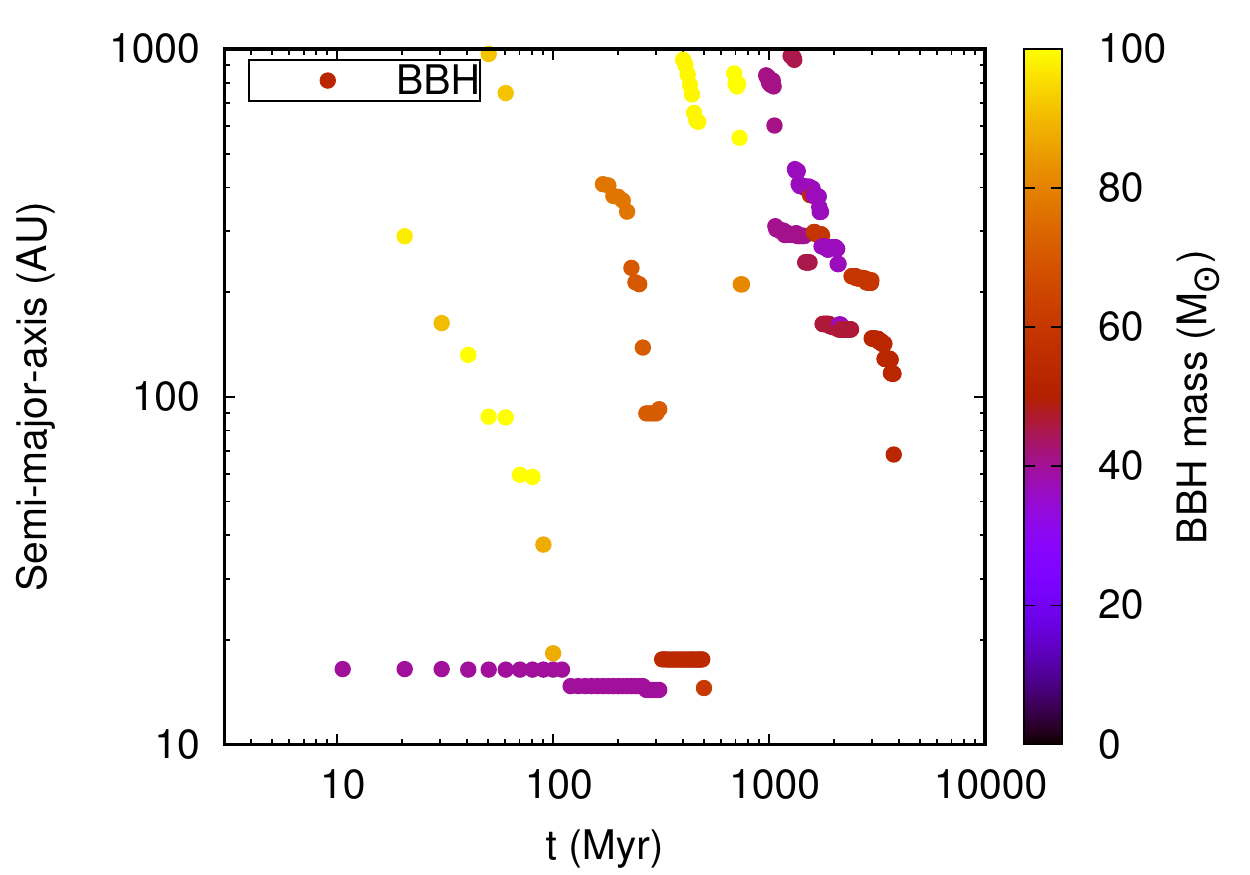}\\
\includegraphics[width=5.94cm,angle=0]{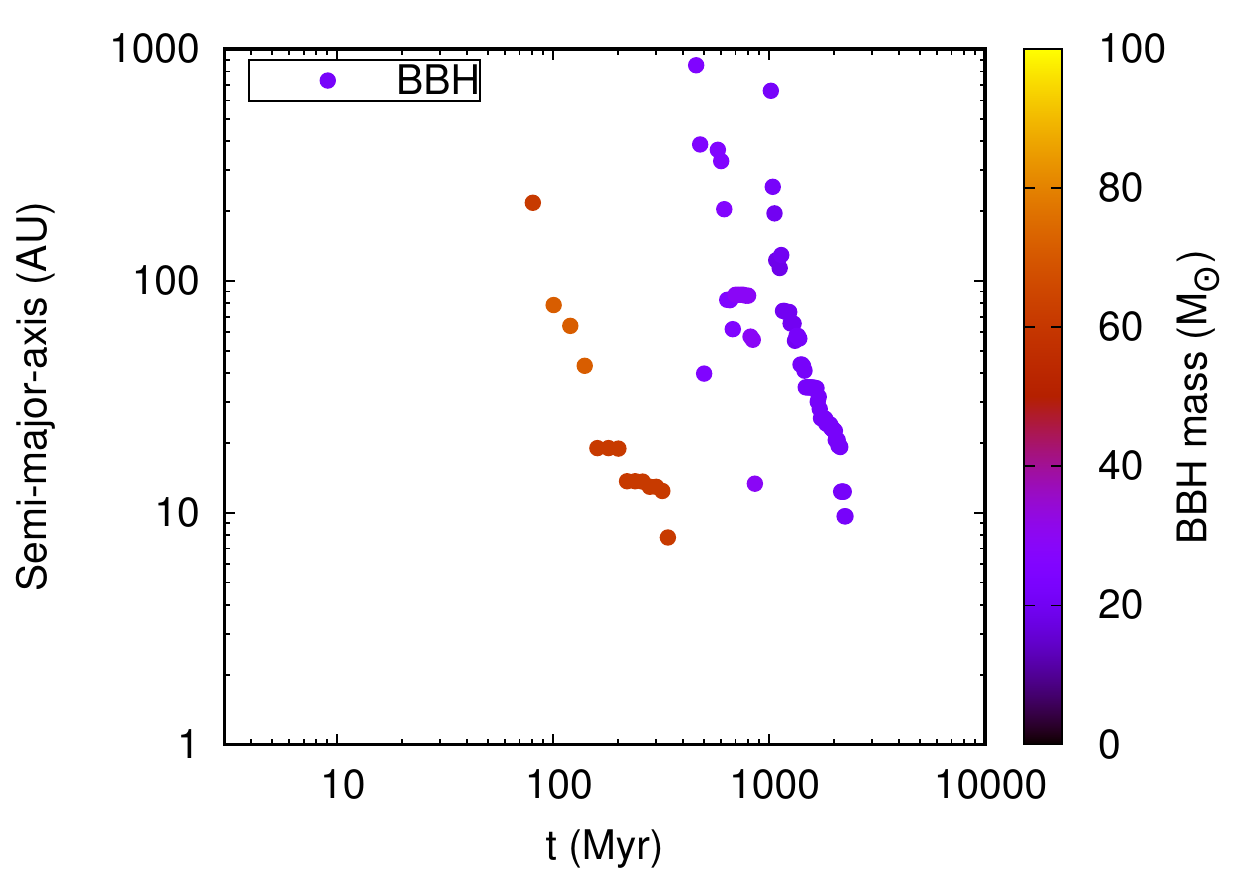}
\hspace{-0.25cm}
\includegraphics[width=5.94cm,angle=0]{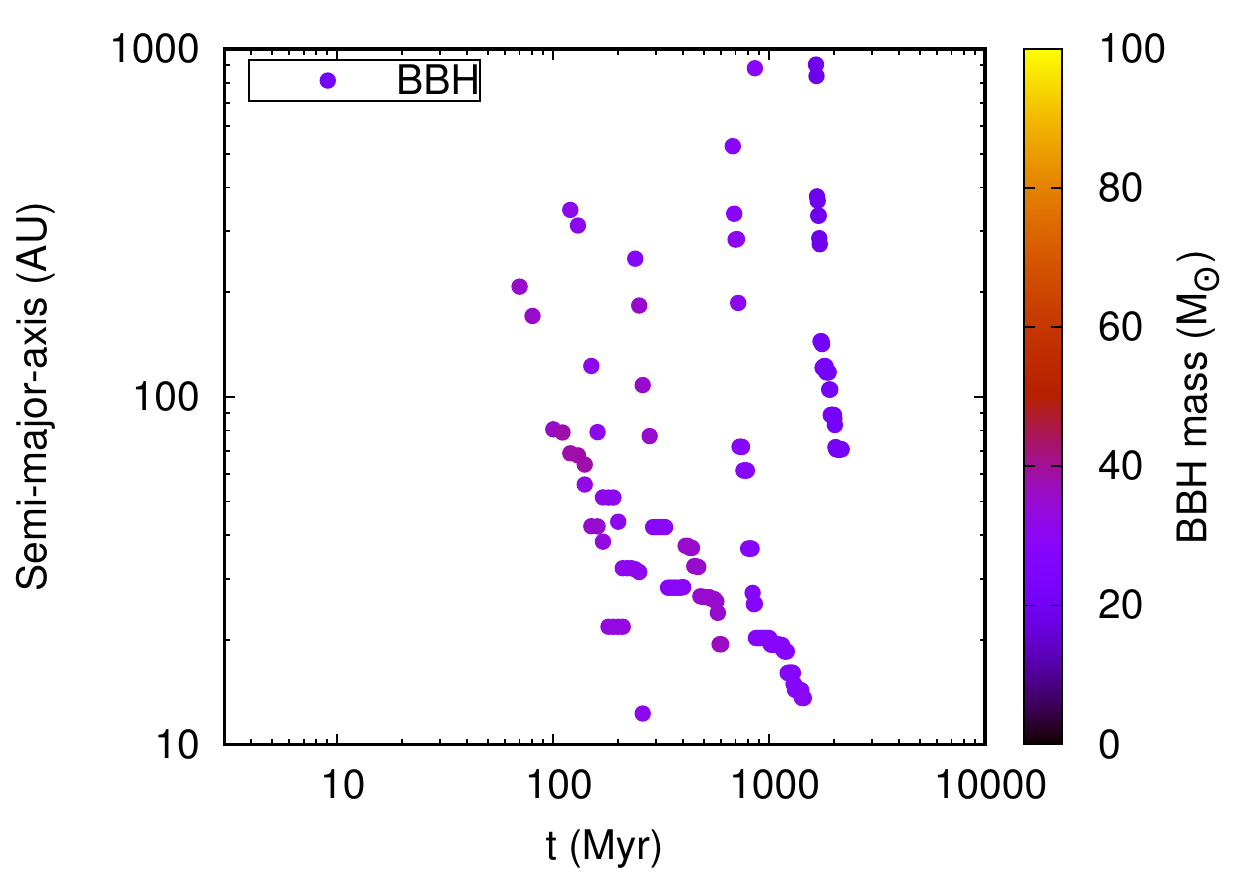}
\hspace{-0.25cm}
\includegraphics[width=5.94cm,angle=0]{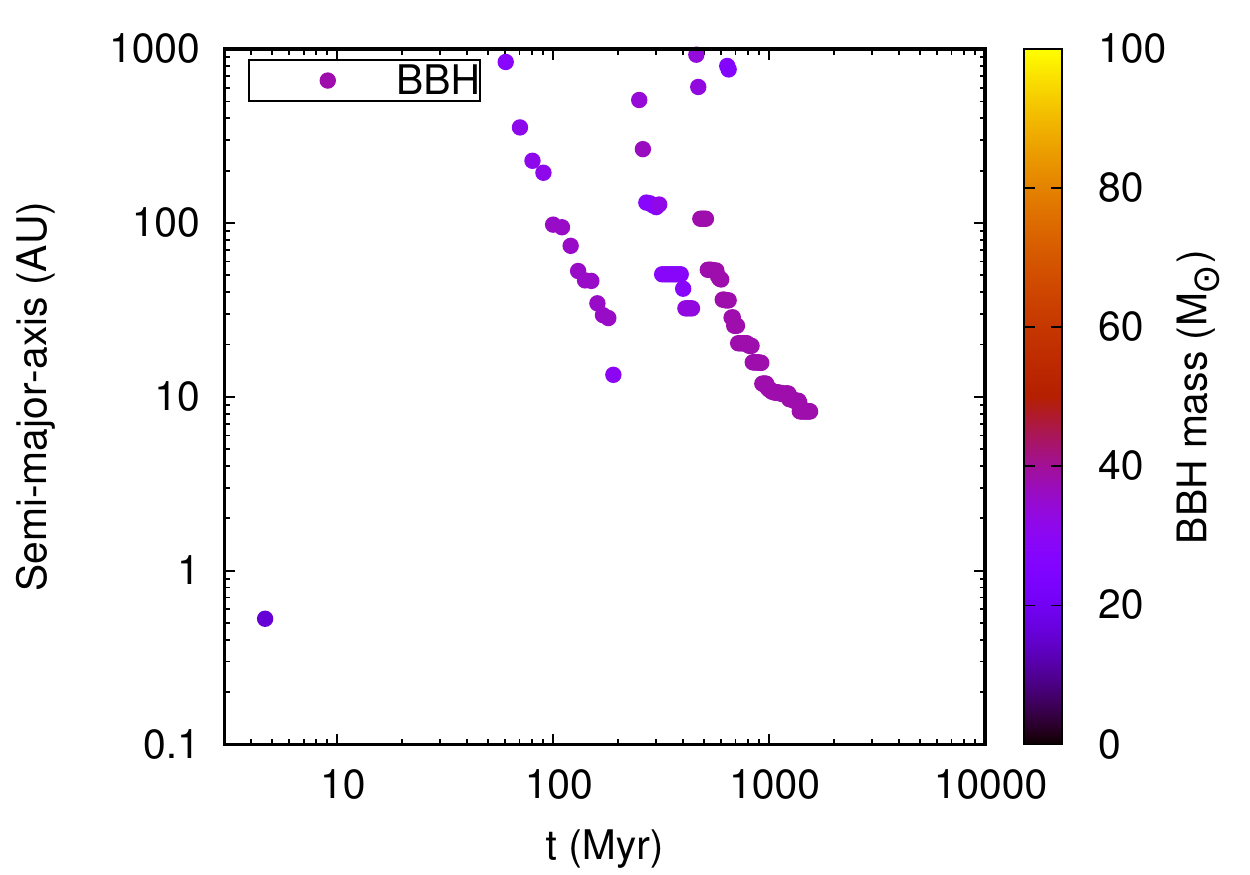}
	\caption{Binary black holes (BBH; filled circles) in the computed $\mcl(0)\approx1.5\times10^4\Ms$
	models with evolutionary time,
	$t$, as a function of increasing primordial binary fraction,
	$\fbin(0)\approx0.05,0.30,0.50$ (left to right panels), and
	for metallicities $Z=0.001$ (top panel) and $0.02$ (bottom panel). On each panel,
	that represents a particular computed model, the vertical axis
	is the instantaneous semi-major-axes ($<1000$ AU) of the BBHs
	that are bound to the cluster snapshot at time $t$. 
        A continuous trail of points represents a specific BH-BH pair.
	The colour coding (colour bars) represents the total BBH mass. In this and
	the following Figs.~\ref{fig:bbh_30k}, \ref{fig:bbh_f05}, \ref{fig:bbh_f50},
	\ref{fig:bbh_ecc}, \ref{fig:bbh_rad}, and \ref{fig:bbhcnt}, the ``test'' runs in Table~\ref{tab1},
	that include tidal interaction (Sec.~\ref{stellrem}), are excluded,
	which are presented separately in Fig.~\ref{fig:bhbin_circ}.}
\label{fig:bbh_15k}
\end{figure*}

\begin{figure*}
\includegraphics[width=8.0cm,angle=0]{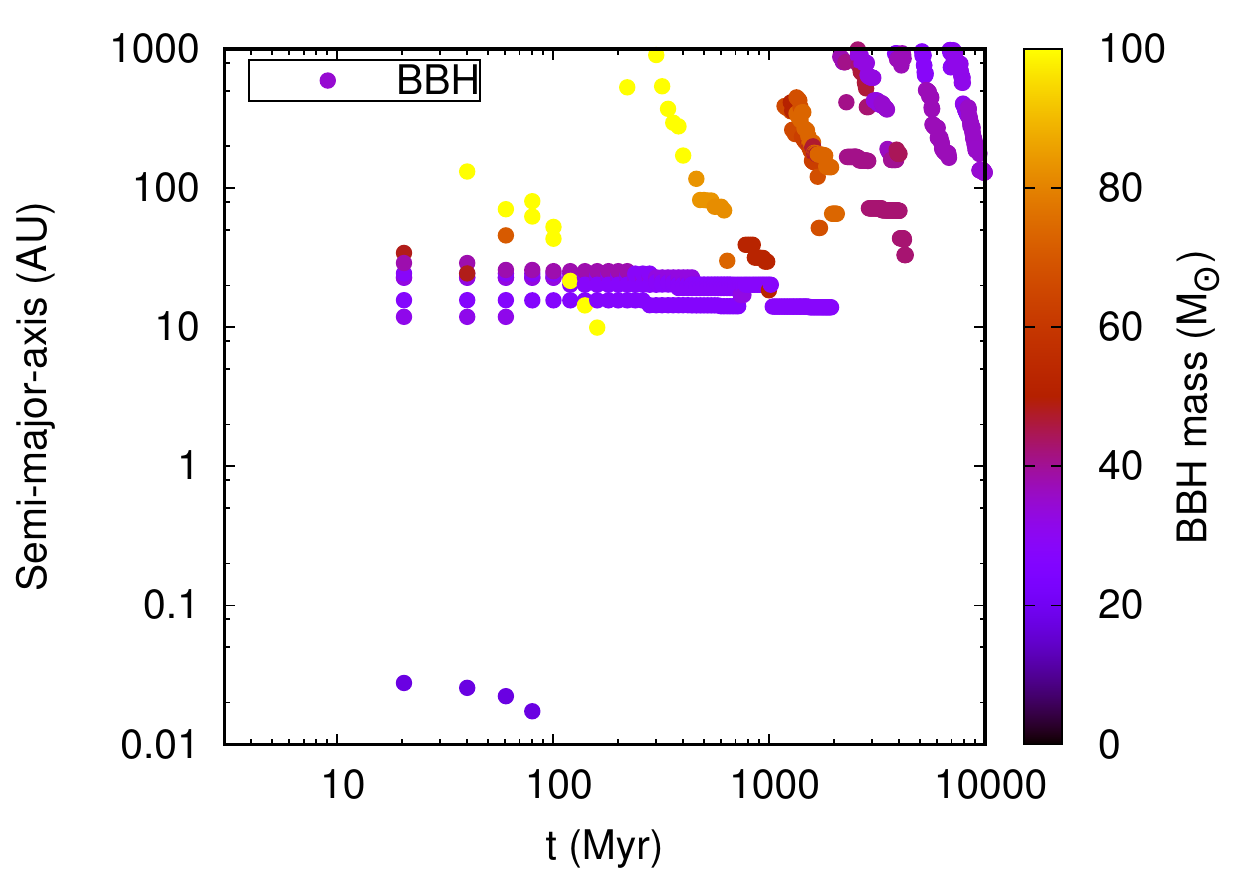}
\includegraphics[width=8.0cm,angle=0]{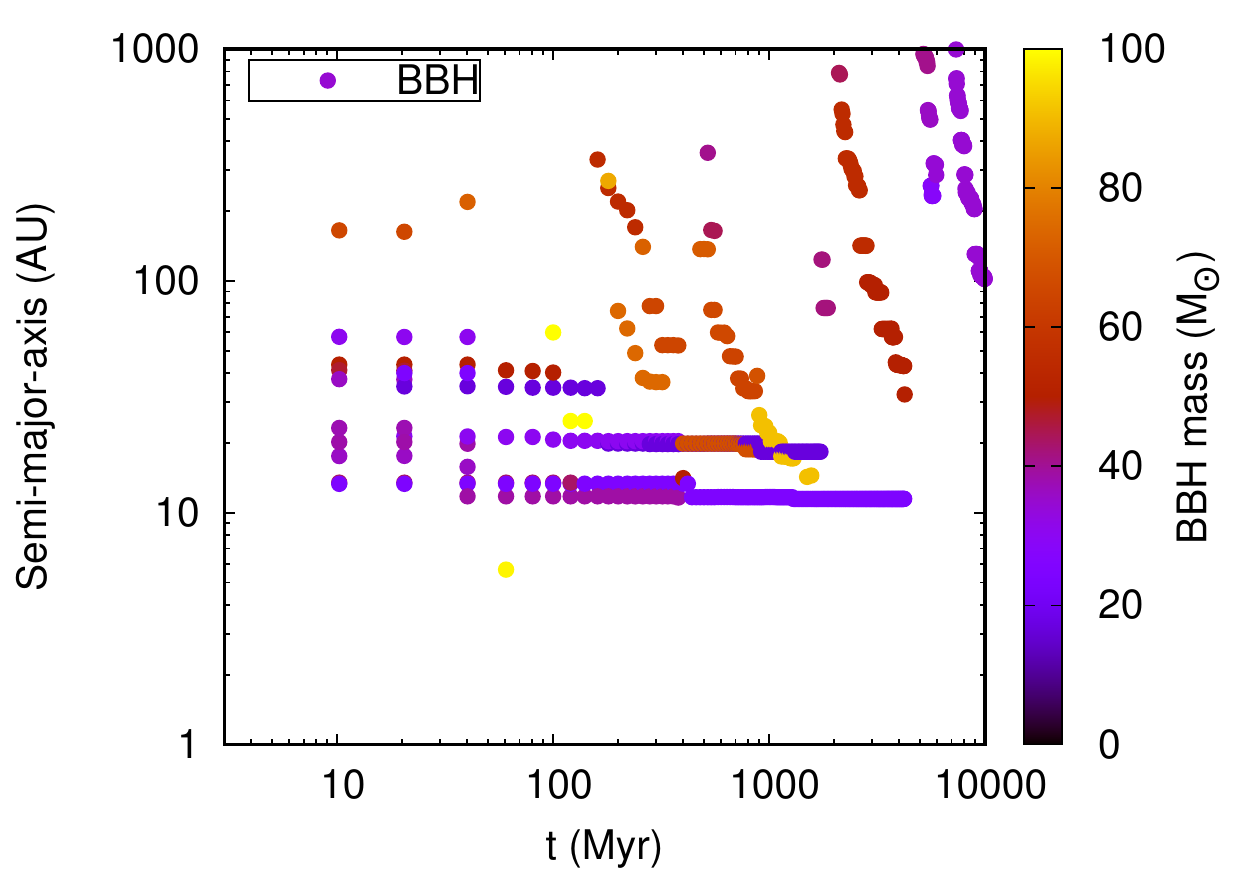}\\
\includegraphics[width=8.0cm,angle=0]{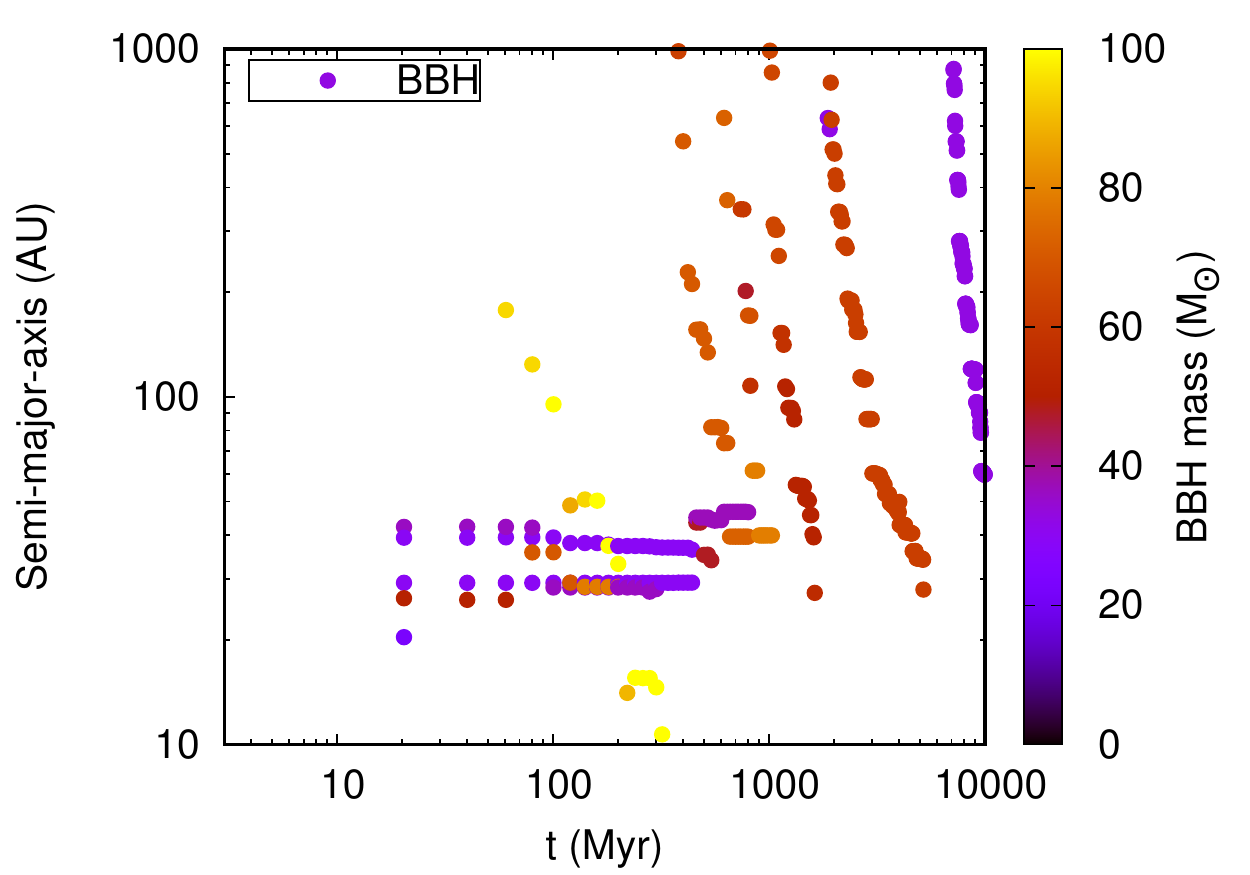}
\includegraphics[width=8.0cm,angle=0]{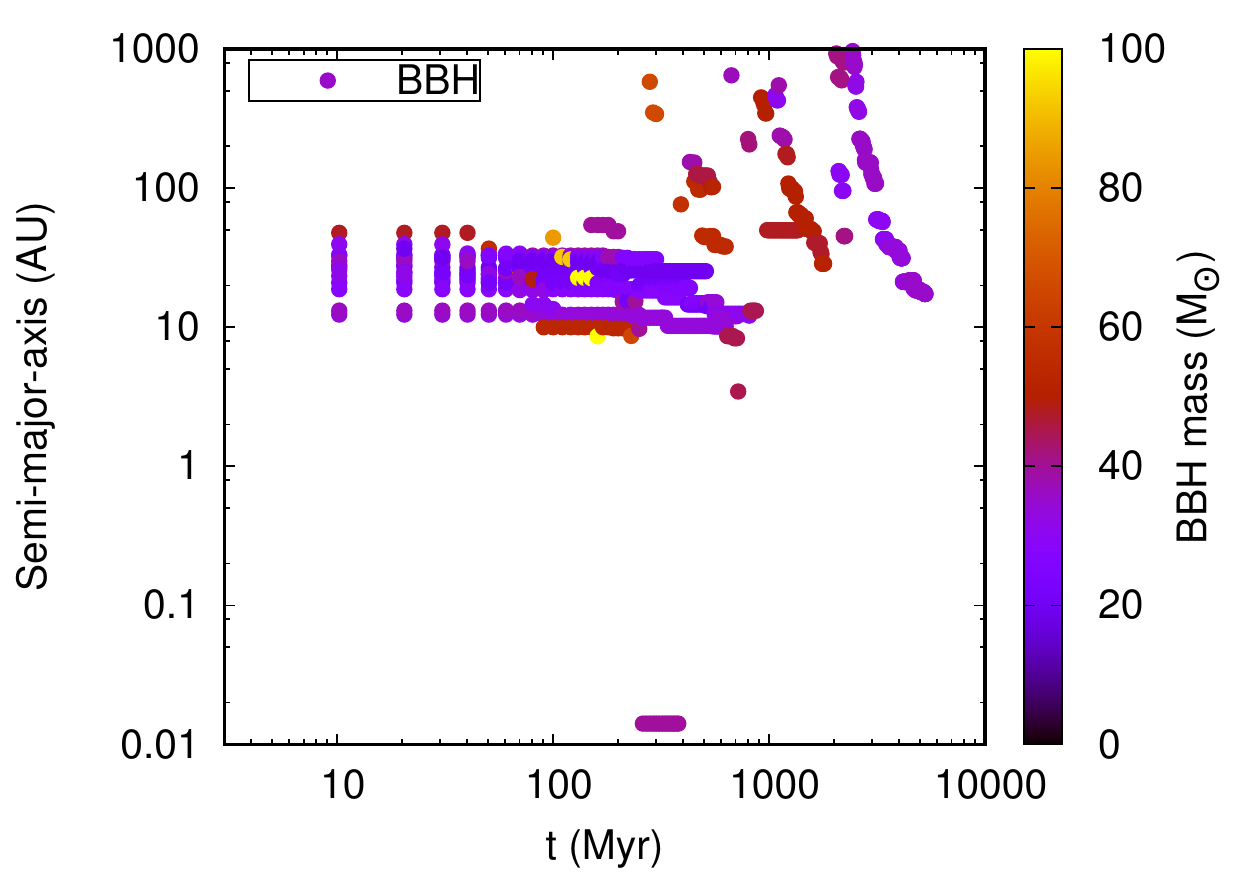}\\
\includegraphics[width=8.0cm,angle=0]{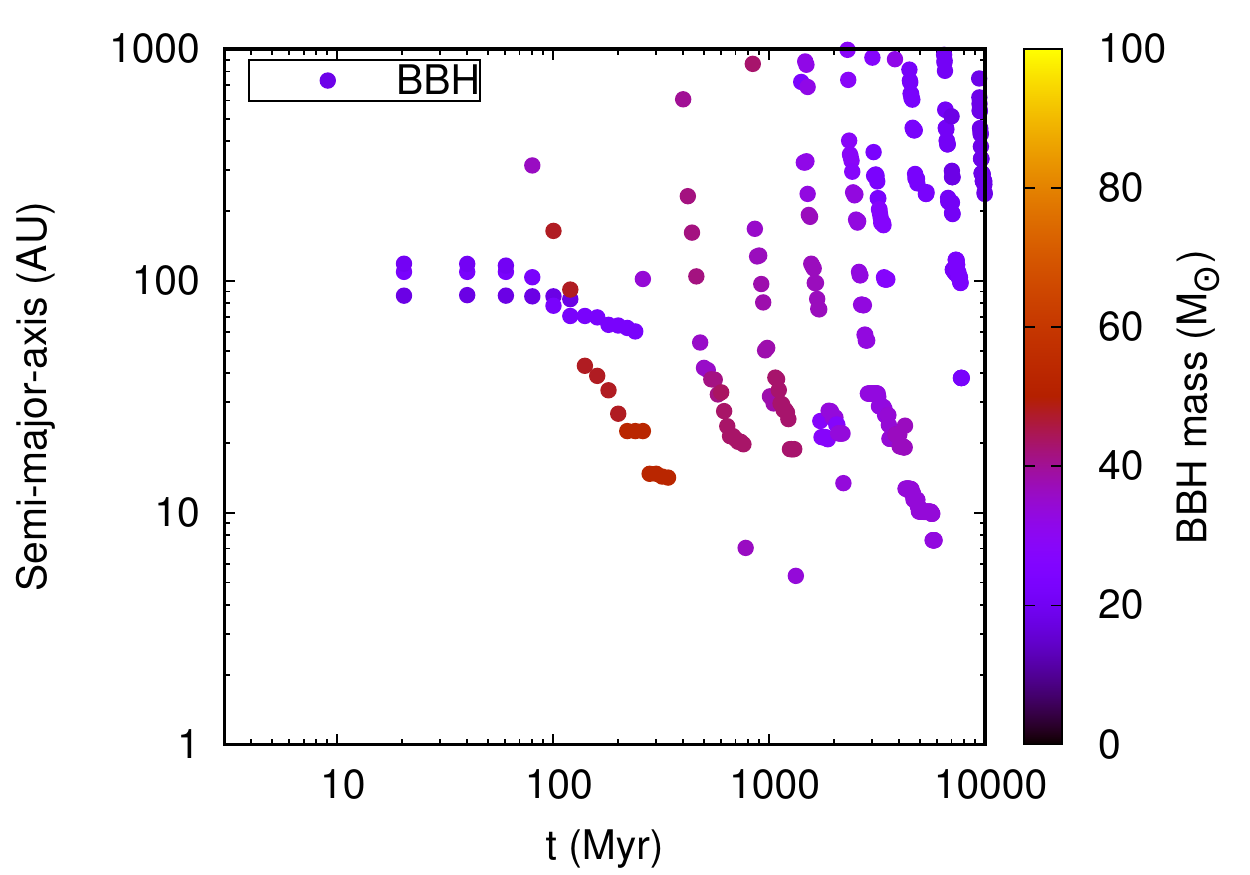}
\hspace{8.0cm}
\caption{The BBH content in the computed $\mcl(0)\approx3.0\times10^4\Ms$ models with evolutionary time,
	$t$, as functions of increasing primordial binary fraction, $\fbin(0)\approx0.05$ (left), $0.10$ (right)
	and metallicity $Z=0.001$ (top), $0.005$ (middle), and $0.02$ (bottom).
	The legends are the same as in Fig.~\ref{fig:bbh_15k}.}
\label{fig:bbh_30k}
\end{figure*}

\begin{figure*}
\centering
\includegraphics[width=5.94cm,angle=0]{figures/BHbin_setC/BBH_15kZ001f05_v2.pdf}
\hspace{-0.25cm}
\includegraphics[width=5.94cm,angle=0]{figures/BHbin_setC/BBH_30kZ001f05_v2.pdf}
\hspace{-0.25cm}
\includegraphics[width=5.94cm,angle=0]{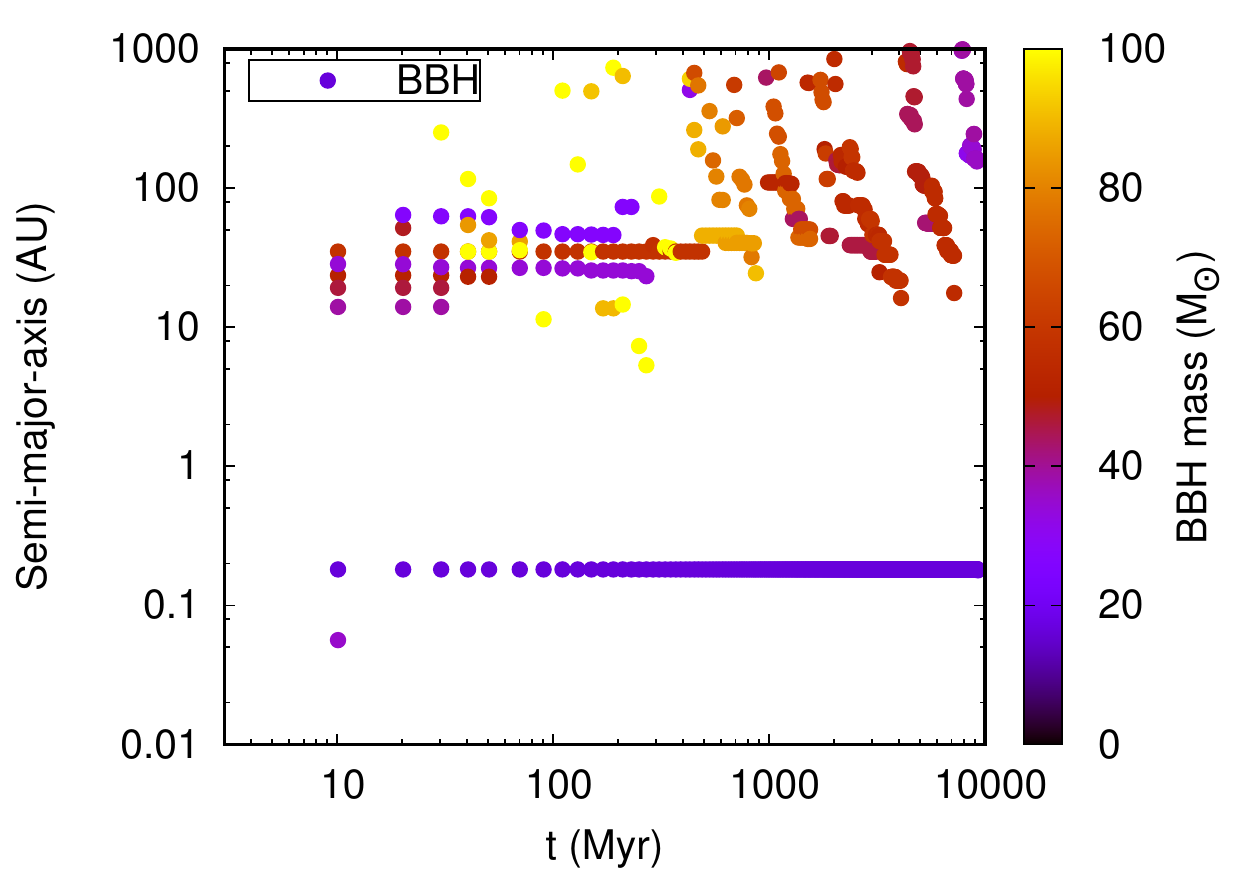}\\
\includegraphics[width=5.94cm,angle=0]{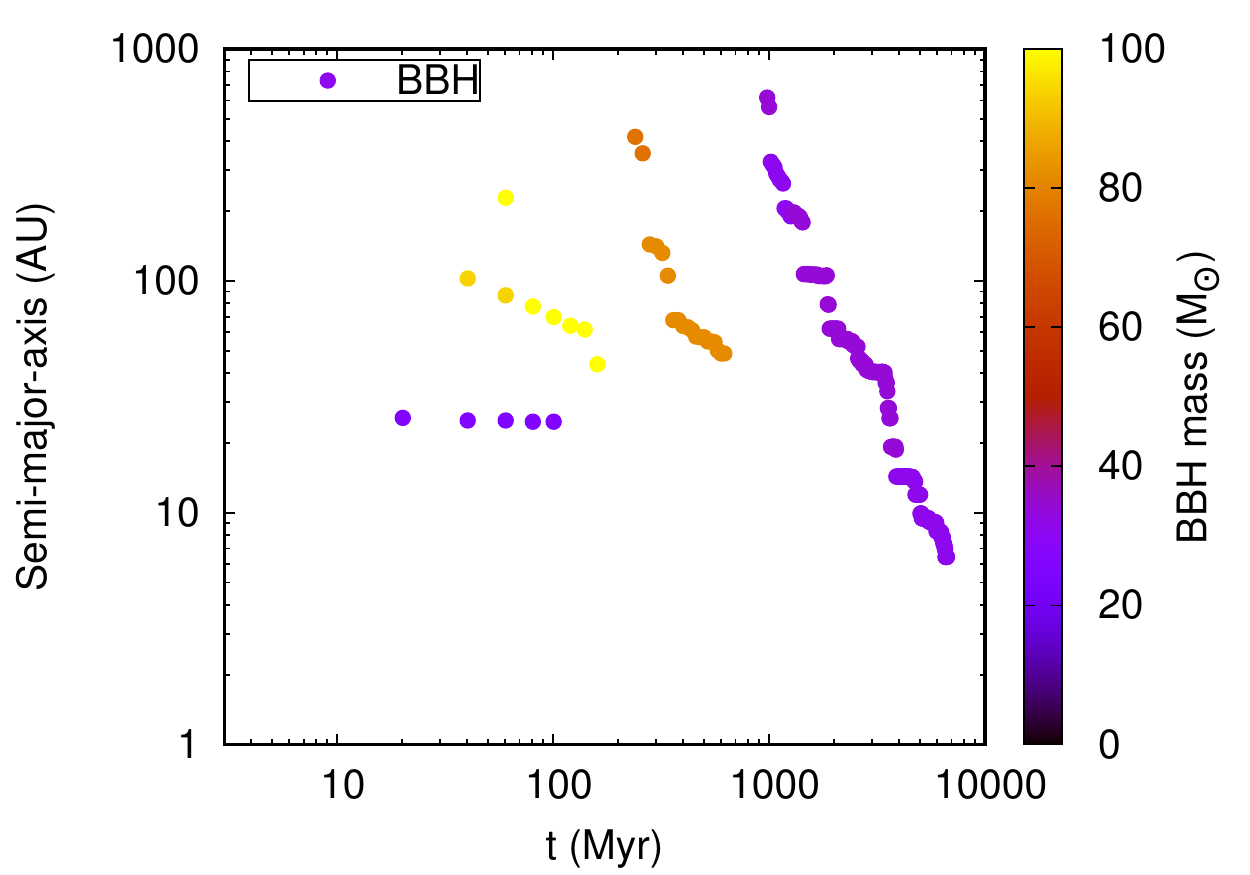}
\hspace{-0.25cm}
\includegraphics[width=5.94cm,angle=0]{figures/BHbin_setC/BBH_30kZ005f05_v2.pdf}
\hspace{-0.25cm}
\includegraphics[width=5.94cm,angle=0]{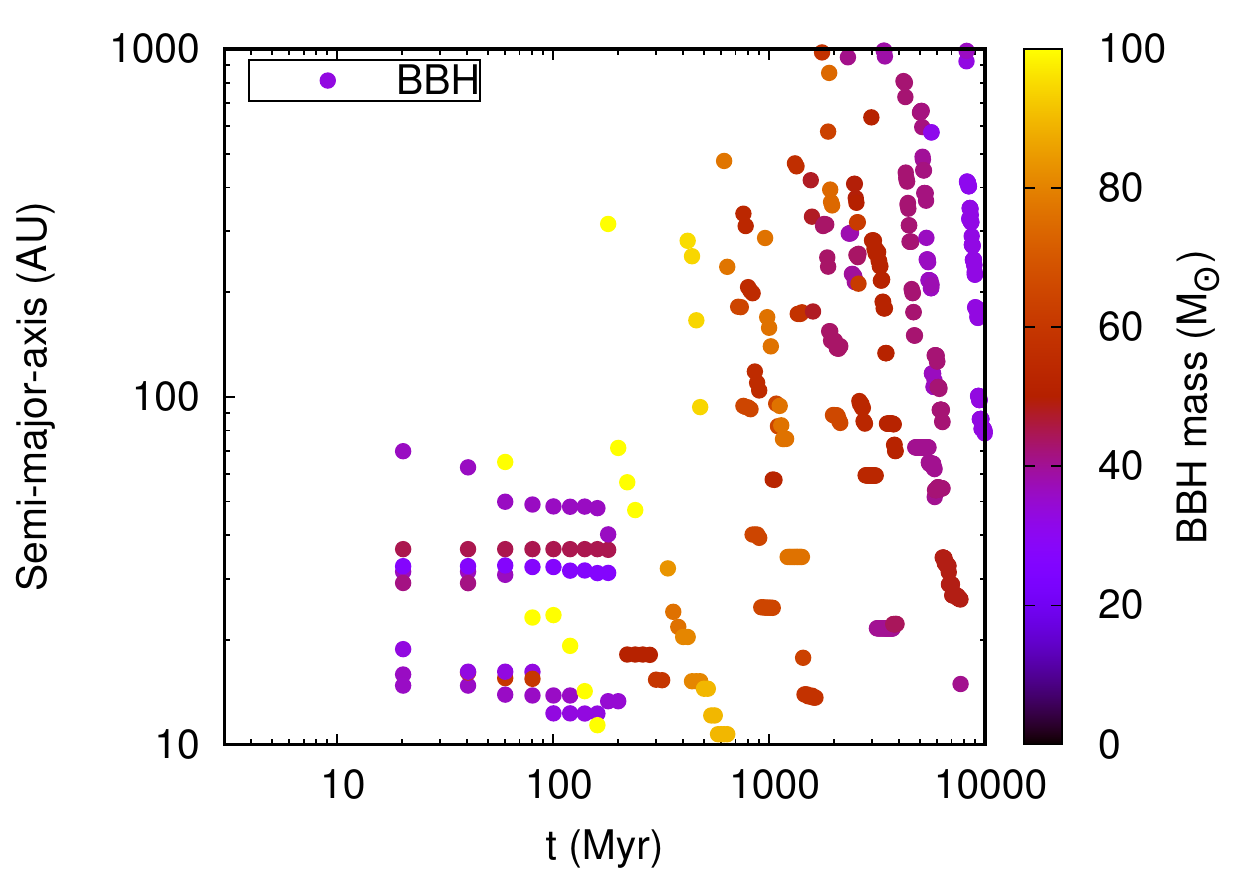}\\
\includegraphics[width=5.94cm,angle=0]{figures/BHbin_setC/BBH_15kZ02f05_v2.pdf}
\hspace{-0.25cm}
\includegraphics[width=5.94cm,angle=0]{figures/BHbin_setC/BBH_30kZ02f05_v2.pdf}
\hspace{-0.25cm}
\includegraphics[width=5.94cm,angle=0]{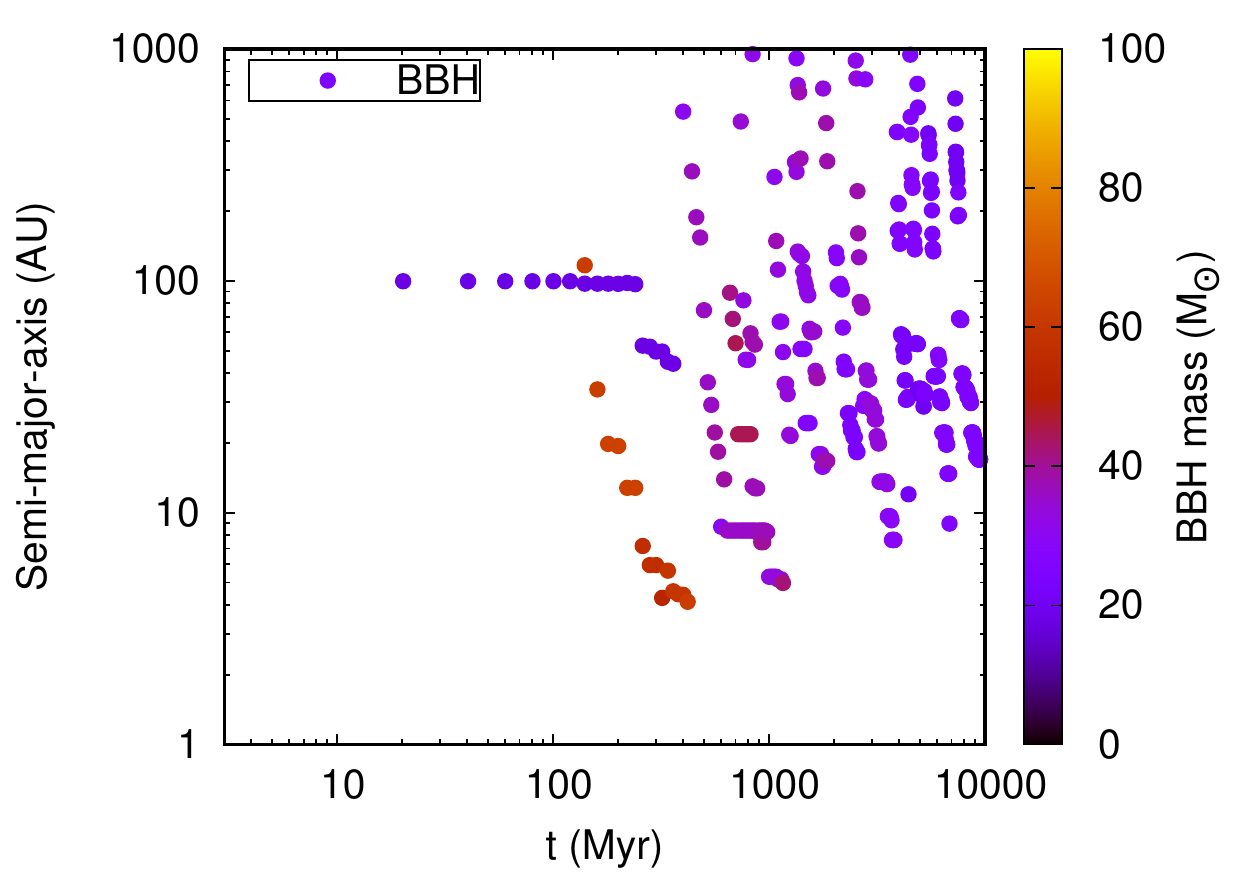}
\caption{The BBH content in the computed $\fbin(0)\approx0.05$ models with evolutionary time,
	$t$, as a function of increasing $\mcl(0)\approx1.5\times10^4\Ms$, $3.0\times10^4\Ms$,
	$5.0\times10^4\Ms$ (left to right) and $Z=0.001$, $0.005$, $0.02$ (top to bottom).
	The legends are the same as in Fig.~\ref{fig:bbh_15k}.}
\label{fig:bbh_f05}
\end{figure*}

\begin{figure*}
\centering
\includegraphics[width=8.0cm,angle=0]{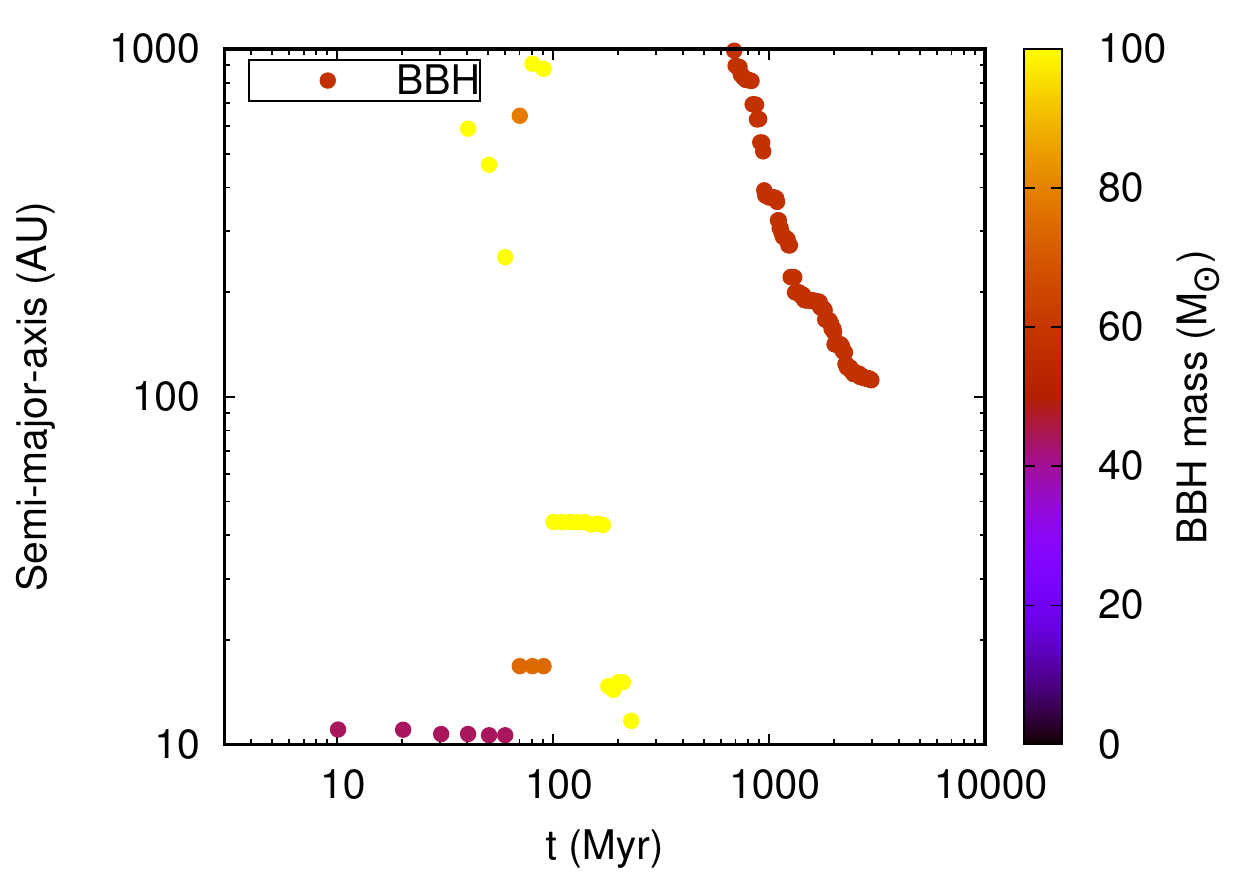}
\includegraphics[width=8.0cm,angle=0]{figures/BHbin_setC/BBH_15kZ001f50_v2.pdf}\\
\includegraphics[width=8.0cm,angle=0]{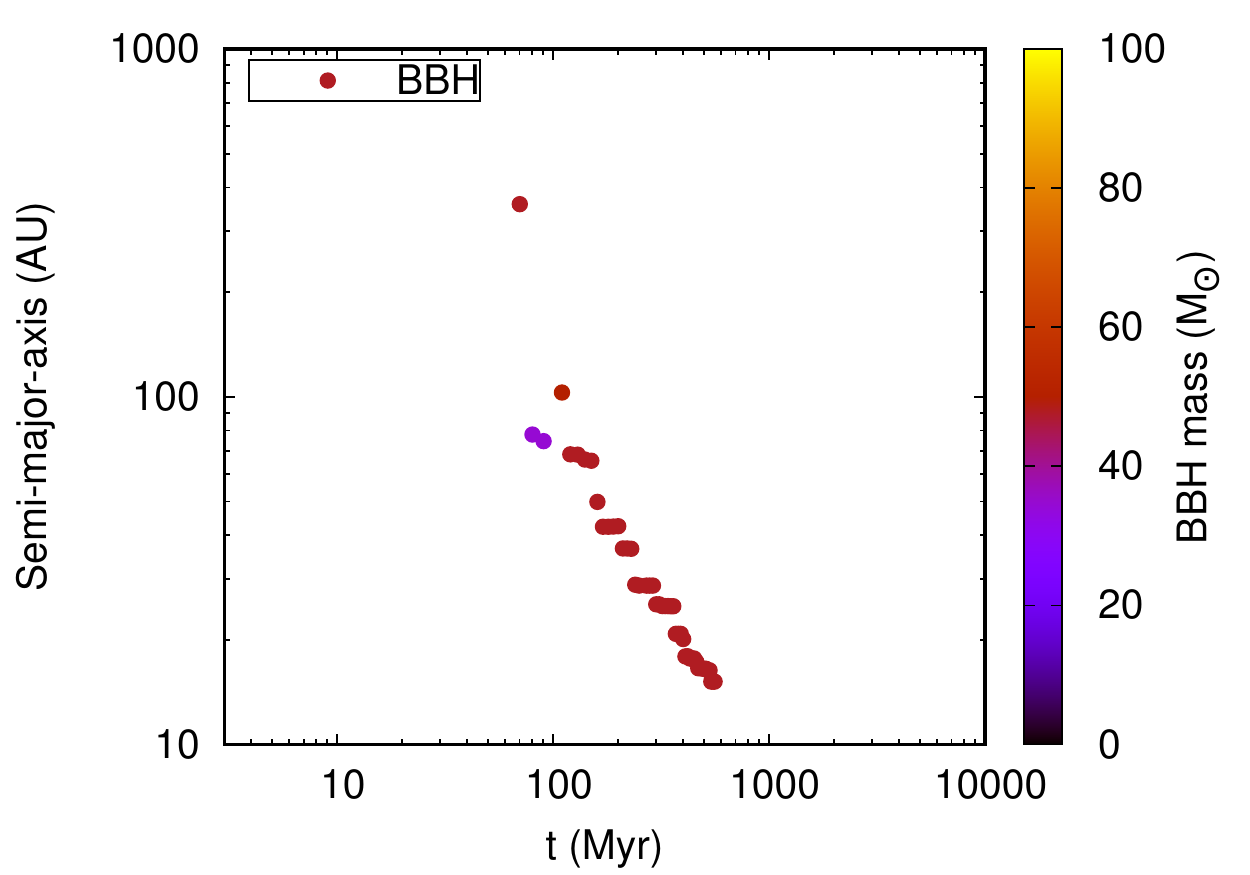}
\includegraphics[width=8.0cm,angle=0]{figures/BHbin_setC/BBH_15kZ02f50_v2.pdf}
\caption{The BBH content in the computed $\fbin(0)\approx0.50$ models with evolutionary time,
	$t$, as a function of increasing $\mcl(0)\approx7.5\times10^3\Ms$ (left) and $1.5\times10^4\Ms$ (right) and
	for metallicities $Z=0.001$ (top) and $0.02$ (bottom).
	The legends are the same as in Fig.~\ref{fig:bbh_15k}.}
\label{fig:bbh_f50}
\end{figure*}

\begin{figure*}
\centering
\includegraphics[width=8.0cm,angle=0]{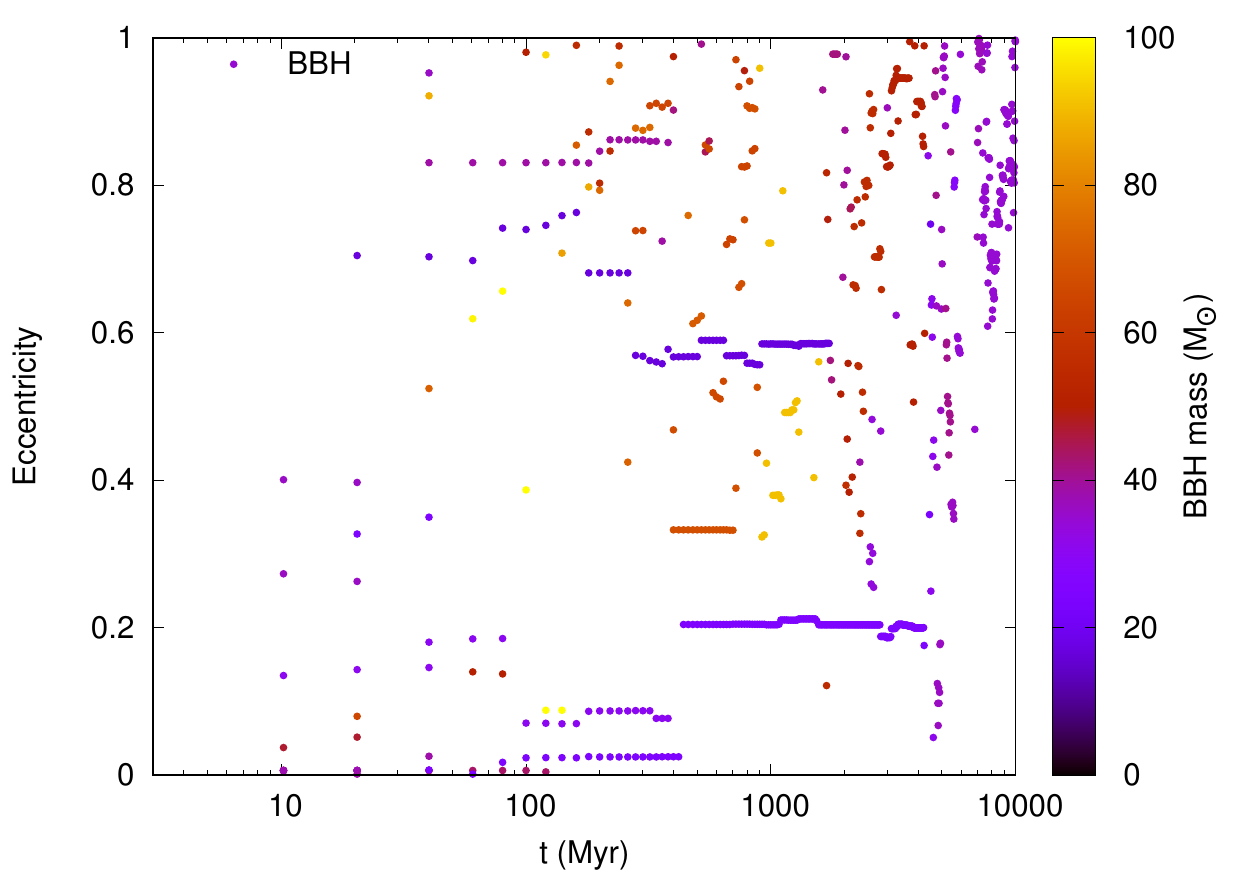}
\includegraphics[width=8.0cm,angle=0]{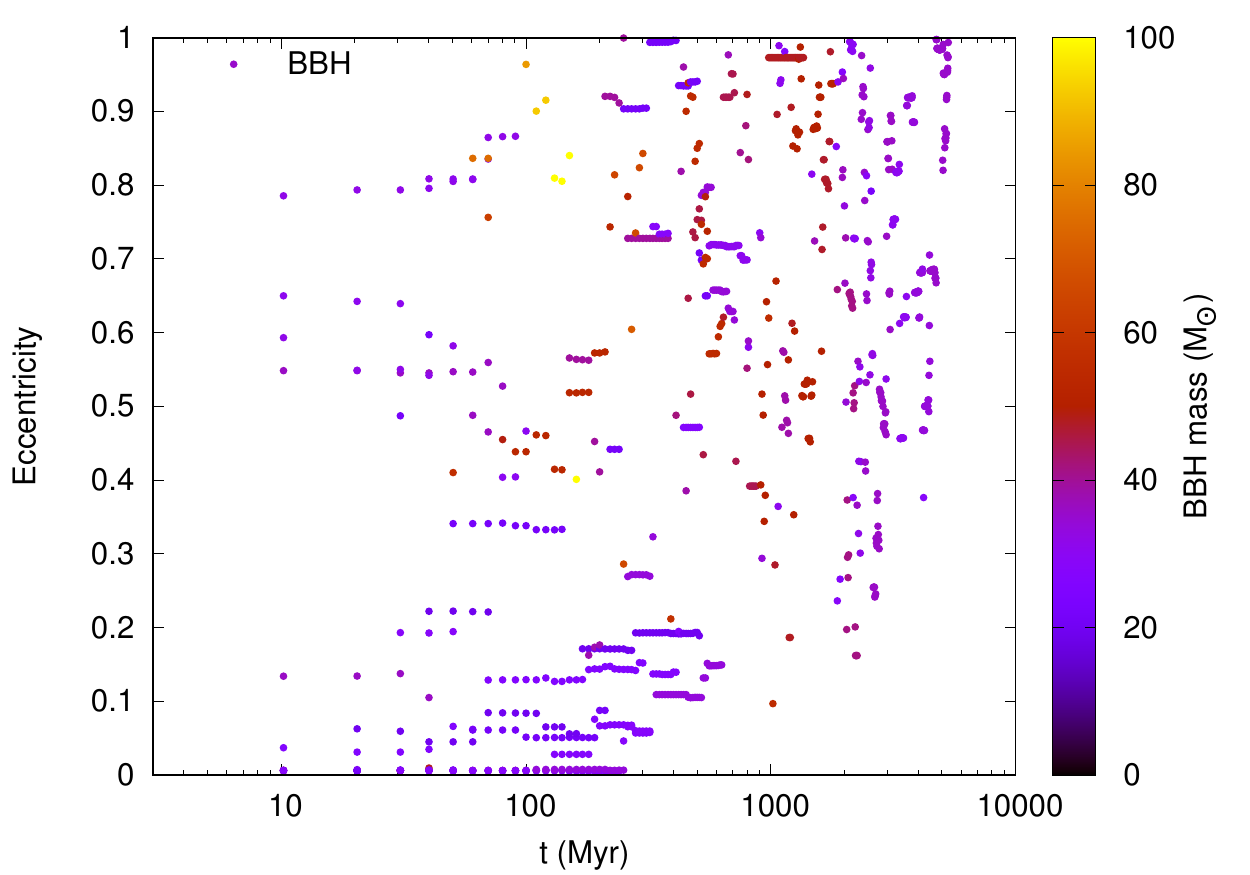}\\
\includegraphics[width=8.0cm,angle=0]{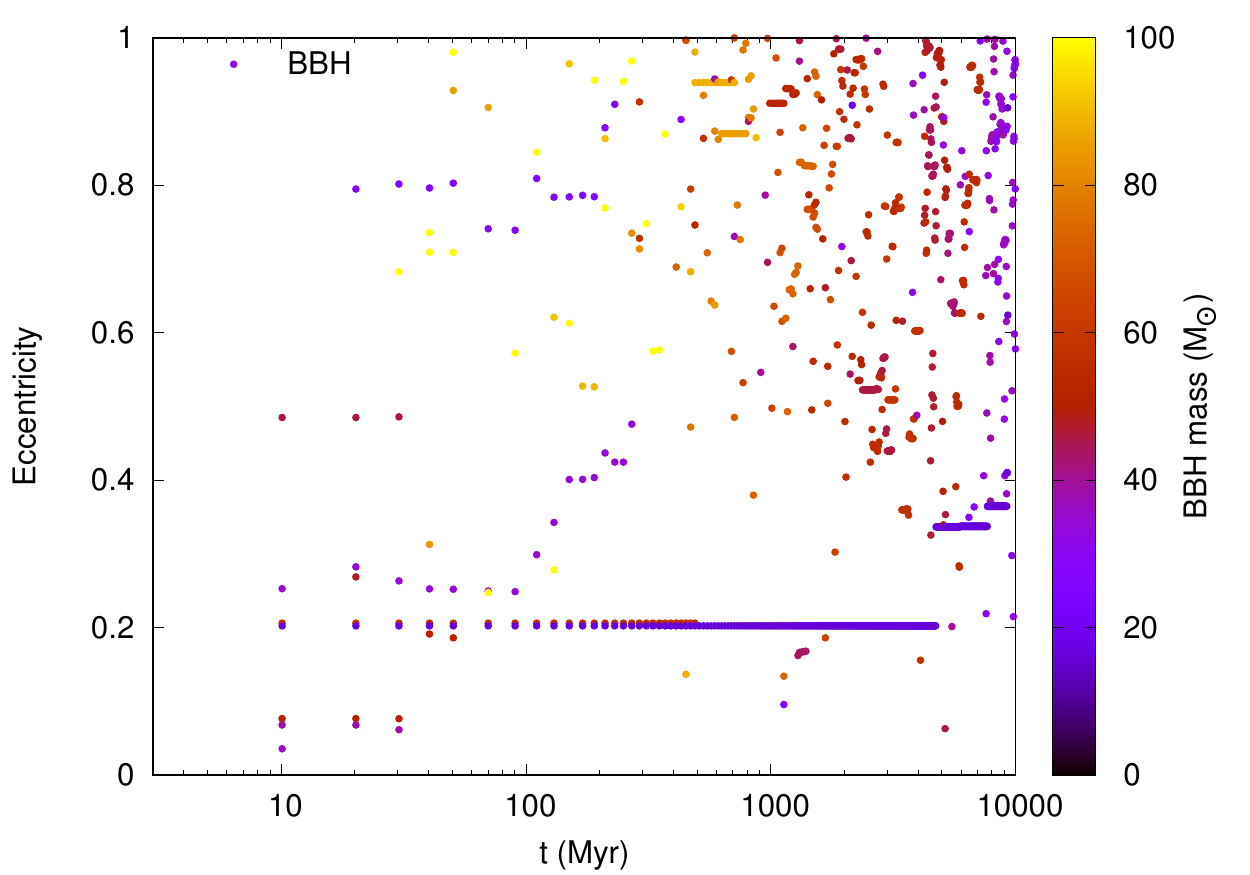}
\includegraphics[width=8.0cm,angle=0]{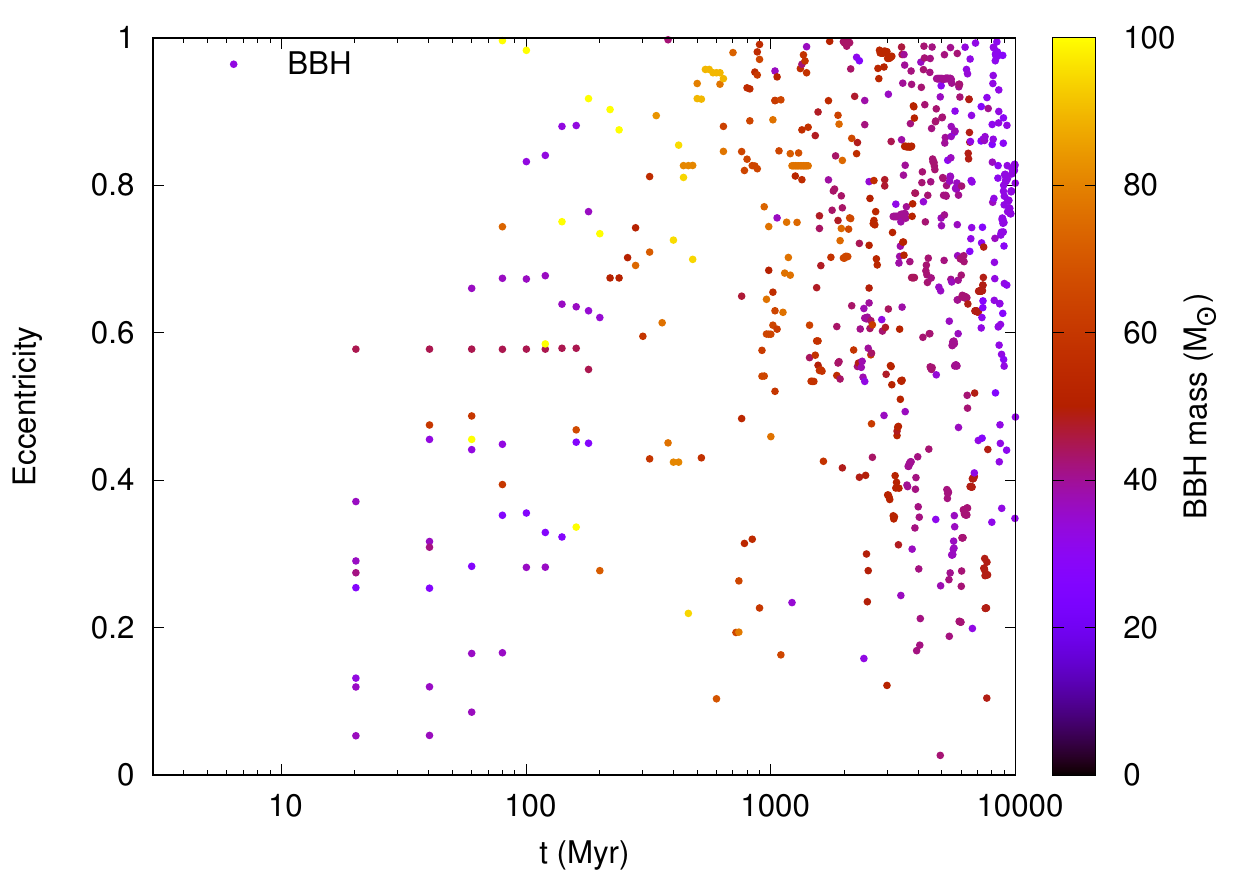}
\caption{The eccentricities of the BBH population bound to the computed models with
	$\mcl(0)\approx3.0\times10^4\Ms$, $\fbin(0)\approx0.10$ (top) and
	$\mcl(0)\approx5.0\times10^4\Ms$, $\fbin(0)\approx0.05$ (bottom) having $Z=0.001$ (left)
	and 0.005 (right). See Table~\ref{tab1} and Sec.~\ref{runs} for these models' details.
	The colour coding (colour bars) represents the total BBH mass.}
\label{fig:bbh_ecc}
\end{figure*}

\begin{figure*}
\centering
\includegraphics[width=8.0cm,angle=0]{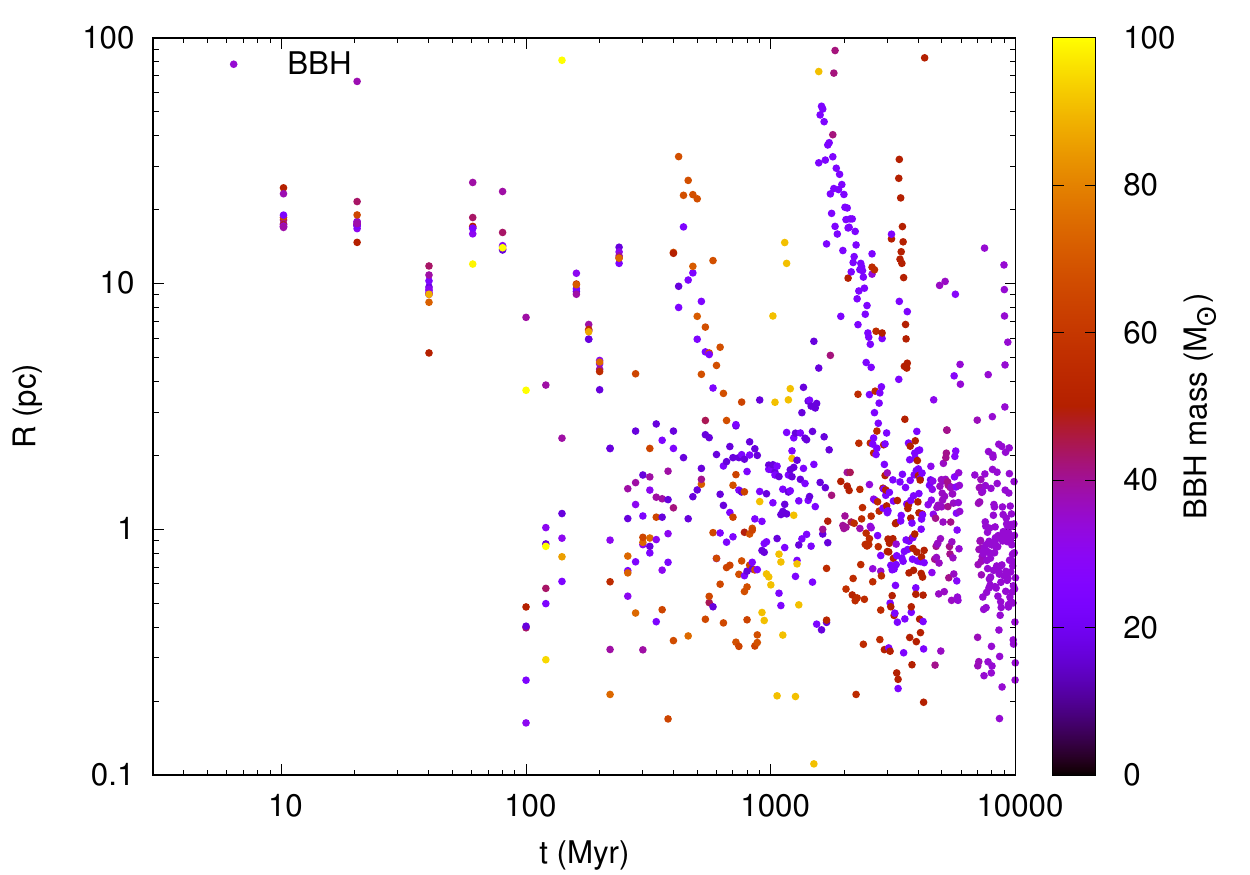}
\includegraphics[width=8.0cm,angle=0]{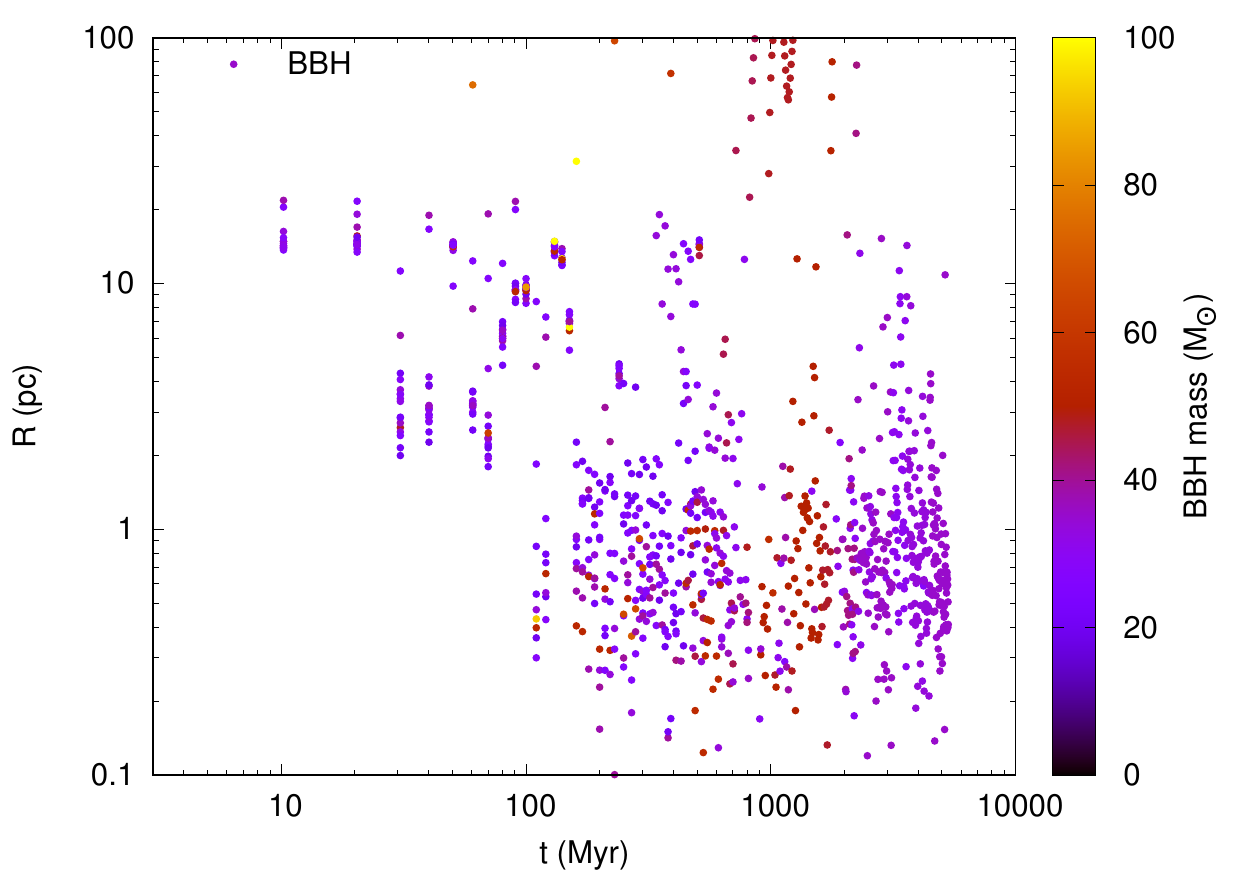}\\
\includegraphics[width=8.0cm,angle=0]{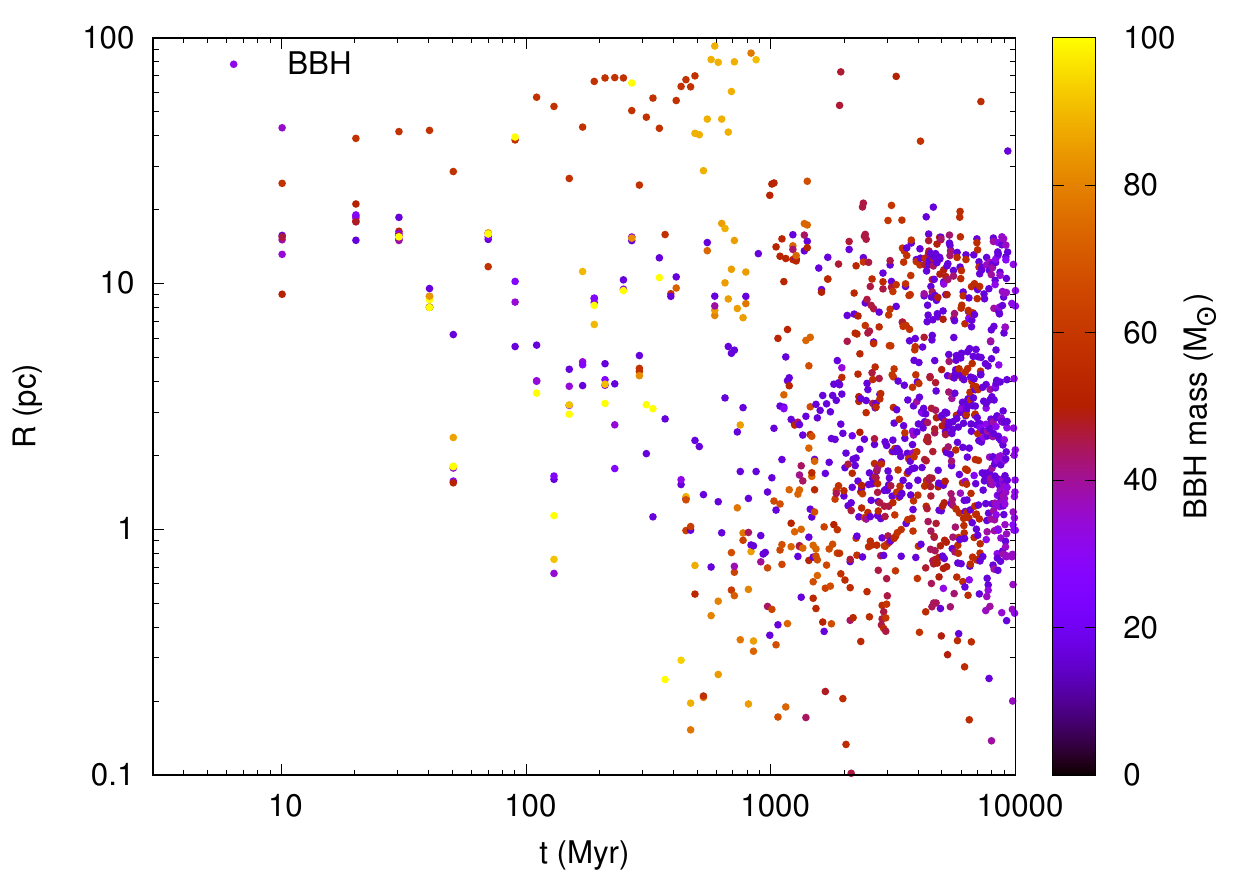}
\includegraphics[width=8.0cm,angle=0]{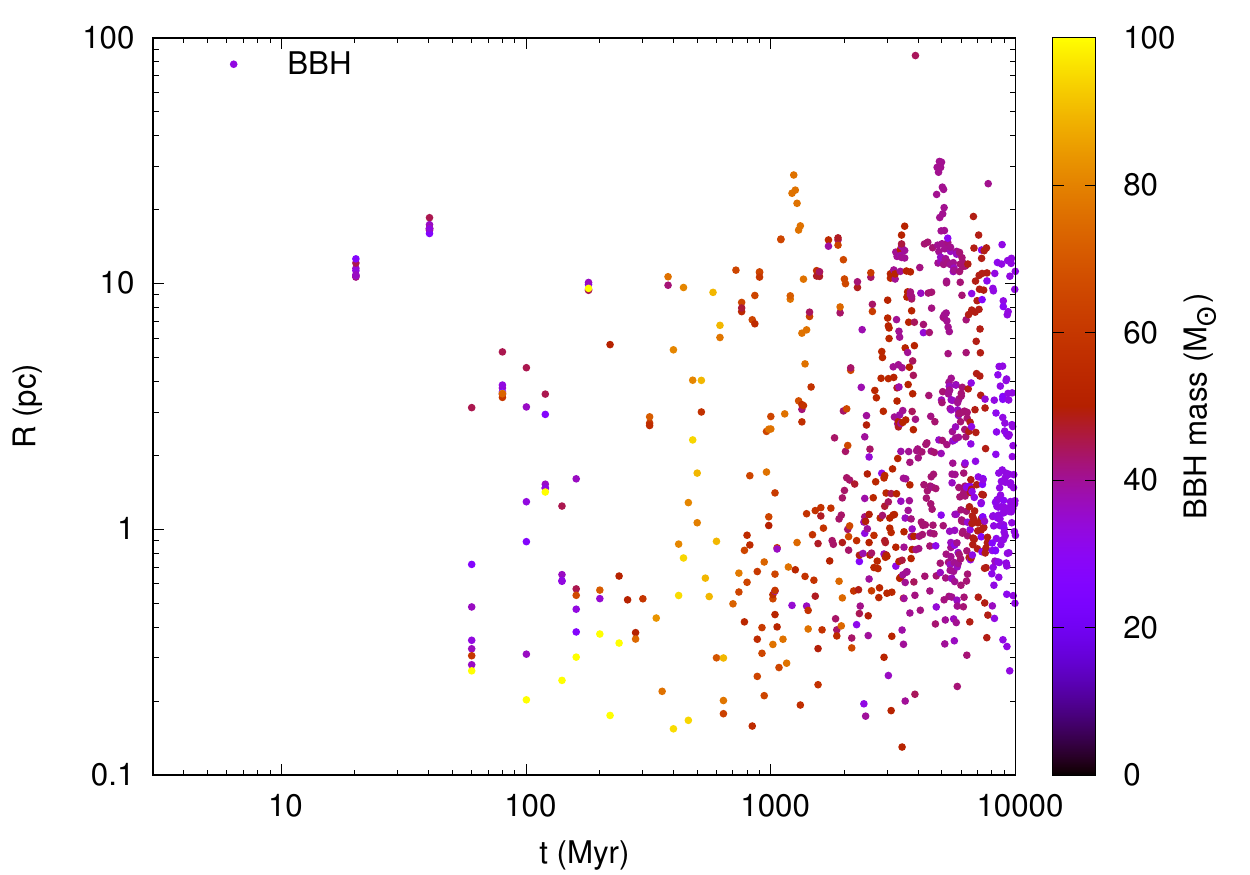}
	\caption{The radial distances w.r.t. the cluster's density center ($R>0.1$ pc)
	of the BBH population bound to the computed models with
	$\mcl(0)\approx3.0\times10^4\Ms$, $\fbin(0)\approx0.10$ (top) and
	$\mcl(0)\approx5.0\times10^4\Ms$, $\fbin(0)\approx0.05$ (bottom) having $Z=0.001$ (left)
	and 0.005 (right). See Table~\ref{tab1} and Sec.~\ref{runs} for these models' details.
	The colour coding (colour bars) represents the total BBH mass.}
\label{fig:bbh_rad}
\end{figure*}

Figs.~\ref{fig:bbh_15k} and \ref{fig:bbh_30k} show the BBHs bound to the evolutionary model clusters 
with $\mcl(0)\approx1.5\times10^4\Ms$ and $3.0\times10^4\Ms$ respectively with varying $Z$ and $\fbin(0)$.
Figs.~\ref{fig:bbh_f05} and \ref{fig:bbh_f50} show those for the models with $\fbin(0)\approx0.05$
and $\fbin(0)\approx0.5$ respectively with varying $\mcl(0)$ and $Z$.
In a particular panel, the semi-major-axes (limited to $a<1000$ AU) of all the BBHs present in the corresponding
model cluster's snapshot at the evolutionary time $t$ (filled squares, colour coded with the
total BBH mass), recovered as described in Sec.~\ref{snap}, are plotted against $t$.
In Fig.~\ref{fig:bbh_ecc}, the eccentricities of the BBHs inside four of the models
are similarly demonstrated and in Fig.~\ref{fig:bbh_rad}, the BBHs' radial
distances from the cluster's density center are plotted for these four models. 

A continuous trail of points, several of which per panel can be easily noticed in these figures, represents the
evolution of a particular BBH. In general, two categories of BBHs exist in a cluster.
Over the earliest evolutionary times ($t\lesssim1$ Gyr), the in-cluster BBH population of a sub-$\Zs$ cluster
is dominated by those that are derived from massive primordial binaries
whose original members continue to remain paired even after they are evolved into BHs (in \nbseven
models, such BBHs necessarily have members with consecutive identity numbers).
These can be noted as the near-horizontal trails in
Figs.~\ref{fig:bbh_15k}, \ref{fig:bbh_30k}, \ref{fig:bbh_f05}, \ref{fig:bbh_f50}.
These BBHs typically have widths $10\lesssim a \lesssim100$ AU and total
masses $\mtot\lesssim40\Ms$ (with a few of them being of $40\Ms\lesssim \mtot\lesssim60\Ms$).
A larger number of such primordial-binary-derived (but potentially dynamically-modified)
BBHs are initially present in models with
higher $\fbin(0)$, higher $\mcl(0)$, and lower $Z$: massive binaries comprising members of lower $Z$
widen to a less extent due to less mass loss in stellar winds and produce more massive (direct-collapse) BHs,
resulting in tighter BBHs that are less susceptible to dynamical dissociation inside a dense environment
\citep{Morscher_2015,2016arXiv160300884C}. The combination of their tightness, relatively
lower mass (hence, weaker gravitational focussing), moderate eccentricities, $e$ (see Fig.~\ref{fig:bbh_ecc}),
and, most importantly, generally outer-region location (see Fig.~\ref{fig:bbh_rad})
where the stellar density is low would
cause them to undergo close interactions rarely and hence to
encounter-harden \citep{1975MNRAS.173..729H} very slowly, mostly
via distant, weak encounters. Although the massive, O-star binaries (see Sec.~\ref{primbin}), that
would potentially evolve into such BBHs, would sink to the cluster's
center via dynamical friction within a few Myr, a fraction of them can get ejected
from the central region to the outer parts of the cluster (but without getting altogether
unbound from the cluster)
via dynamical encounters over a similar period due to the concentration of the
most massive entities within the innermost region. Additionally, the wind mass
loss from such a dynamically-kicked massive binary, which is still substantial at low $Z$ (according
to \citealt{Vink_2001} as adopted here; see Paper I), would further
drive it into outer orbits, apart from widening it at the same time.
Admittedly, the presence and the properties of such primordially-paired BBHs
would depend on the choice of the primordial massive-binary population (Sec.~\ref{primbin}).

In contrast, the dynamically-formed BBHs (recognizable by non-consecutive member identities)
begin to appear from $t\gtrsim100$ Myr with generally much wider $a\gtrsim100$ AU and 
higher $e$ (Fig.~\ref{fig:bbh_ecc}) and much more centrally (Fig.~\ref{fig:bbh_rad}),
enabling them to intercept
encounters much more frequently and undergo encounter hardening efficiently
\citep[][and references therein]{Banerjee_2006}.
These are
the slant trails in Figs.~\ref{fig:bbh_15k}, \ref{fig:bbh_30k}, \ref{fig:bbh_f05}, \ref{fig:bbh_f50}
\citep[\cf, Fig.~3 in][]{2010MNRAS.402..371B}.
As can be noted in these figures, in general, more massive dynamical BBHs appear and harden at earlier evolutionary
times which is expected since the most massive BHs in the system would
tend to pair up first and hence get dynamically processed earlier
\citep{Morscher_2015,2016arXiv160906689C,Banerjee_2017b}. The eccentricities of these frequently-interacting
BBHs alter widely with time, which is why it is difficult to locate such trails in Fig.~\ref{fig:bbh_ecc}.

However, at any given time, there is necessarily only a few dynamically-assembled BBHs existing in a cluster,
irrespective of the cluster's mass, stellar-binary fraction, and BH population. This is demonstrated
in Fig.~\ref{fig:bbhcnt} where, on each panel, the time evolutions of the total bound BBH population
and of the dynamically-formed BBH population are shown separately.
The majority of both types
of BBHs ultimately get ejected from the clusters via a close dynamical encounter, when the trails in
Figs.~\ref{fig:bbh_15k}, \ref{fig:bbh_30k}, \ref{fig:bbh_f05}, \ref{fig:bbh_f50} terminate, typically
at $a\sim10$ AU, depending on the escape speed at the location of the encounter within
the cluster (likely close to the cluster's center) at the time of the ejection
\citep{2010MNRAS.402..371B,2016arXiv160300884C}. In particular, for a given cluster and over similar
cluster-evolutionary times, those BBHs that generally reside in the inner regions
(mostly, the dynamically-formed ones) are hardened dynamically to tighter $a$s before
getting ejected by a final encounter: that is why, as in, \eg, Figs.~\ref{fig:bbh_30k} \&
\ref{fig:bbh_f05}, there are (more massive) BBHs that are being hardened efficiently,
at early as $t\sim100$ Myr,
unlike most of the (primordially-paired) others that aren't, over similar evolutionary times (\cf, the
corresponding panels in Fig.~\ref{fig:bbh_rad}).
The properties of such ejected BBHs are discussed in detail in
Papers I \& II.
A few of them terminate by undergoing triple-induced GR coalescences as discussed
in Sec.~\ref{bbhmrg}. As can be read from Figs.~\ref{fig:bbh_15k}, \ref{fig:bbh_30k},
\ref{fig:bbh_f05}, \ref{fig:bbh_f50}, the characteristic lifetime of the
primordially-paired BBHs is $\sim100$ Myr and that for the dynamical BBHs is
$\sim$ Gyr.

\subsection{Black hole-main sequence binaries inside clusters}\label{bhmspop}

At late evolutionary times, BH---non-BH binaries form in these model clusters
exclusively via exchange interactions between BHs and normal stellar binaries.
Since, motivated by observations, the BH-progenitor O-type stars are here
paired separately (see Sec.~\ref{primbin}), only BH---O-star binaries, that exist over
the first $\approx10$ Myr, can be obtained from the initial pairings, among
the BH---non-BH pairs in the cluster.

Fig.~\ref{fig:binevol} demonstrates the evolution of the stellar-binary population in
a representative set of the models from Table~\ref{tab1}. As expected, the total
number of binaries, $\nbin(t)$
(normalized w.r.t. the initial number of binaries, $\nbin(0)$; right
column in Fig.~\ref{fig:binevol}), decreases monotonically with time as they get dissociated or
ejected from the clusters due to dynamical interactions. The instantaneous
binary fraction, $\fbin(t)$, decreases for the first 1-3 Gyr after which
$\fbin(t)$ begins to increase again (left column in Fig.~\ref{fig:binevol}).
As the BHs deplete from a cluster due to encounters
(see Fig.~\ref{fig:bbhcnt}; Paper I), their
dynamical heating weakens, aiding the stellar binaries to resume their mass segregation
(see Papers I \& II) via two-body relaxation. When, due to its expansion by BH and binary
heating (see Papers I \& II; Fig.~\ref{fig:lagr}), the cluster begins to get trimmed,
through the tidal boundary, from its outer, lower-binary-fraction
regions (as per the mass segregation of the binaries), the binary fraction
within its remaining, bound part increases. The growth of $\fbin(t)$
is to a larger extent, closer the cluster is to its dissolution at the end
of the computation, and can even exceed $\fbin(0)$ close to its disruption (see Fig.~\ref{fig:binevol}).

Figs.~\ref{fig:bhms_15k}, \ref{fig:bhms_30k}, \ref{fig:bhms_f05}, and \ref{fig:bhms_f50}
show the BH-MS (filled circles) and NS-MS (filled squares) binaries ($a<1000$ AU), in the $a-t$ plane,
formed inside the computed model clusters of Table~\ref{tab1}.
Figs.~\ref{fig:bhms_15k} and \ref{fig:bhms_30k} show the BH-MSs/NS-MSs inside the clusters 
with $\mcl(0)\approx1.5\times10^4\Ms$ and $3.0\times10^4\Ms$ respectively with varying $Z$ and $\fbin(0)$.
Figs.~\ref{fig:bhms_f05} and \ref{fig:bhms_f50} show those for the models with $\fbin(0)\approx0.05$
and $\fbin(0)\approx0.5$ respectively with varying $\mcl(0)$ and $Z$.
In these figures, which
are obtained from the computed models' snapshots as described in Sec.~\ref{snap},
the points are colour coded according to the MS companions' mass (at time $t$).

As seen in these figures, nearly all models produce intermediate- ($t\gtrsim10$ Myr)
and late-time ($t\gtrsim1$ Gyr) BH-MS binaries (filled circles)
of widths $a\sim100$ AU. However, the high-$\fbin(0)$ and/or lower-$Z$ models contain
BH-MSs of $a<1$ AU to $a\sim100$ AU at times as early as $t\leq10$ Myr, which
are evolved pairs of massive stars from $t=0$ (that may thereafter become
BBHs derived from primordial pairs; see Sec.~\ref{bbhpop}). As explained above,
all the BH-MS binaries at intermediate and late times are necessarily dynamically
assembled.
Some of the models form tighter BH-MSs at intermediate and late times,
of $a<1$ AU to $a\sim10$ AU, which are prospective candidates to be
identified through radial-velocity measurements of the MS companion.
Typically, these are the models that have high $\fbin(0)$ ($>0.3$) 
or expand moderately or undergo contraction at late times (see Fig.~\ref{fig:lagr})
or have a combination of these properties, which favour exchange
interactions of the BHs with the relatively tight stellar binaries.
A few these tighter BH-MS binaries
have their MS companion's mass $\gtrsim0.8\Ms$ (colour coding in
Figs.~\ref{fig:bhms_15k}, \ref{fig:bhms_30k}, \ref{fig:bhms_f05},
and \ref{fig:bhms_f50}), which makes them similar to the radial-velocity-identified
(detached) BH-MS binary in the GC NGC 3201 \citep{Giesers_2018}.
The overall mass range of the MS companion in these dynamically-formed BH-MS binaries
is, however, wide: from $\approx0.1\Ms$ to $\gtrsim1.0\Ms$.

Some of these models produce NS-MS binaries (filled squares in
Figs.~\ref{fig:bhms_15k}, \ref{fig:bhms_30k}, \ref{fig:bhms_f05}, and \ref{fig:bhms_f50})
which are generally wide, of $a\sim100$ AU. Two of the NS-MSs
are tighter with $a\sim10$ AU (see top-right \& bottom-left panels of Fig.~\ref{fig:bhms_15k}).
As pointed out in Paper II \citep[see also][]{Fragione_2018},
the BH heating keeps the NS population within a cluster
from mass-segregating strongly towards the cluster's center so that
the NSs, although comparable to or larger than the BHs in number,
intercept lower-density environments in the cluster compared to the
BHs where the
chances of exchanging with a tight stellar binary is smaller
(the two tight NS-MS binaries form in less-expanded and high-stellar binary fraction models). 

The recent observations by \citet{Shishkovsky_2018} make it tempting to look
for binaries in the computed models in which a BH or an NS is paired with
a giant or an even more evolved stellar companion. Such binaries could be
identified in a few of the models in Table~\ref{tab1} which are shown
in Fig.~\ref{fig:bhall_misc}. Here, such binaries containing BHs are
formed with $a\sim1-\sim10$ AU. Also, those containing NSs are formed with
$a\sim10-\sim100$ AU. None of these are, however, tight enough to be
RLO, as may be the case with the BH candidate in  
\citet{Shishkovsky_2018}.

The lifetimes of the in-cluster BH-MS/BH---evolved-star binaries in
Figs.~\ref{fig:bhms_15k}, \ref{fig:bhms_30k}, \ref{fig:bhms_f05}, \ref{fig:bhms_f50},
and \ref{fig:bhall_misc}
are typically $\sim 100$ Myr; they cover a wide range of values
from $\sim 10$ Myr (seen as a dot) to $\sim$ Gyr. The
same can be said for the NS binaries in these figures.
These binaries either get dynamically ejected from the cluster or
get their stellar member exchanged by an NS or a BH. 
Indeed, the $\sim$ Gyr-long
binaries are hosted only by the lower-mass clusters, where
the chances of exchange or ejection are generally lower due to
lower stellar densities and fewer numbers of NSs and BHs.
Only a few  BH-MS/BH---evolved-star binaries are ejected from 
the present models, taken together. Likewise the BBHs (Sec.~\ref{bbhpop}),
at most a few BH-MS retain in a cluster at a given time irrespective
of the cluster's properties or the BH population within it,
as demonstrated in Fig.~\ref{fig:bhmscnt}. Such sparsity of BBH/BH-MS/BH---evolved-star
systems is also inferred in recent Monte Carlo studies of GC models
\citep[\eg,][]{ArcaSedda_2018,Askar_2018}.

All the BBH- and BH-MS---forming models discussed until now do not materialize
the orbital energy dissipation due to tidal interactions which may influence
binary orbits and aid in tightening dynamically-formed compact---non-compact
binaries and also potentially result in tighter double-compact
binaries; see Sec.~\ref{stellrem}. A few relatively short duration test
computations, with otherwise similar type of cluster models (Sec.~\ref{runs}), are performed
with the tidal interaction activated (see Table~\ref{tab1}). Fig.~\ref{fig:bhbin_circ}
shows the BBHs (upper panels) and the BH-MSs (lower panels) formed in these models.
The BBHs exhibit similar properties as in the previous models. In two
of these three models, very tight ($<1$ AU) BH-MSs are formed well within 1 Gyr of
cluster evolution that live for $\sim100$ Myr and in the third model, several $\sim10$ AU 
BH-MSs of similar longevity are formed within 1 Gyr. 
Compared to the frequency of occurrences of similarly-tight binaries
in the previous models spanning a similar range of parameters,
it then appears that the inclusion of tidal energy dissipation (which, currently,
is implemented in \nbseven according to \citealt{Mardling_2001} which recipes are valid for
low-mass MS stars only) may indeed help to form tighter BH-MSs, which
will be investigated in detail in a forthcoming work.
None of the models presented in this work, however, form a stable RLO binary: longer-term
computations including tidal interaction may be the key to form them in model clusters
\citep[as in, \eg,][]{Kremer_2017} which will be investigated in the near future.

\subsection{Remark on the contribution from open clusters
on the dynamical binary black hole merger rate}\label{remark}

In Paper II, it has been shown, based on minimalistic estimates on BBH merger
generation from young massive- and open-type clusters, that such clusters
would collectively produce present-day (dynamical) BBH mergers at a rate comparable to that
from the GCs (and hence to that from the nuclear clusters since the latter
two can also be similar; see, \citealt{Hoang_2017}). The key to this
estimate (but see Sec.~3.3 of Paper II for the details)
is the observationally-motivated \citep[\eg,][]{Gieles_2006b,Larsen_2009}
assumption that the progenitors
of open clusters and GCs follow a power-law (cluster) initial mass function of
index $-2$ and that the lower birth-mass limit of the merger-producing clusters
is $\mcl(0)\approx10^4\Ms$, inferred based on the computations in Paper II.

The set of model computations presented here extends down to $\mcl(0)\approx7.5\times10^3\Ms$ 
which still produces BBH mergers, at least for high $\fbin(0)$ and low $Z$
(see Table~\ref{tab1}). Although, several models of $\mcl(0)\approx5.0\times10^3\Ms$ has also been
explored under various conditions, \eg, with low and high $\fbin(0)$ and $Z$ and also
under solar-neighborhood-like and dwarf-galaxy-like external fields, no BBH or
other types of GR coalescences, either in-cluster or
ejected, could be obtained from them. Such models, however, can
still be interesting for the production of temporarily-relativistic systems
(Sec.~\ref{sling}).

Hence, based on the computations until now and given the adopted
recipe for stellar-remnant retention (Sec.~\ref{stellrem}),
if $\mcl(0)\approx7.5\times10^3\Ms$ is plausibly taken to be the working lower mass limit
of the progenitors of open clusters (\ie, of young massive clusters)
that would give rise to dynamical GR coalescences, then
it only strengthens the conclusion that the young massive and open
clusters would collectively produce present-day (dynamical) BBH mergers,
visible by the LISA and the LIGO-Virgo,
at a rate that is comparable to those from the GCs and the galactic nuclei. 

\section{Summary}\label{summary}

\begin{figure*}
\centering
\includegraphics[width=5.94cm,angle=0]{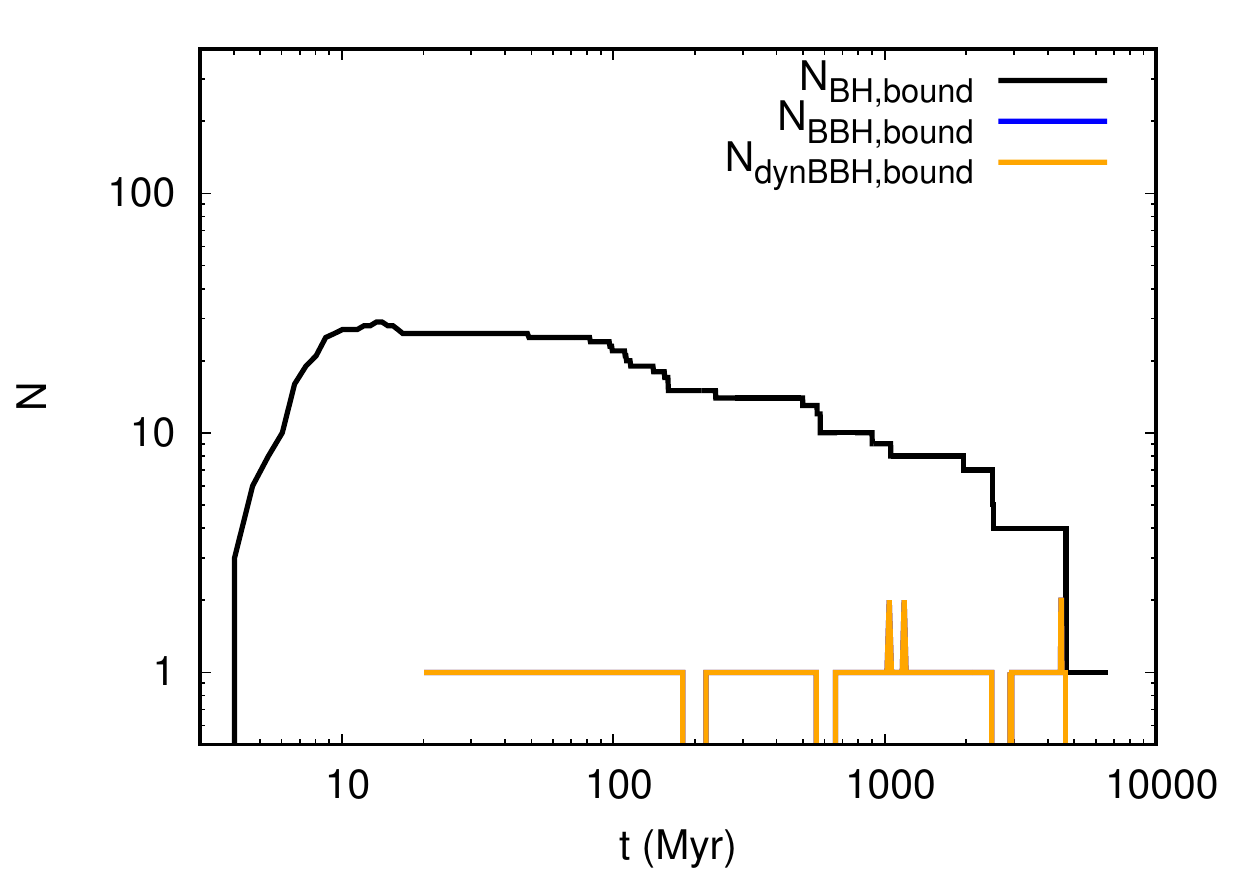}
\hspace{-0.25cm}
\includegraphics[width=5.94cm,angle=0]{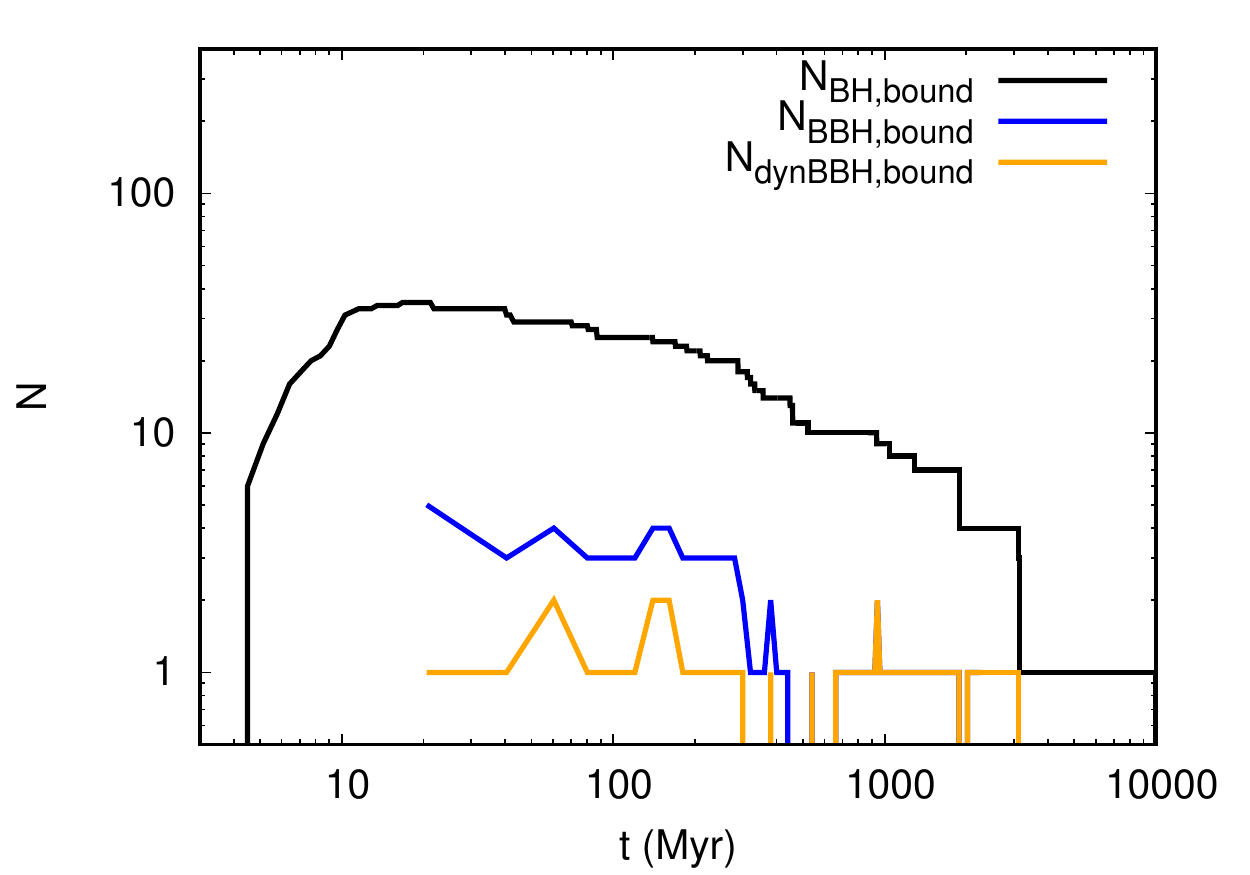}
\hspace{-0.25cm}
\includegraphics[width=5.94cm,angle=0]{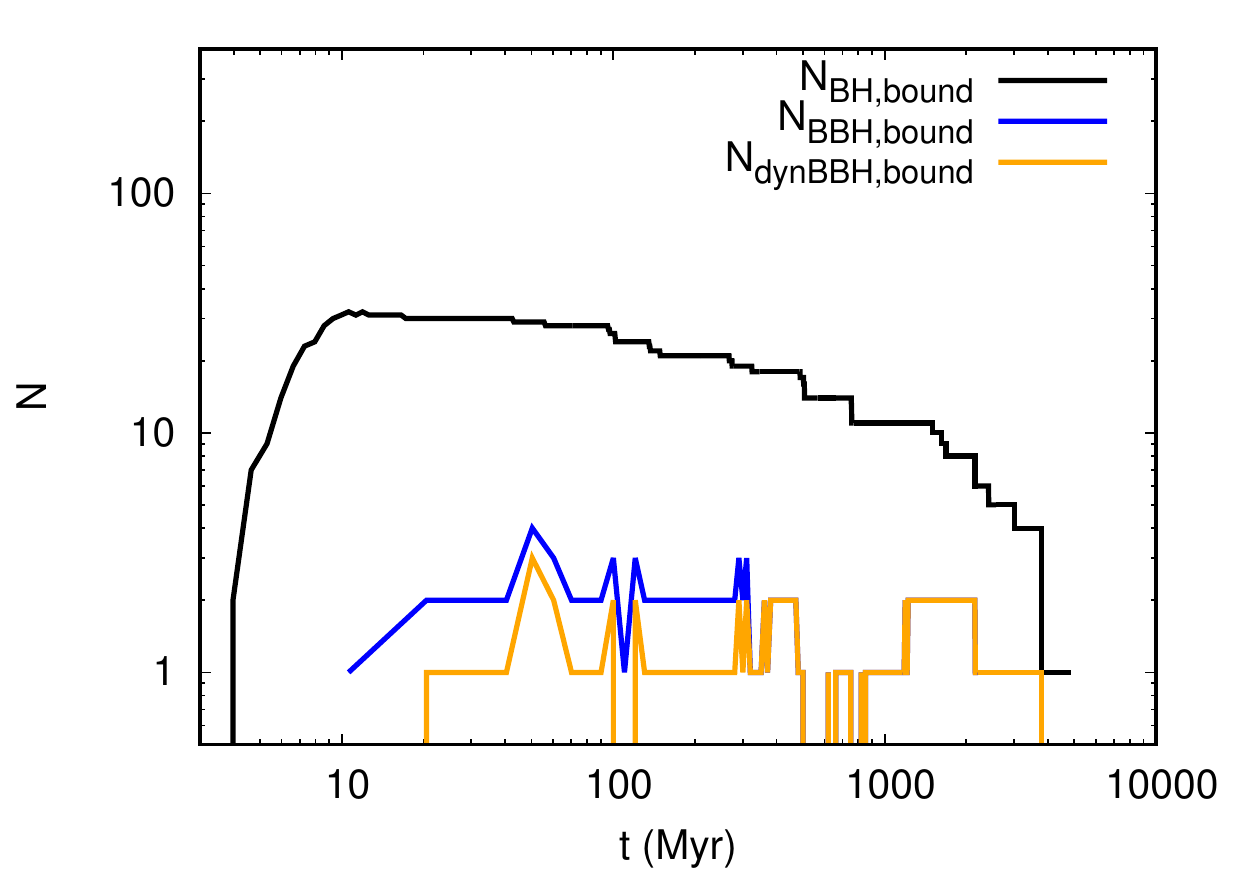}\\
\includegraphics[width=5.94cm,angle=0]{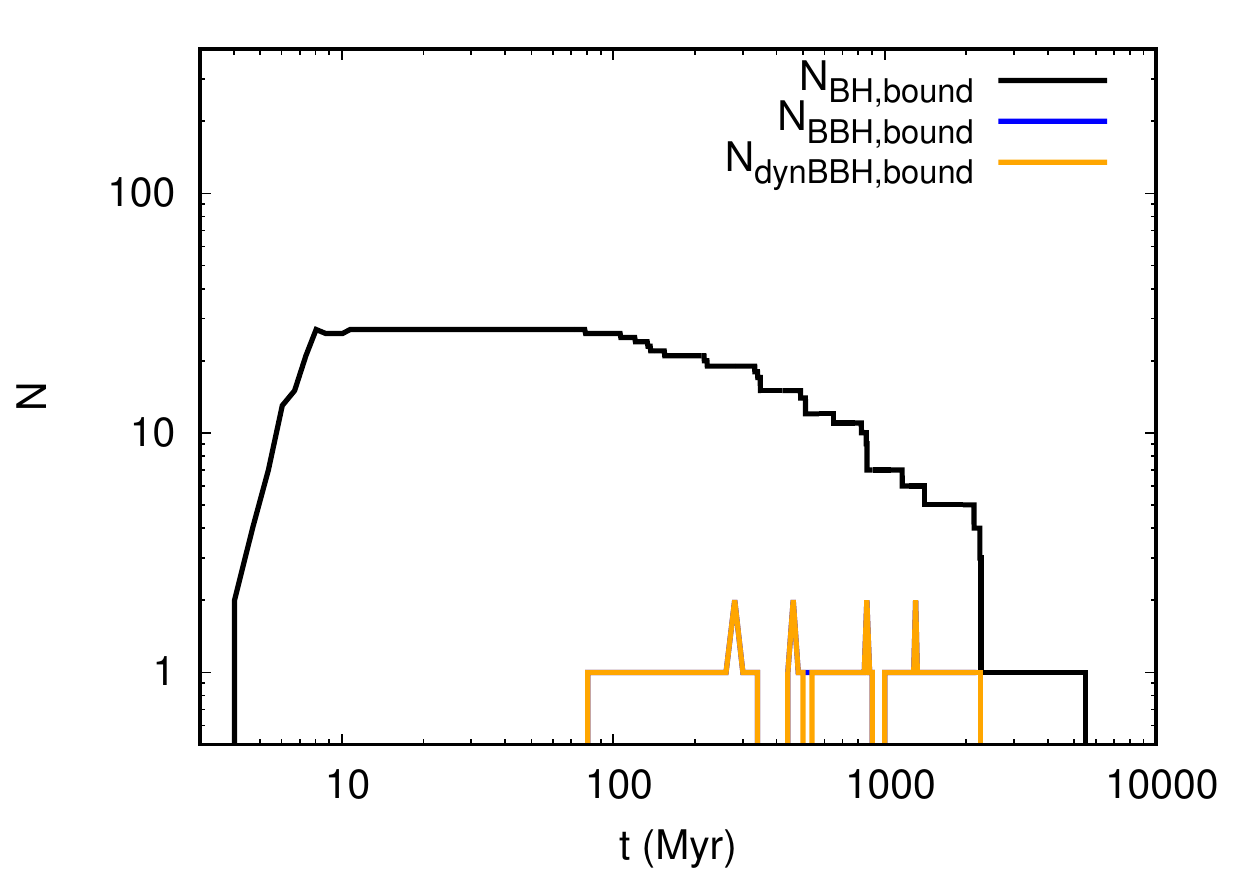}
\hspace{-0.25cm}
\includegraphics[width=5.94cm,angle=0]{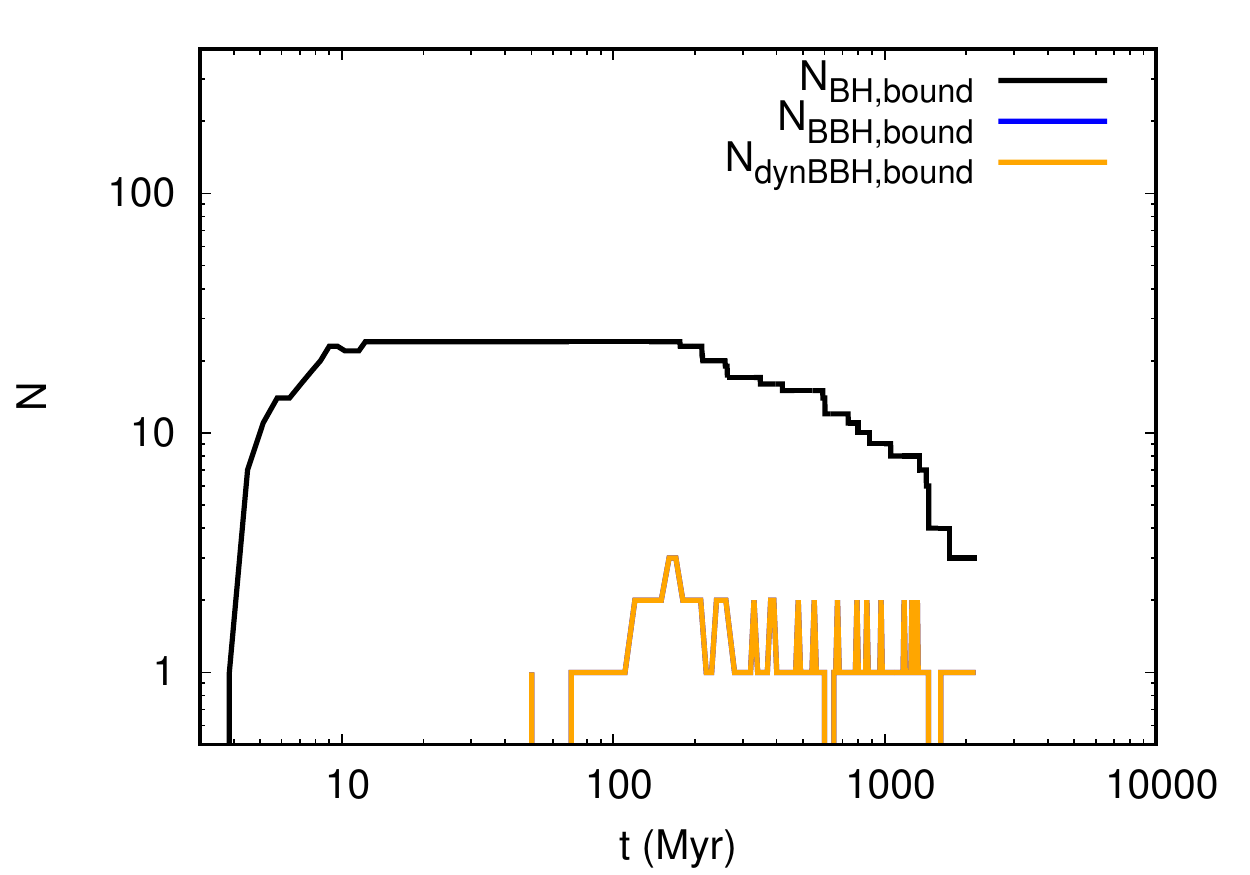}
\hspace{-0.25cm}
\includegraphics[width=5.94cm,angle=0]{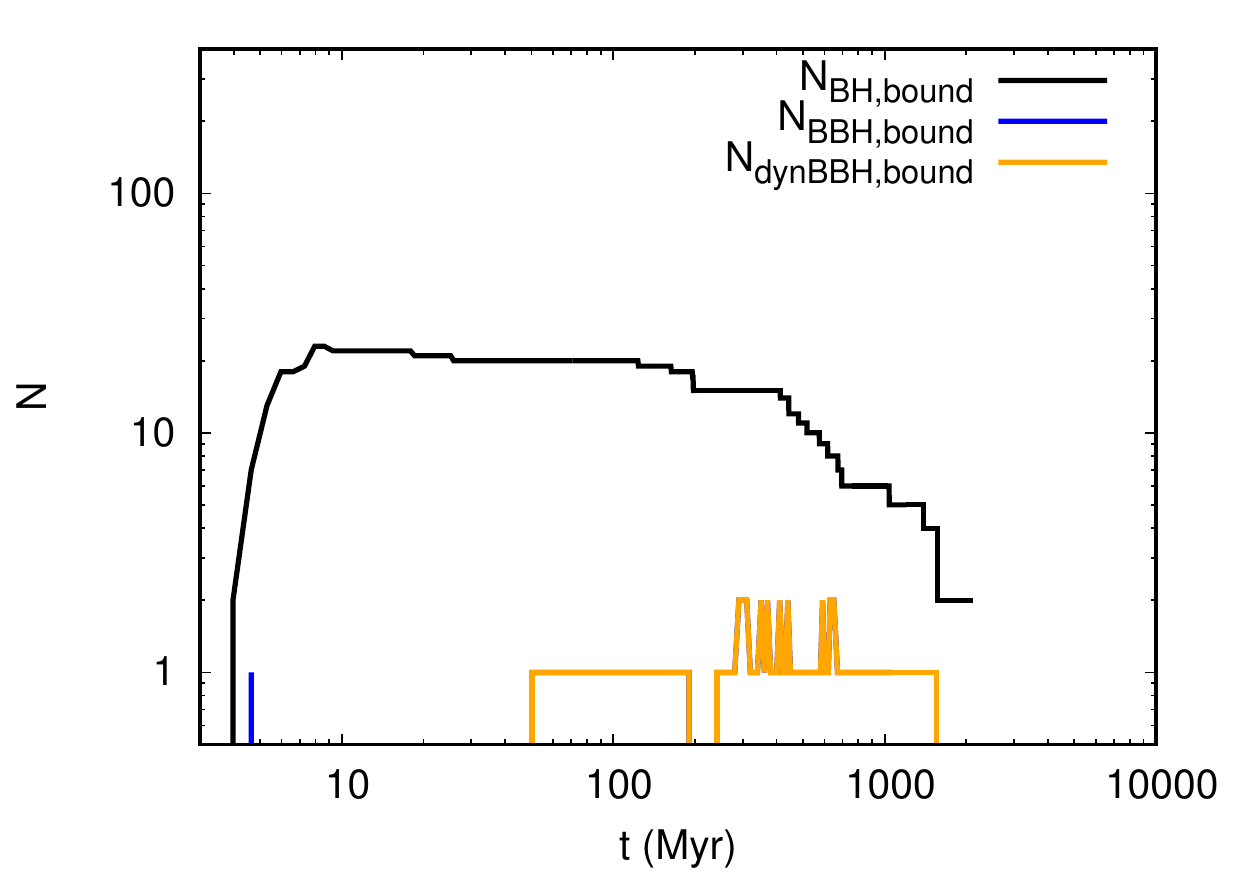}\\
\includegraphics[width=5.94cm,angle=0]{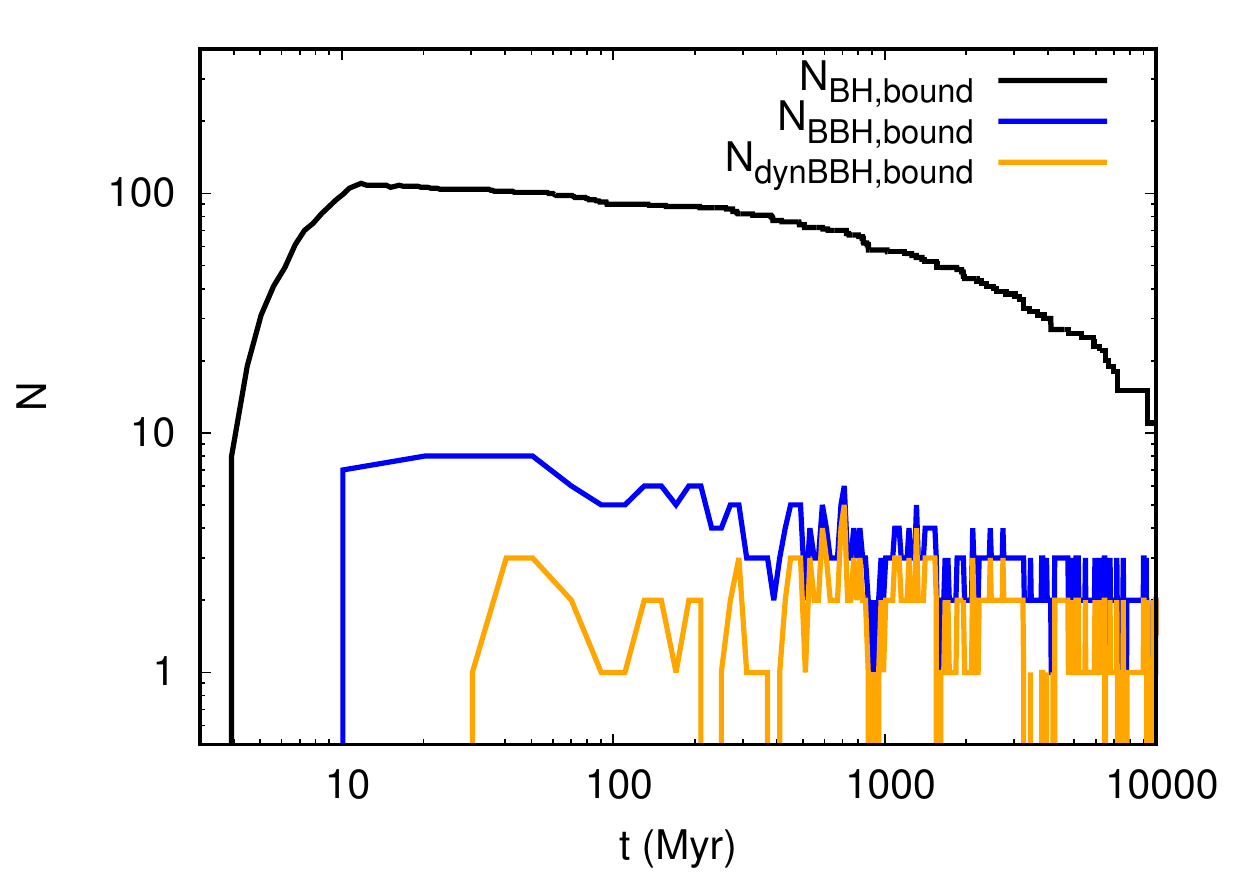}
\hspace{-0.25cm}
\includegraphics[width=5.94cm,angle=0]{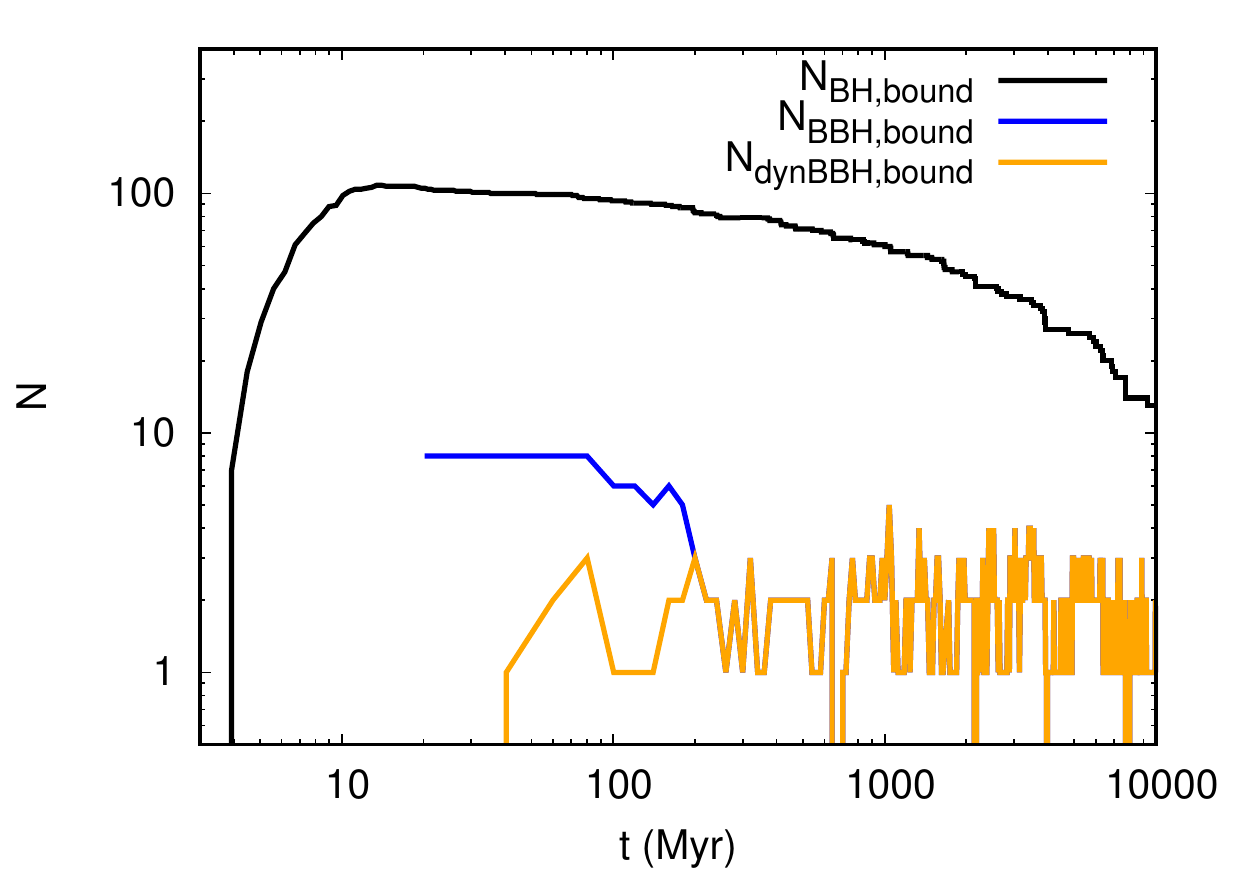}
\hspace{-0.25cm}
\includegraphics[width=5.94cm,angle=0]{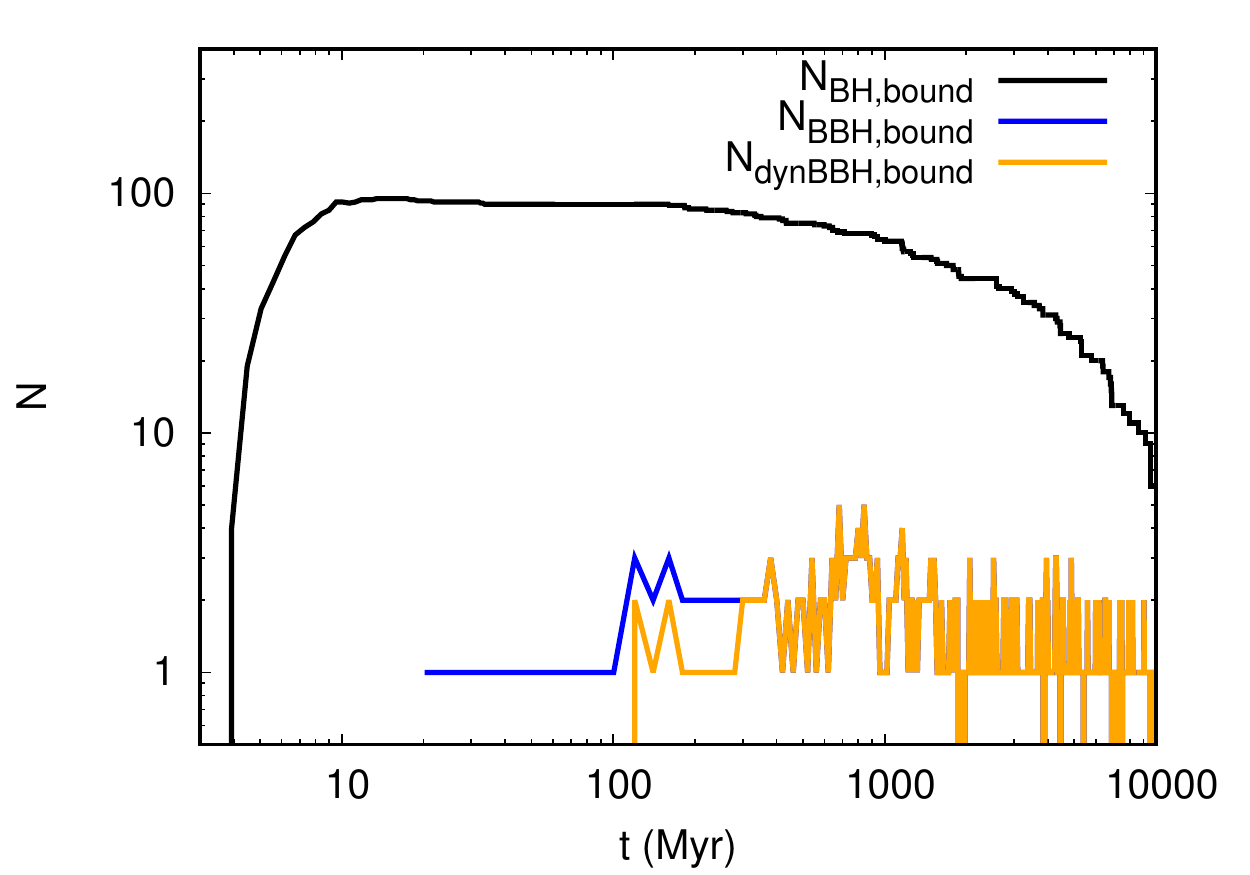}
\caption{Each of the panels represents the time evolutions of the total number of BHs, $\nbhbound$ (black line),
of the total number of BBHs, $\nbbhbound$ (blue line), and of the number of dynamically-formed BBHs,
$\ndynbbhbound$ (orange line), that are bound to the cluster.
{\bf Top and middle:} for the same models as in Fig.~\ref{fig:bbh_15k} shown in the same order, \ie,
those with
$\mcl(0)\approx1.5\times10^4\Ms$, $\fbin(0)\approx0.05,0.30,0.50$ (left to right), and $Z=0.001$ (top),
$0.02$ (middle). {\bf Bottom:} for the models with $\mcl(0)\approx5.0\times10^4\Ms$, $\fbin(0)\approx0.05$,
$Z=0.001$, $0.005$, $0.02$ (left to right).}
\label{fig:bbhcnt}
\end{figure*}

\begin{figure*}
\centering
\includegraphics[width=8.0cm,angle=0]{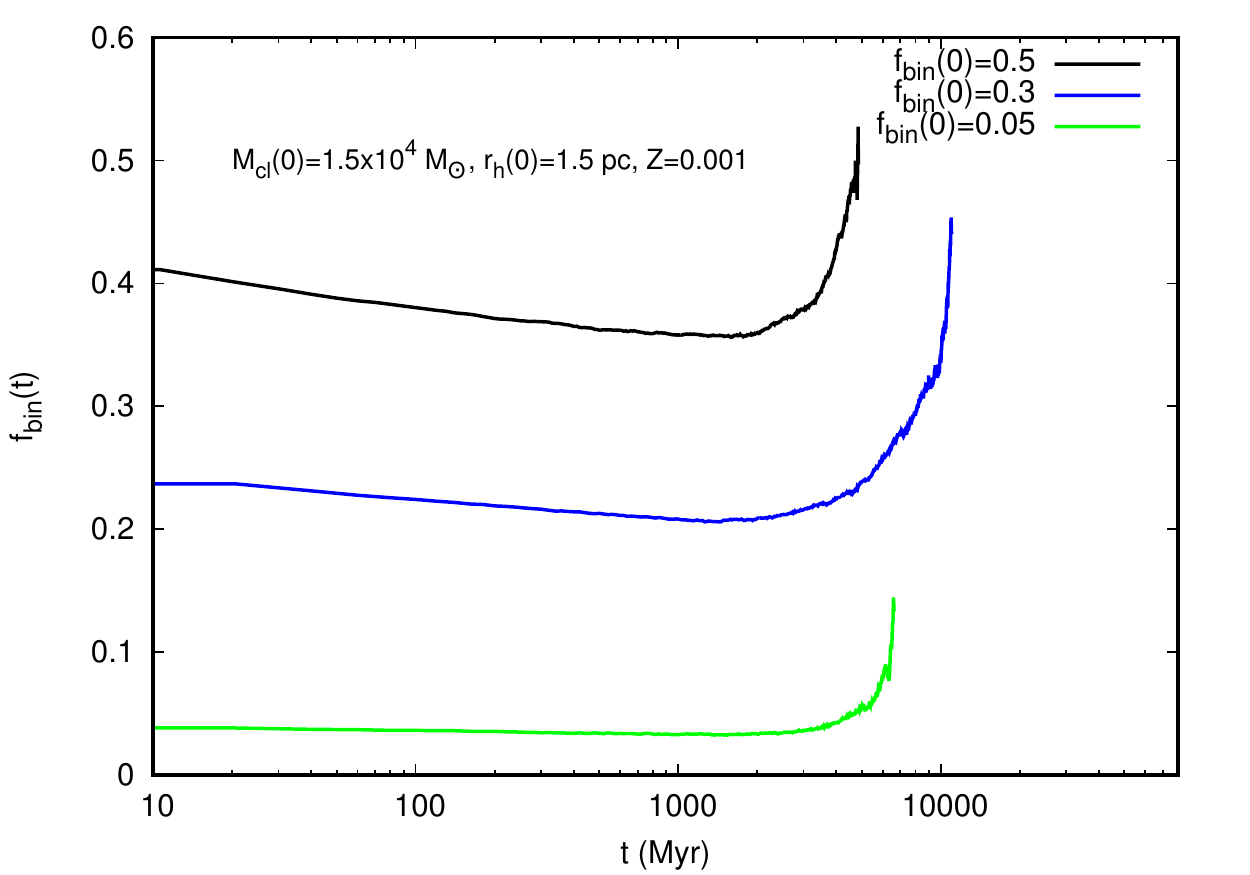}
\includegraphics[width=8.0cm,angle=0]{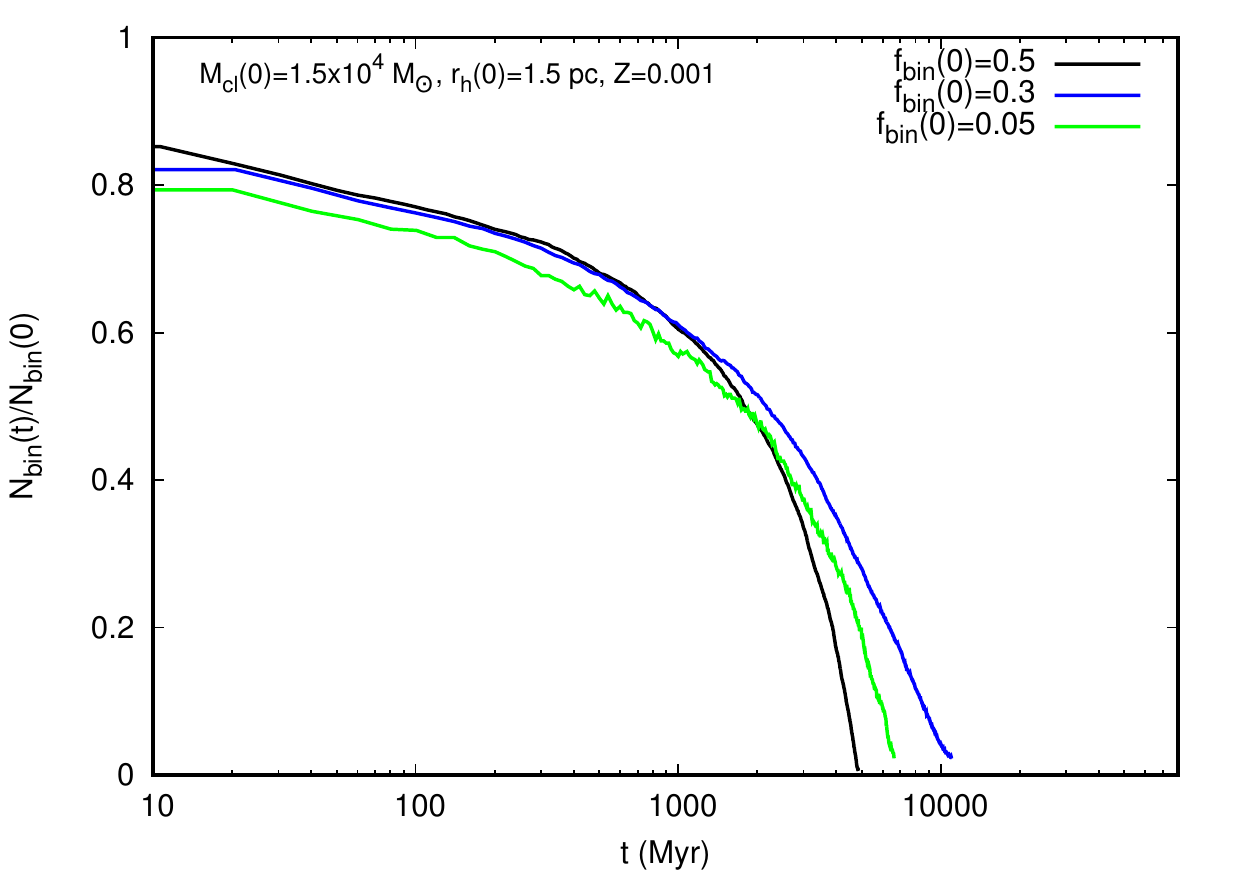}\\
\includegraphics[width=8.0cm,angle=0]{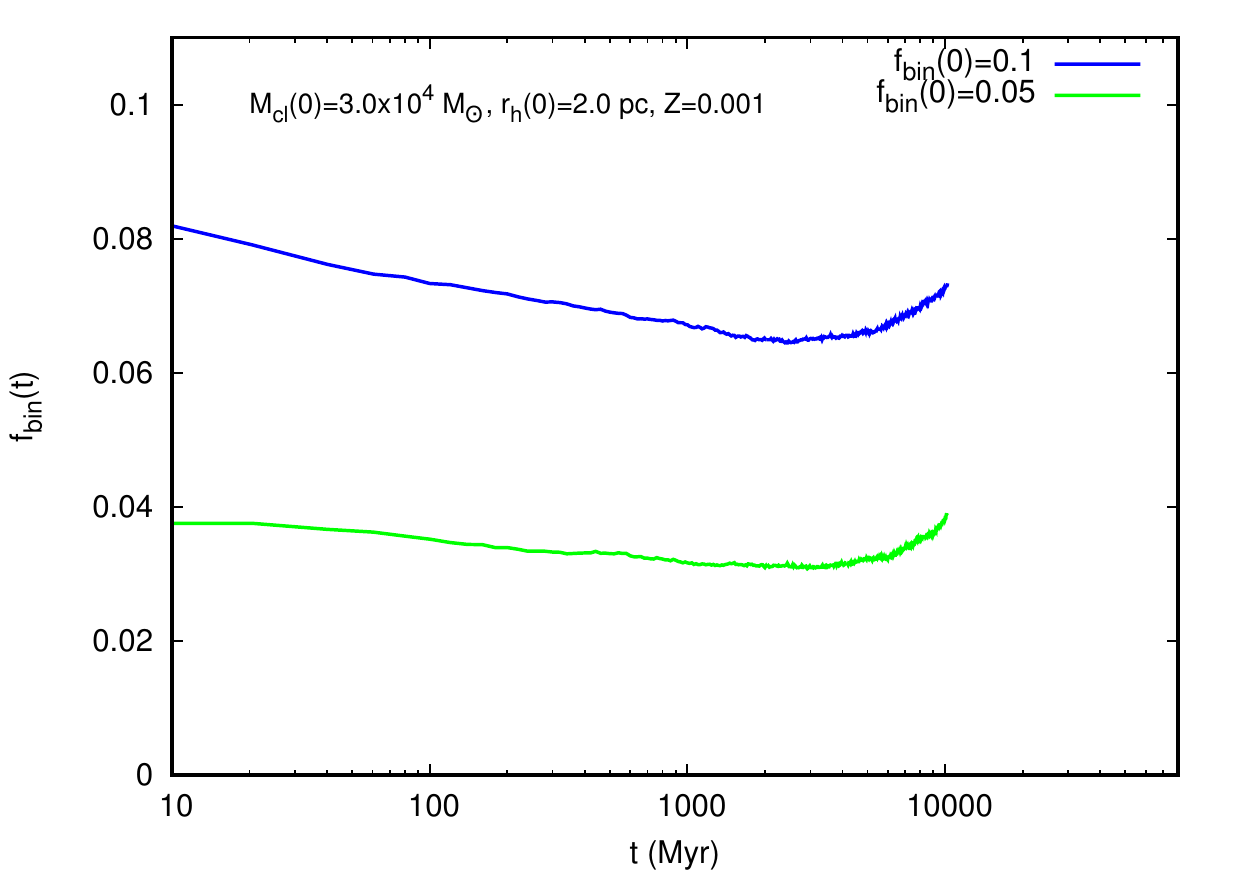}
\includegraphics[width=8.0cm,angle=0]{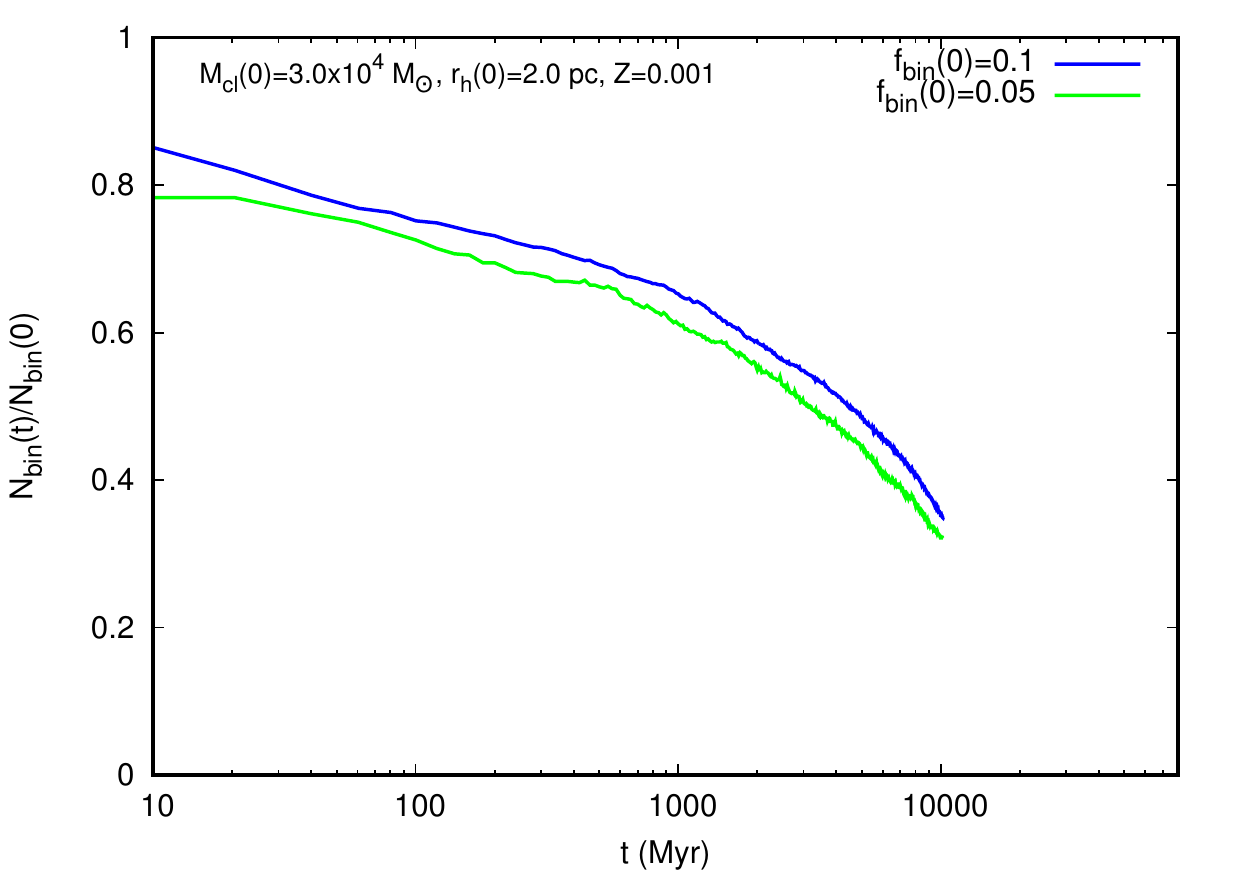}\\
\includegraphics[width=8.0cm,angle=0]{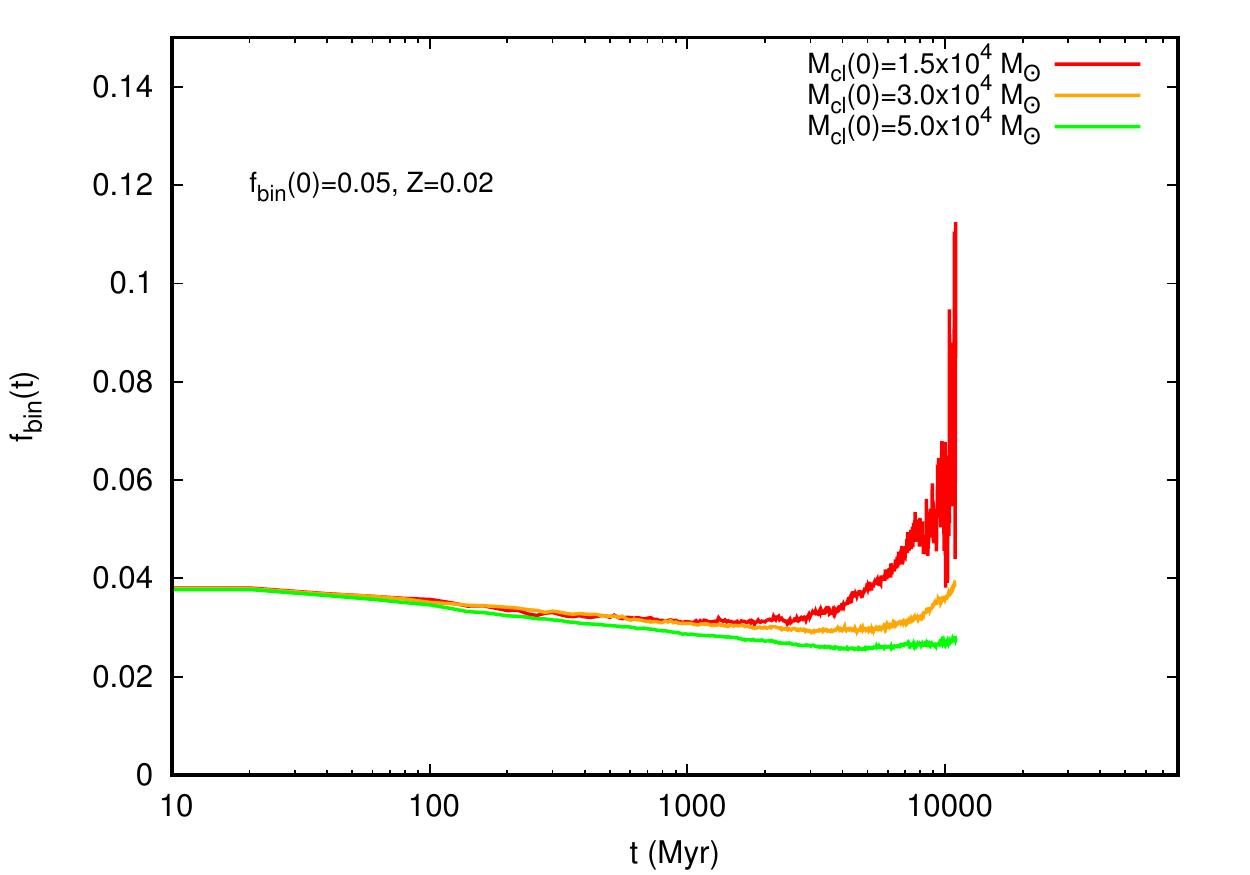}
\includegraphics[width=8.0cm,angle=0]{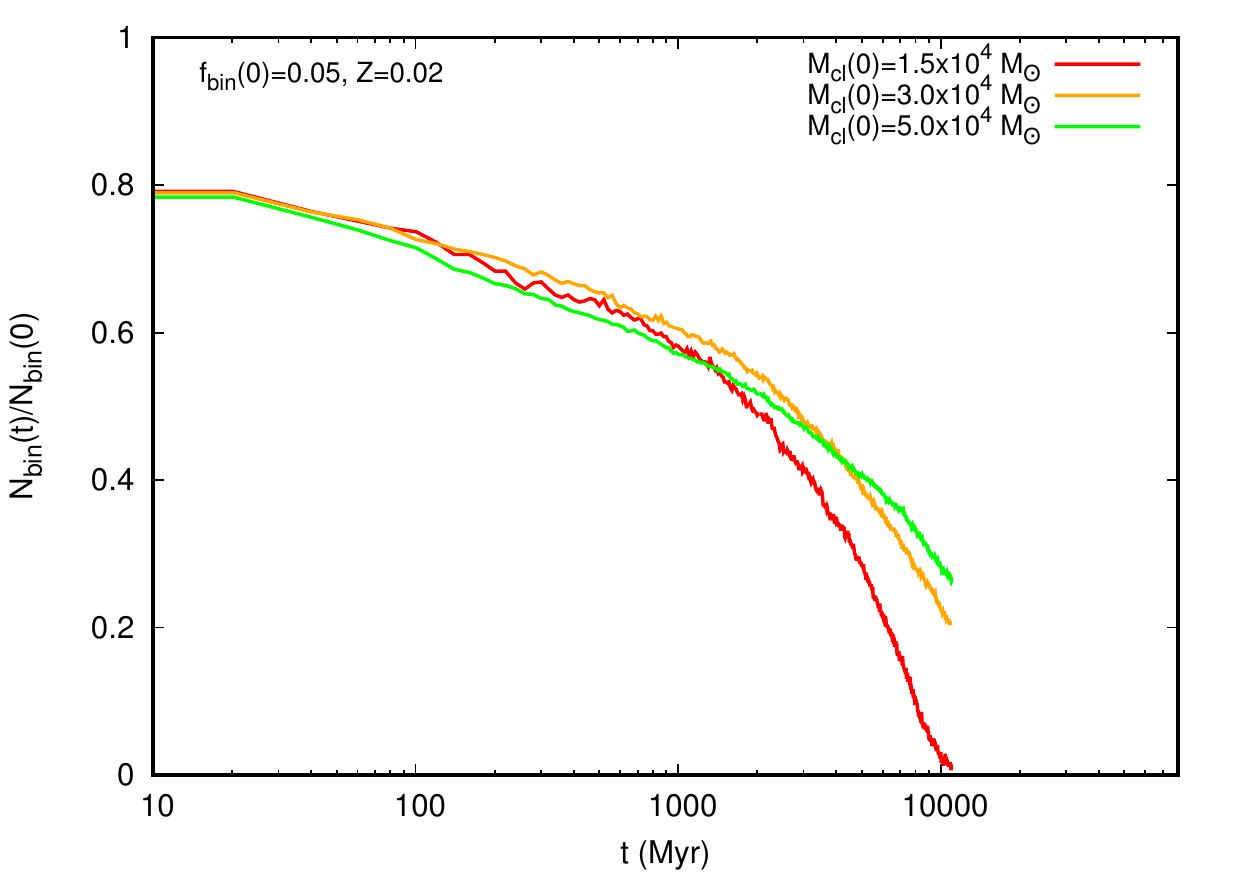}
	\caption{The time evolution of the instantaneous binary fraction, $\fbin(t)$ (left),
	and the total number of binaries, $\nbin(t)$, normalized w.r.t.
	the initial number of binaries, $\nbin(0)$ (right),
	in a representative set of computed models with initial parameters as indicated in the
	legends (see Table~\ref{tab1}).}
\label{fig:binevol}
\end{figure*}

\begin{figure*}
\centering
\includegraphics[width=5.94cm,angle=0]{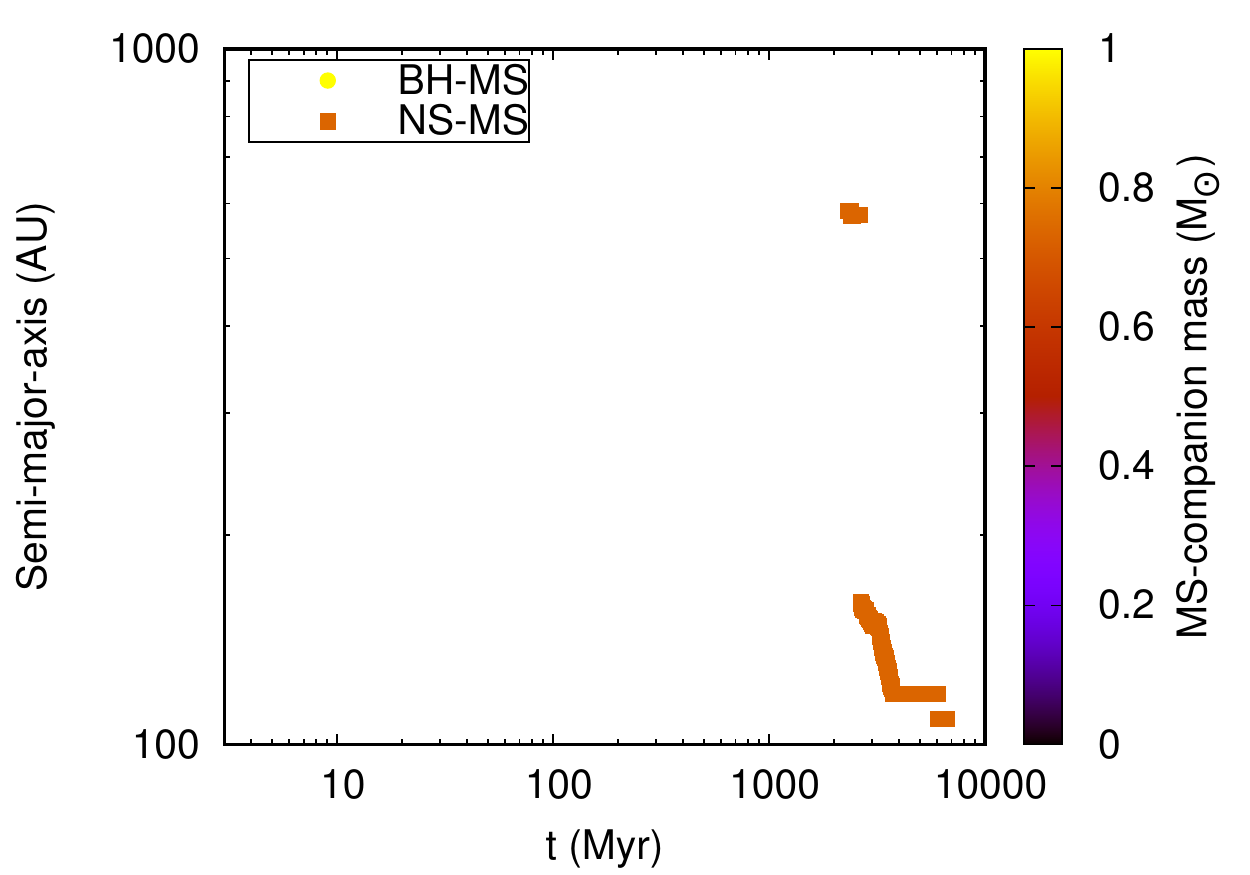}
\hspace{-0.25cm}
\includegraphics[width=5.94cm,angle=0]{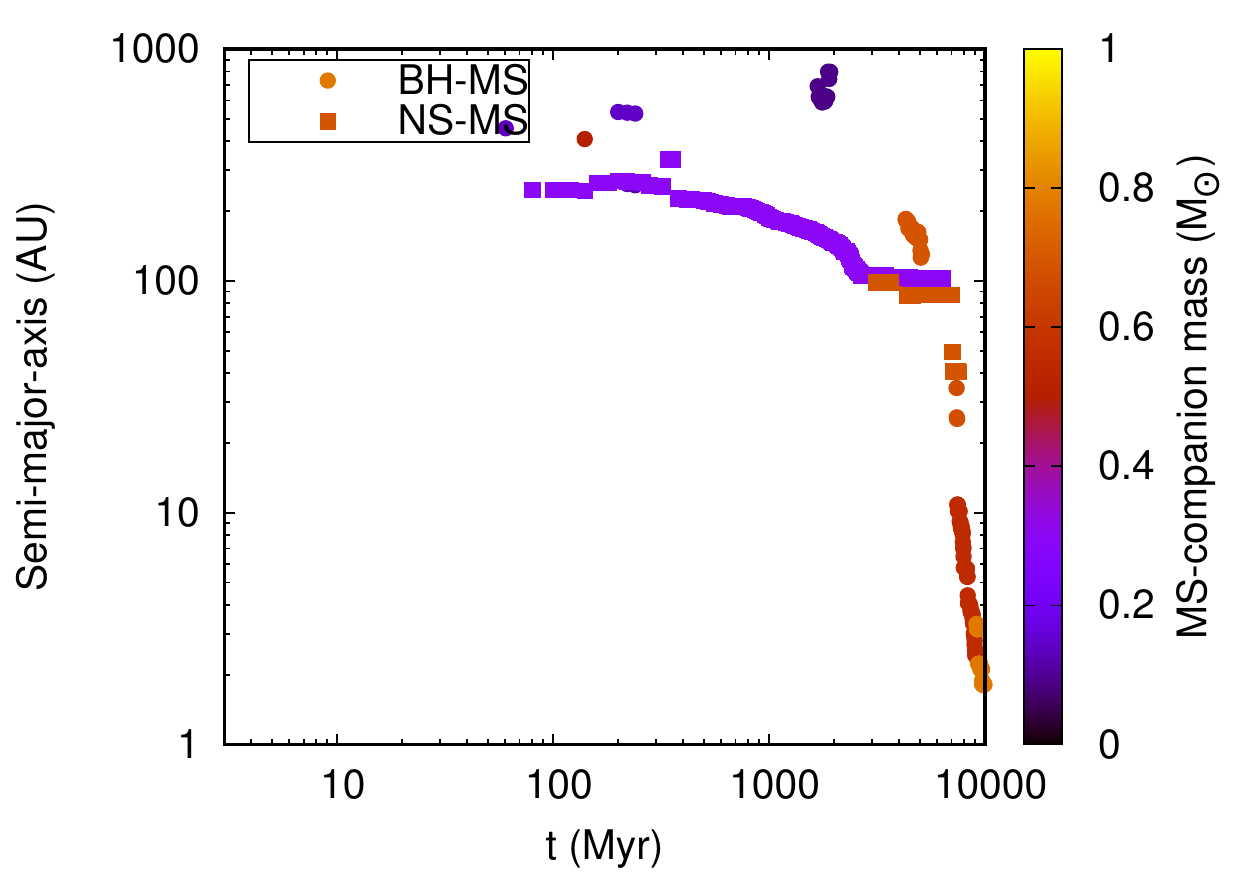}
\hspace{-0.25cm}
\includegraphics[width=5.94cm,angle=0]{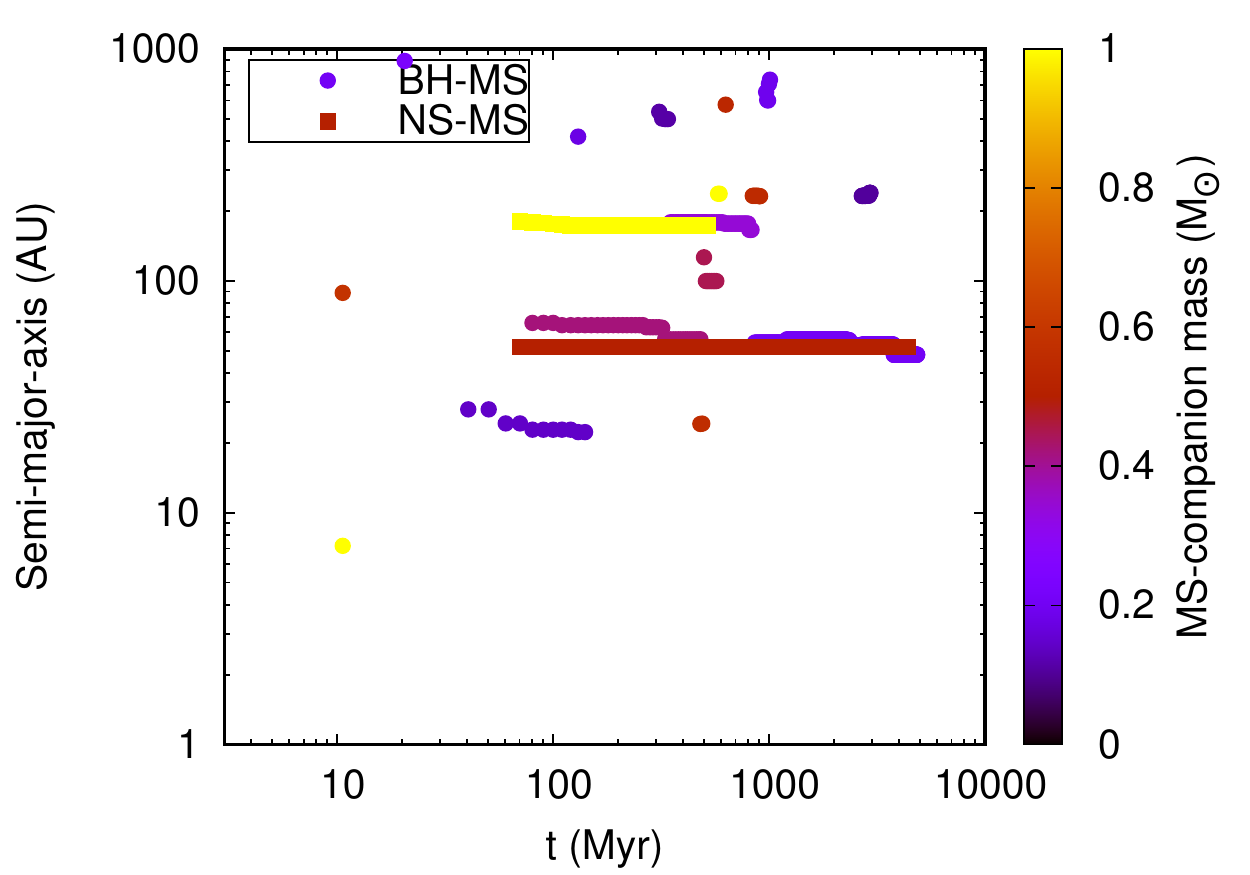}\\
\includegraphics[width=5.94cm,angle=0]{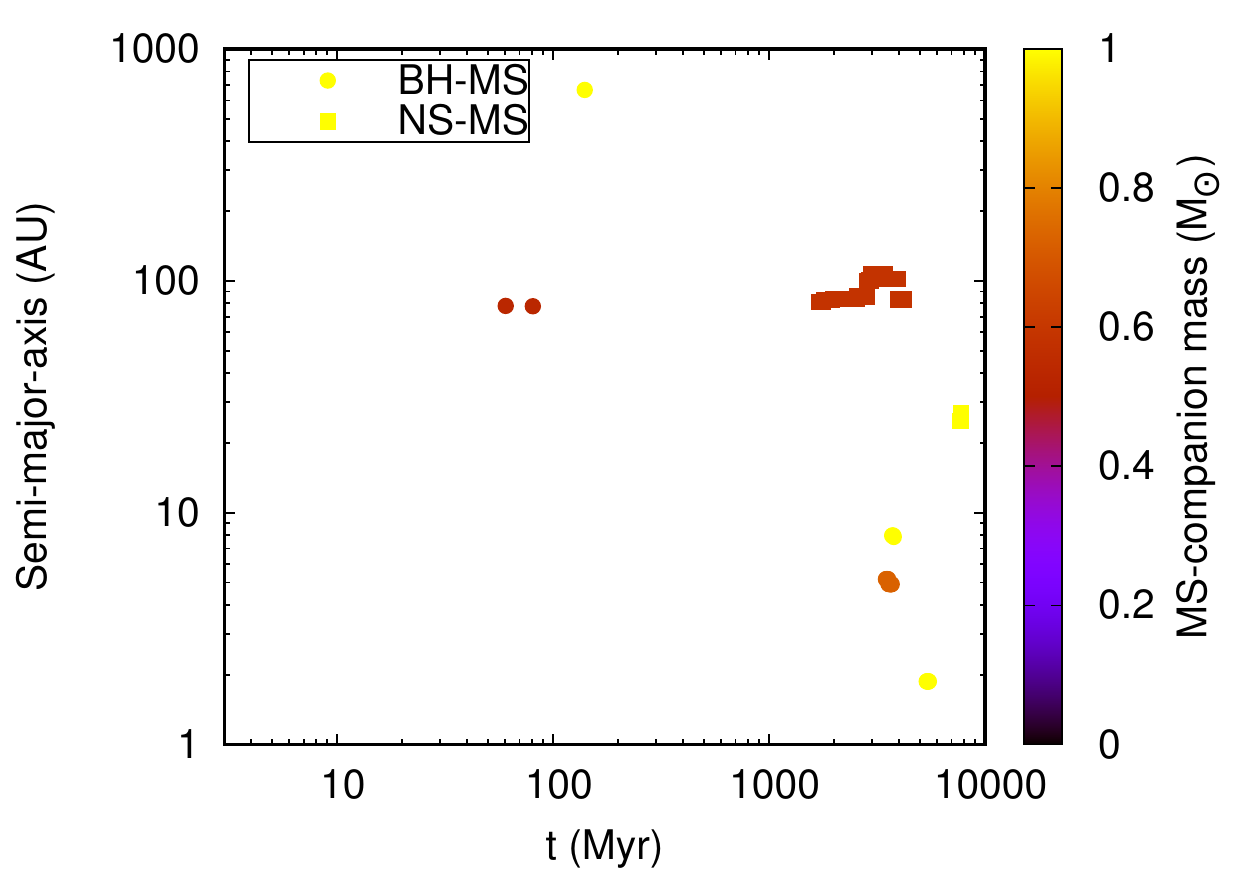}
\hspace{-0.25cm}
\includegraphics[width=5.94cm,angle=0]{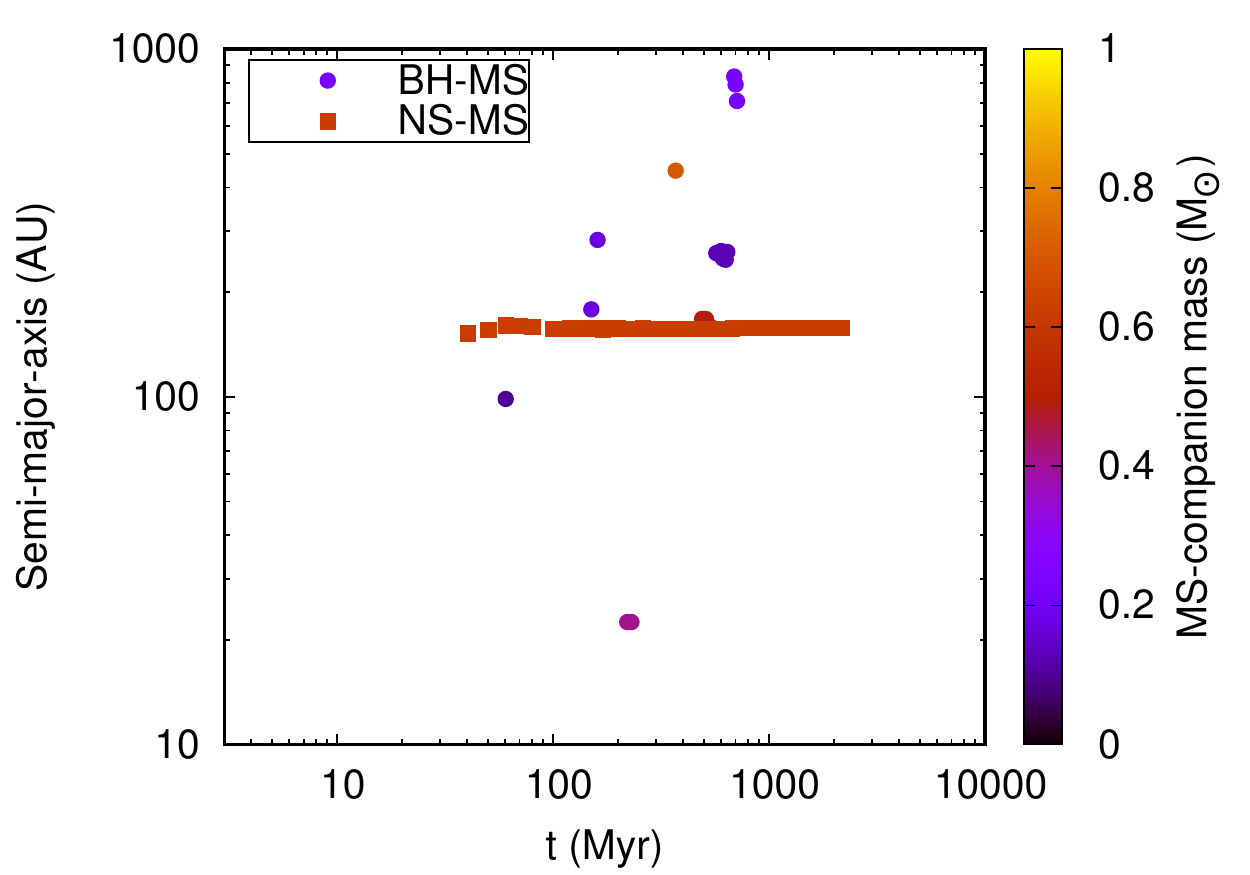}
\hspace{-0.25cm}
\includegraphics[width=5.94cm,angle=0]{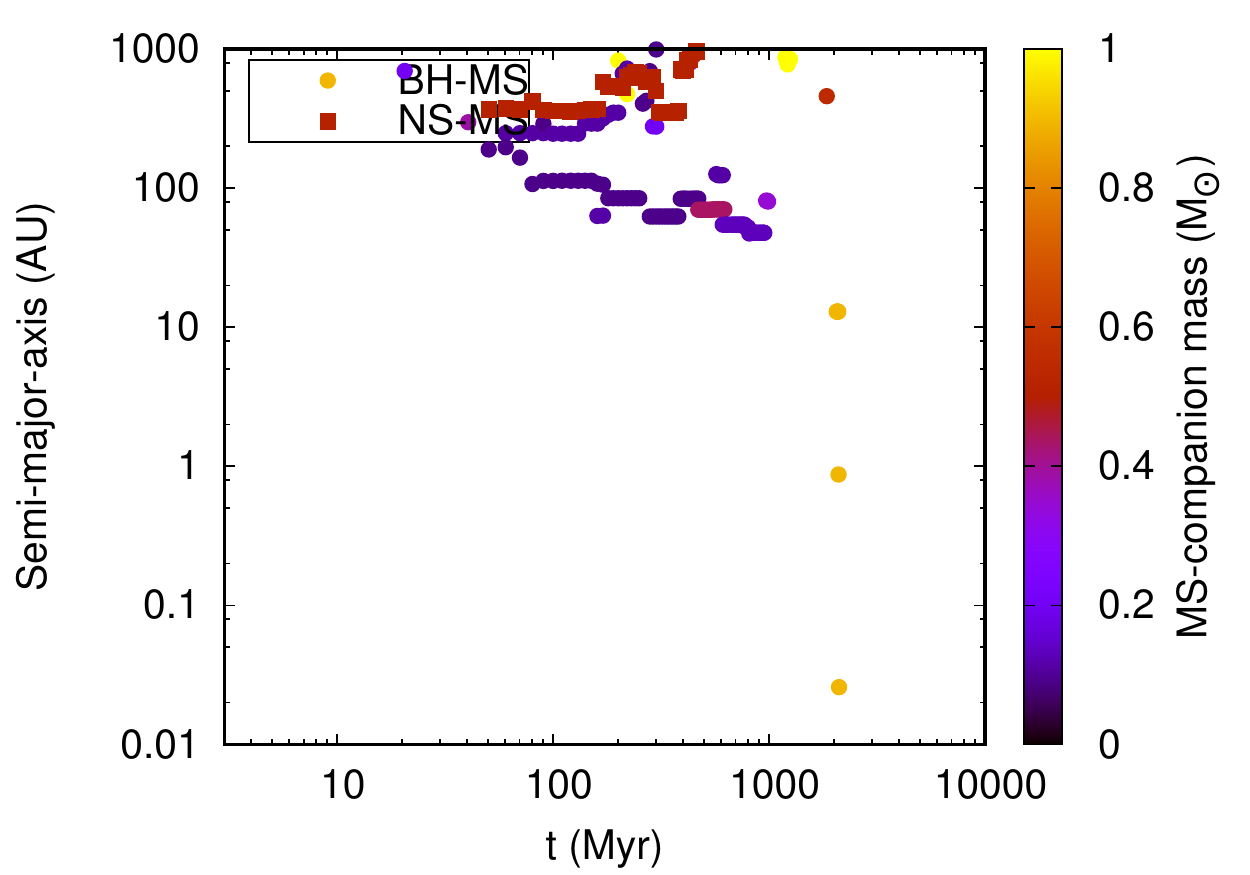}
	\caption{Black hole-main sequence (BH-MS; filled circles) and neutron star-main sequence
	(NS-MS; filled squares) binaries in the computed
	$\mcl(0)\approx1.5\times10^4\Ms$ models with evolutionary time,
	$t$, as a function of increasing primordial binary fraction,
	$\fbin(0)\approx0.05,0.30,0.50$ (left to right panels), and
	for metallicities $Z=0.001$ (top panel) and $0.02$ (bottom panel). On each panel,
	that represents a particular computed model, the vertical axis
	is the instantaneous semi-major-axes ($<1000$ AU) of the BH-MS/NS-MS binaries
	that are bound to the cluster snapshot at time $t$.
	A continuous trail of points represents a specific BH-MS/NS-MS pair. 
	The colour coding (colour bars) represents the mass of the MS companion.
	In this and
	the following Figs.~\ref{fig:bhms_30k}, \ref{fig:bhms_f05}, \ref{fig:bhms_f50}, \ref{fig:bhall_misc},
	\ref{fig:bhmscnt}, and \ref{fig:lagr}, the ``test'' runs in Table~\ref{tab1},
	that include tidal interaction (Sec.~\ref{stellrem}), are excluded,
	which are presented separately in Fig.~\ref{fig:bhbin_circ}.}
\label{fig:bhms_15k}
\end{figure*}

\begin{figure*}
\includegraphics[width=8.0cm,angle=0]{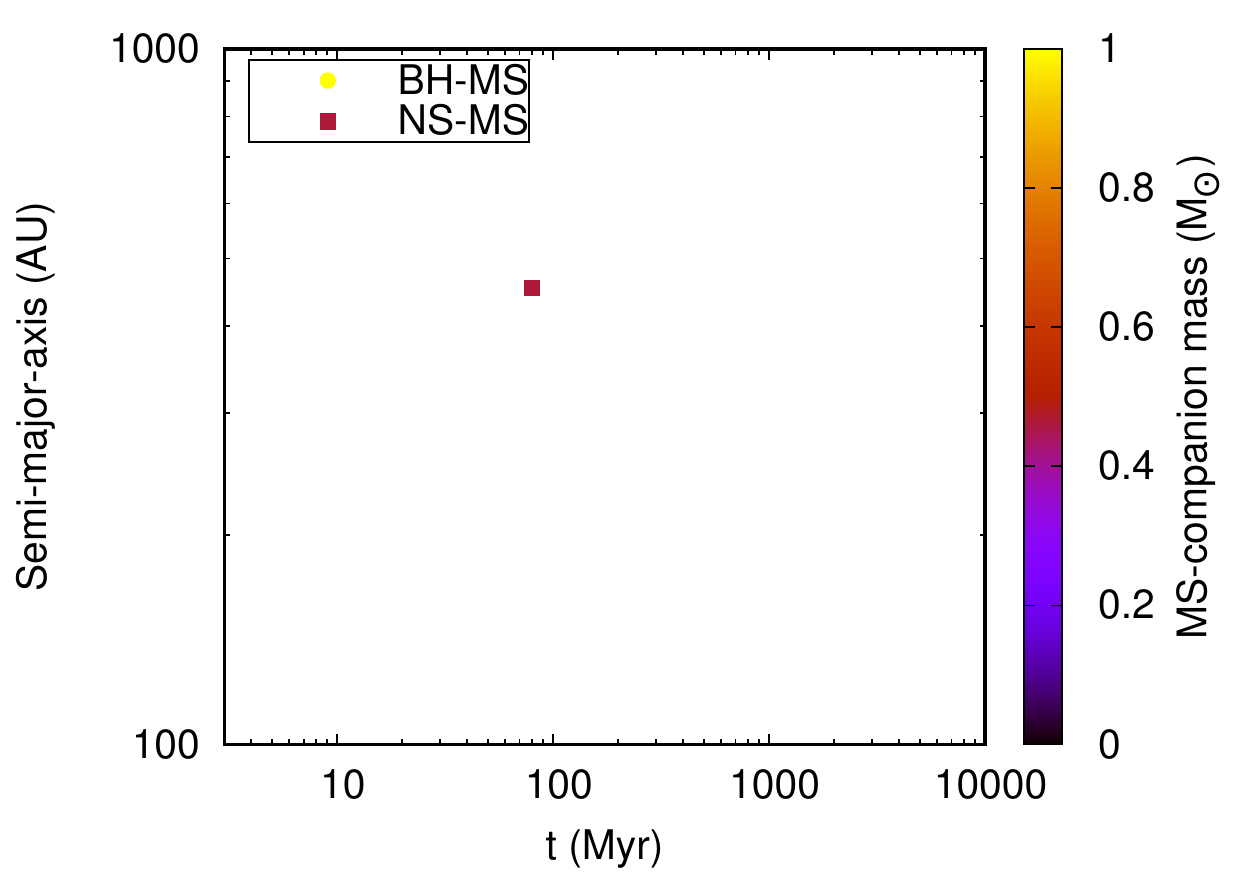}
\includegraphics[width=8.0cm,angle=0]{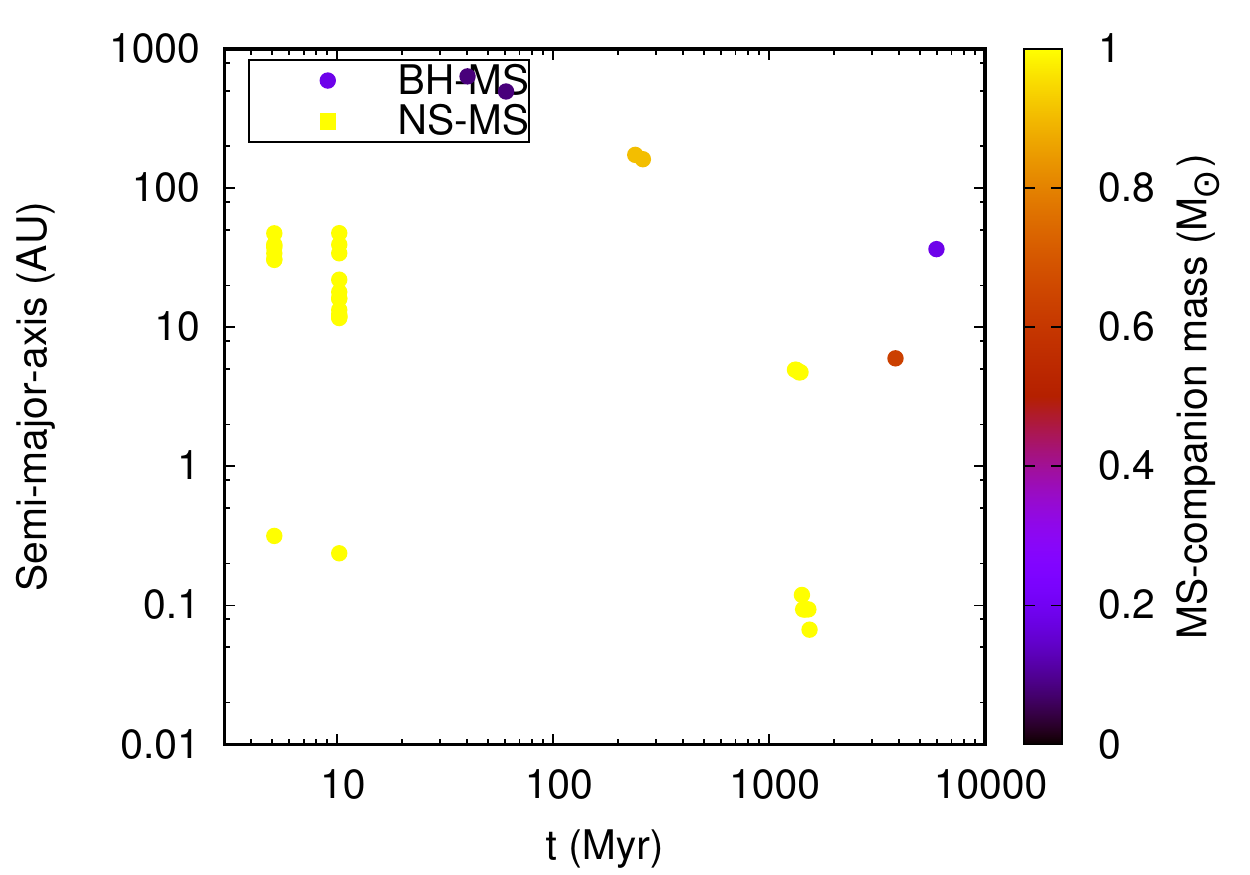}\\
\includegraphics[width=8.0cm,angle=0]{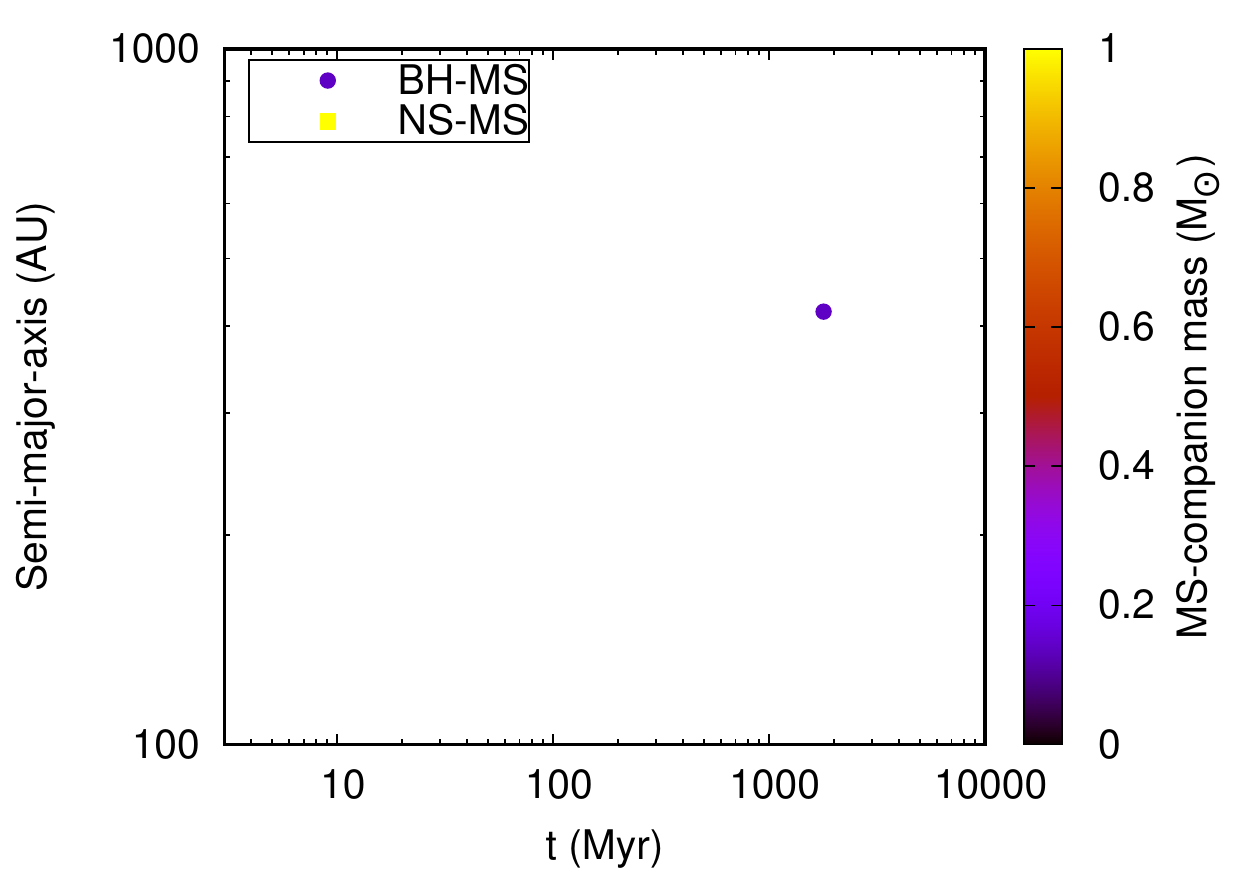}
\includegraphics[width=8.0cm,angle=0]{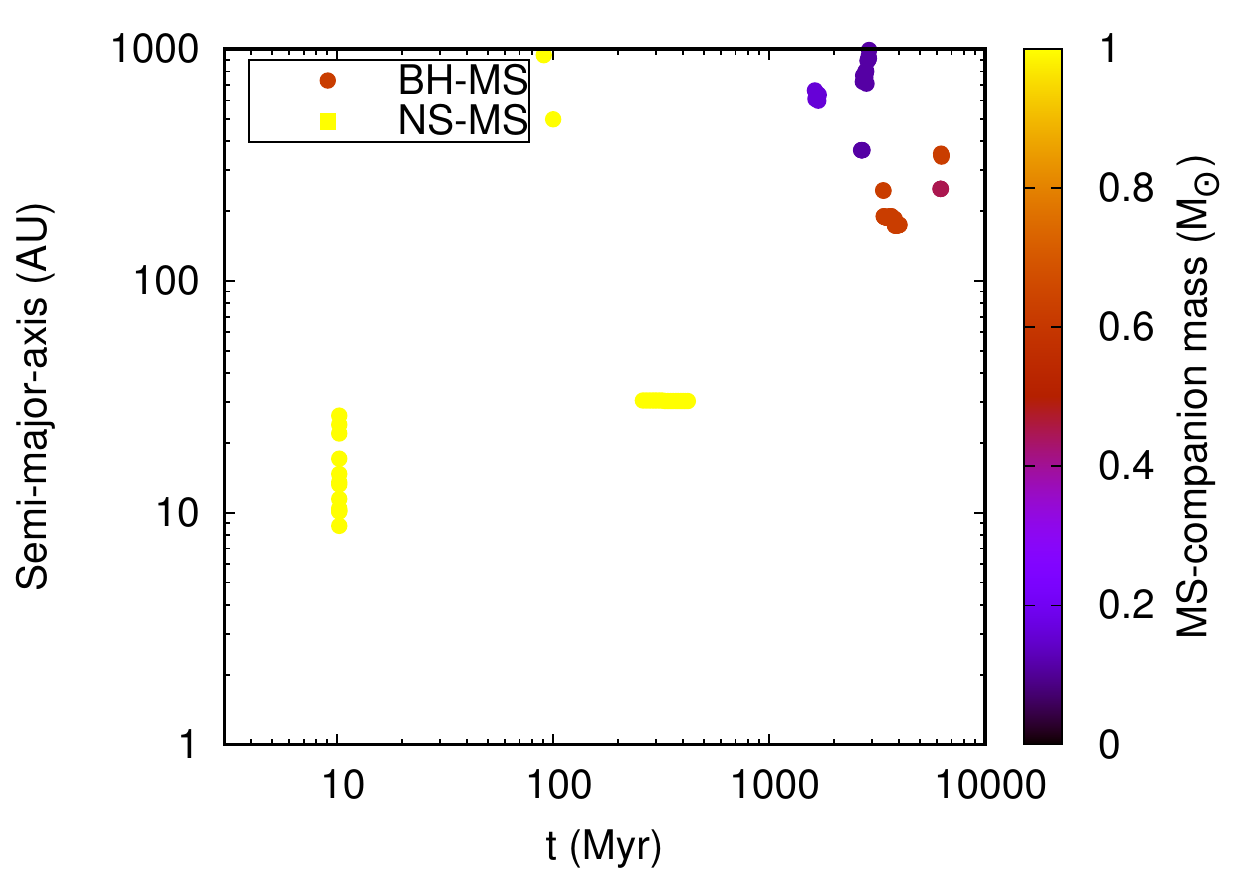}\\
\includegraphics[width=8.0cm,angle=0]{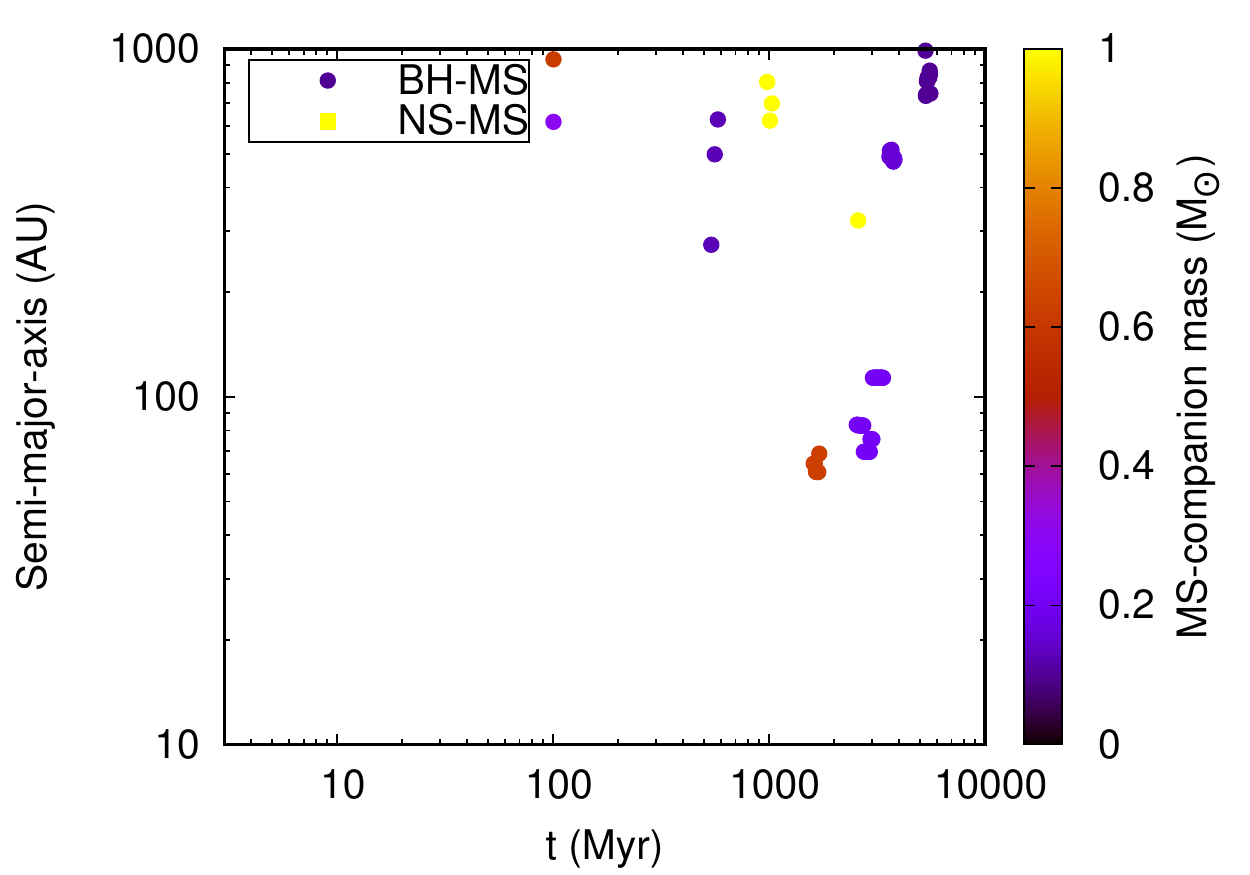}
\hspace{8.0cm}
\caption{BH-MS/NS-MS binaries in the computed $\mcl(0)\approx3.0\times10^4\Ms$ models with evolutionary time,
	$t$, as functions of increasing primordial binary fraction, $\fbin(0)\approx0.05$ (left), $0.10$ (right)
	and metallicity $Z=0.001$ (top), $0.005$ (middle), and $0.02$ (bottom).
	The legends are the same as in Fig.~\ref{fig:bhms_15k}.}
\label{fig:bhms_30k}
\end{figure*}

\begin{figure*}
\centering
\includegraphics[width=5.94cm,angle=0]{figures/BHbin_setC/BH-MS_15kZ001f05_v2.pdf}
\hspace{-0.25cm}
\includegraphics[width=5.94cm,angle=0]{figures/BHbin_setC/BH-MS_30kZ001f05_v2.pdf}
\hspace{-0.25cm}
\includegraphics[width=5.94cm,angle=0]{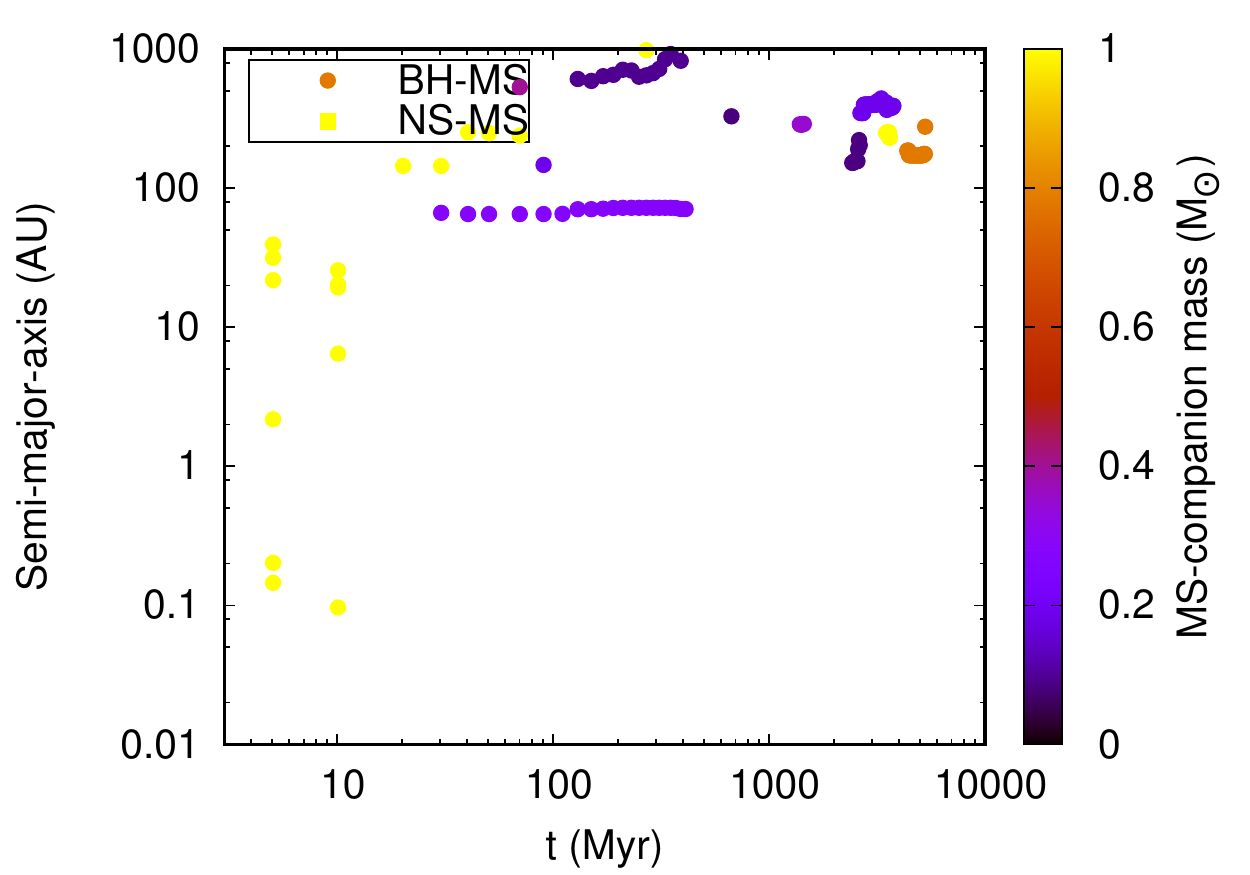}\\
\includegraphics[width=5.94cm,angle=0]{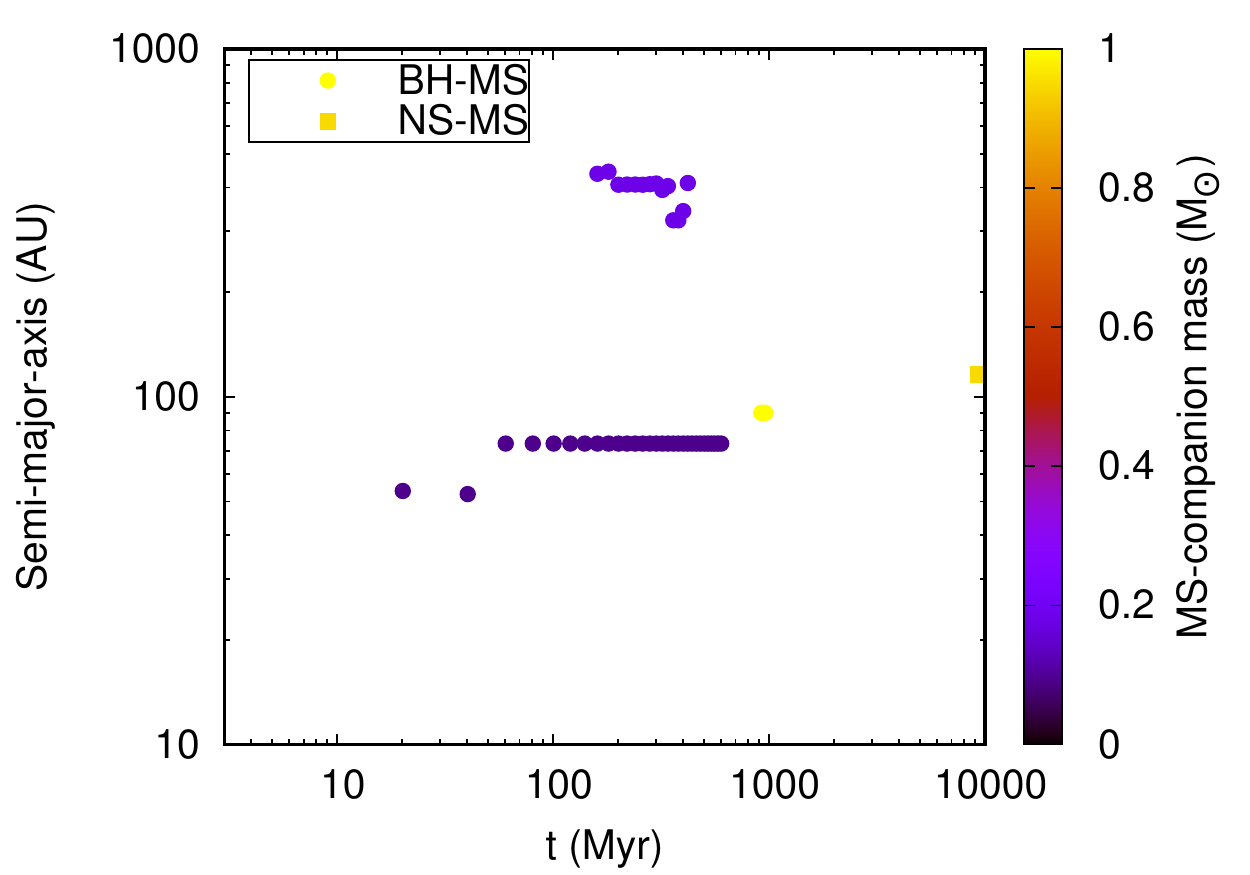}
\hspace{-0.25cm}
\includegraphics[width=5.94cm,angle=0]{figures/BHbin_setC/BH-MS_30kZ005f05_v2.pdf}
\hspace{-0.25cm}
\includegraphics[width=5.94cm,angle=0]{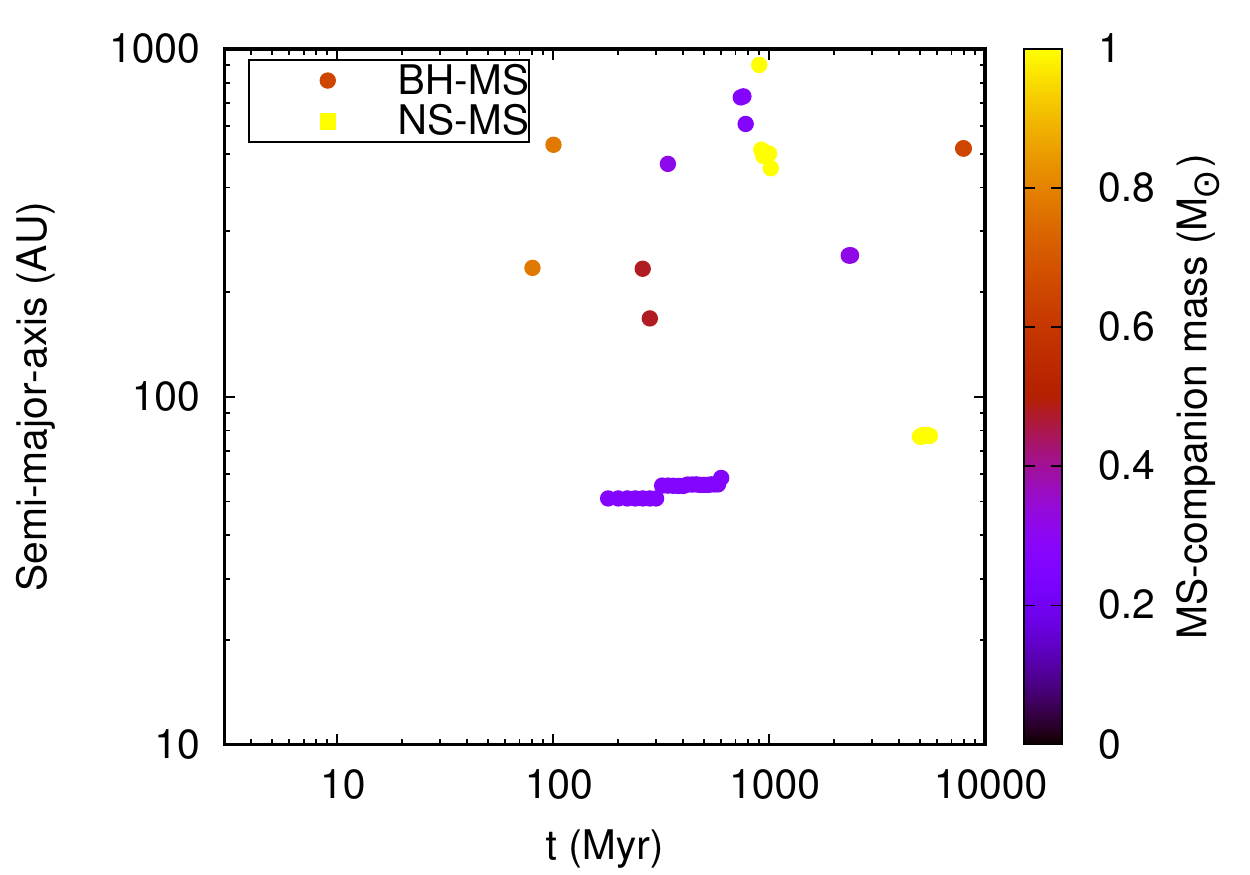}\\
\includegraphics[width=5.94cm,angle=0]{figures/BHbin_setC/BH-MS_15kZ02f05_v2.pdf}
\hspace{-0.25cm}
\includegraphics[width=5.94cm,angle=0]{figures/BHbin_setC/BH-MS_30kZ02f05_v2.pdf}
\hspace{-0.25cm}
\includegraphics[width=5.94cm,angle=0]{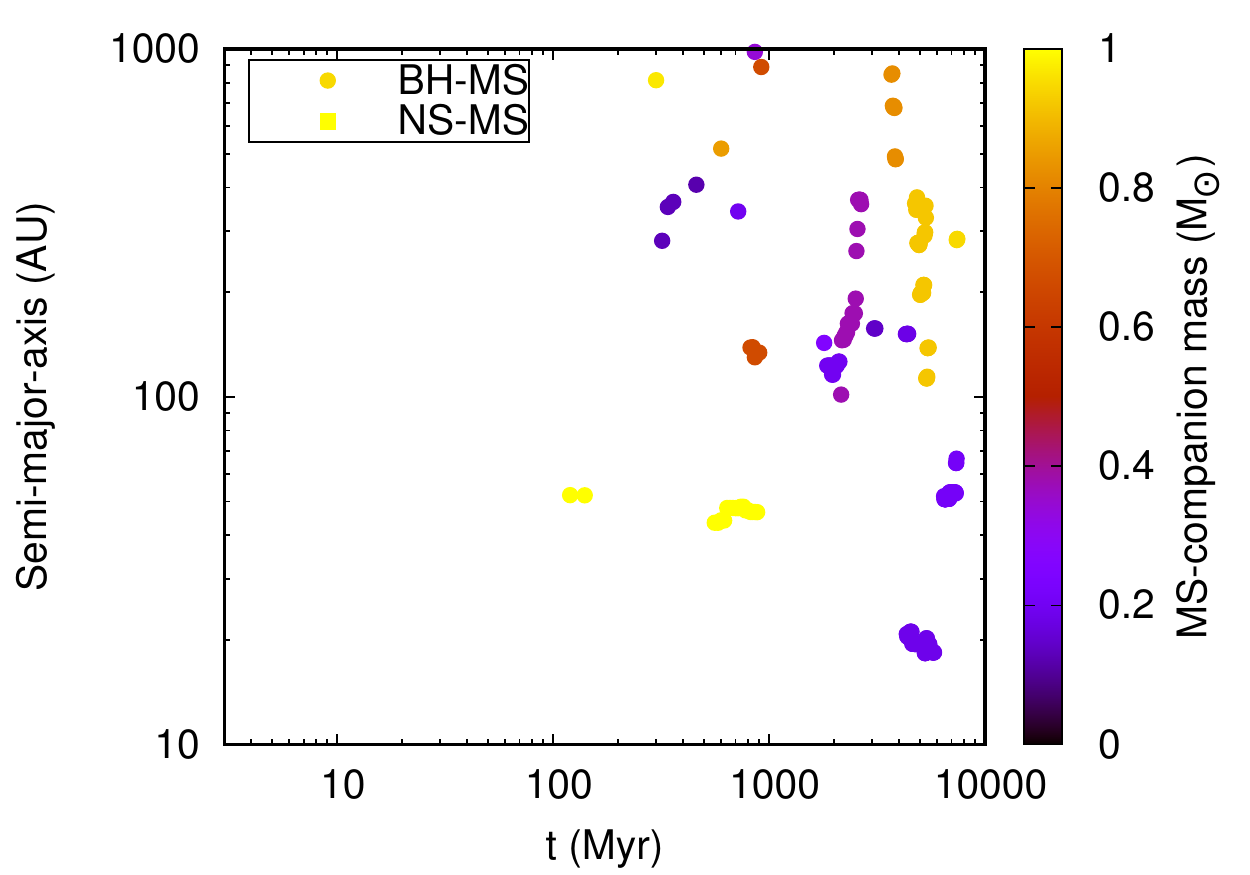}
\caption{BH-MS/NS-MS binaries in the computed $\fbin(0)\approx0.05$ models with evolutionary time,
	$t$, as a function of increasing $\mcl(0)\approx1.5\times10^4\Ms$, $3.0\times10^4\Ms$,
	$5.0\times10^4\Ms$ (left to right) and $Z=0.001$, $0.005$, $0.02$ (top to bottom).
	The legends are the same as in Fig.~\ref{fig:bhms_15k}.}
\label{fig:bhms_f05}
\end{figure*}

\begin{figure*}
\centering
\includegraphics[width=8.0cm,angle=0]{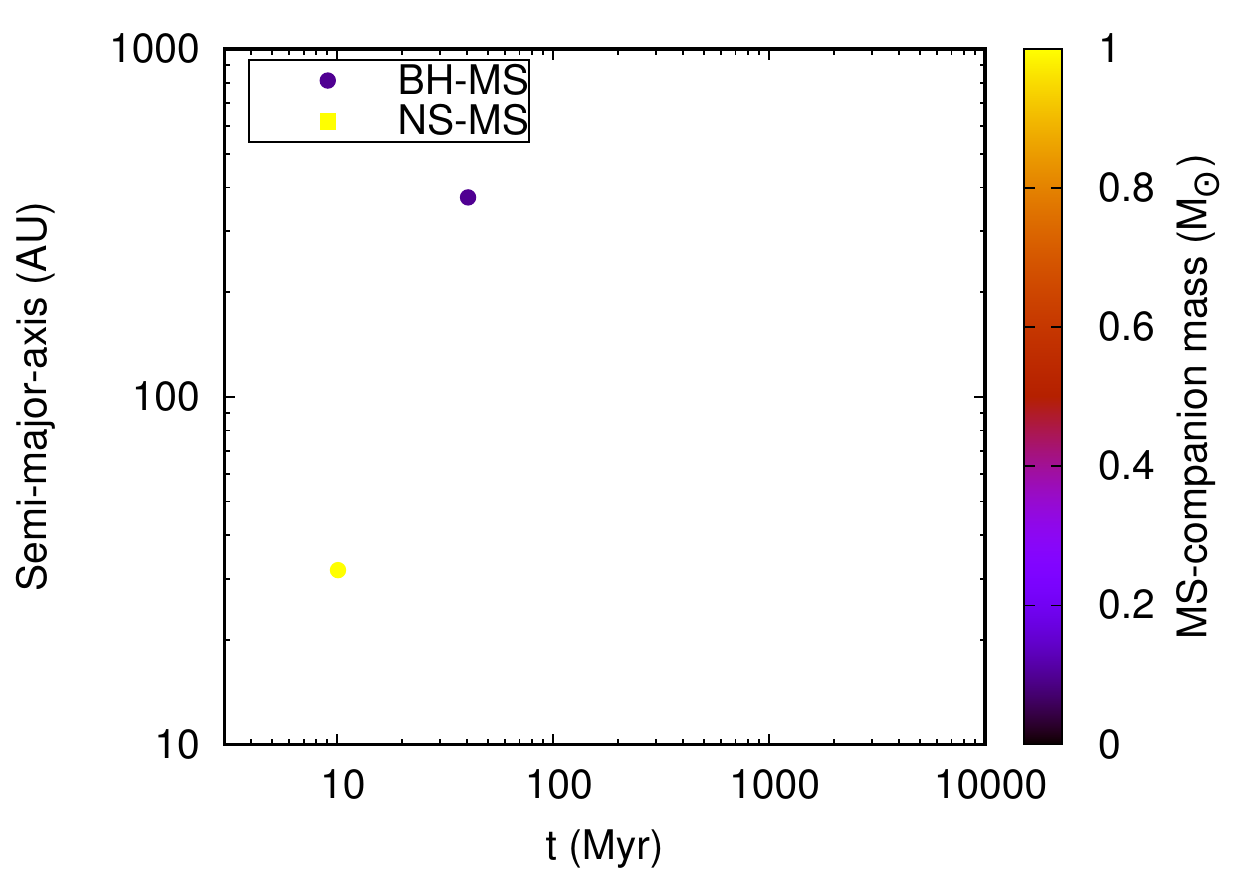}
\includegraphics[width=8.0cm,angle=0]{figures/BHbin_setC/BH-MS_15kZ001f50_v2.pdf}\\
\includegraphics[width=8.0cm,angle=0]{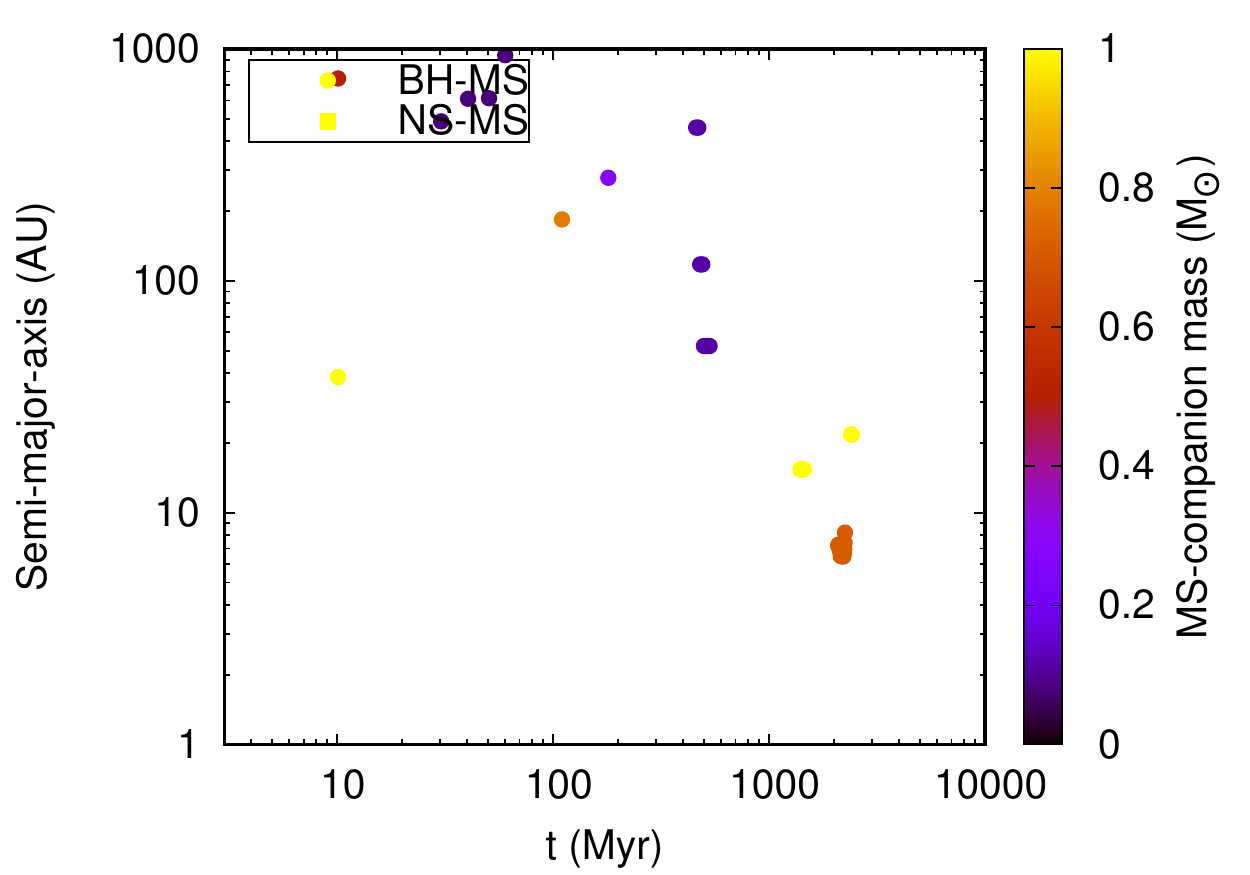}
\includegraphics[width=8.0cm,angle=0]{figures/BHbin_setC/BH-MS_15kZ02f50_v2.pdf}
	\caption{BH-MS/NS-MS binaries in the computed $\fbin(0)\approx0.50$ models with evolutionary time,
	$t$, as a function of increasing $\mcl(0)\approx7.5\times10^3\Ms$ (left) and $1.5\times10^4\Ms$ (right) and
	for metallicities $Z=0.001$ (top) and $0.02$ (bottom).
	The legends are the same as in Fig.~\ref{fig:bhms_15k}.}
\label{fig:bhms_f50}
\end{figure*}

\begin{figure*}
\centering
\includegraphics[width=8.0cm,angle=0]{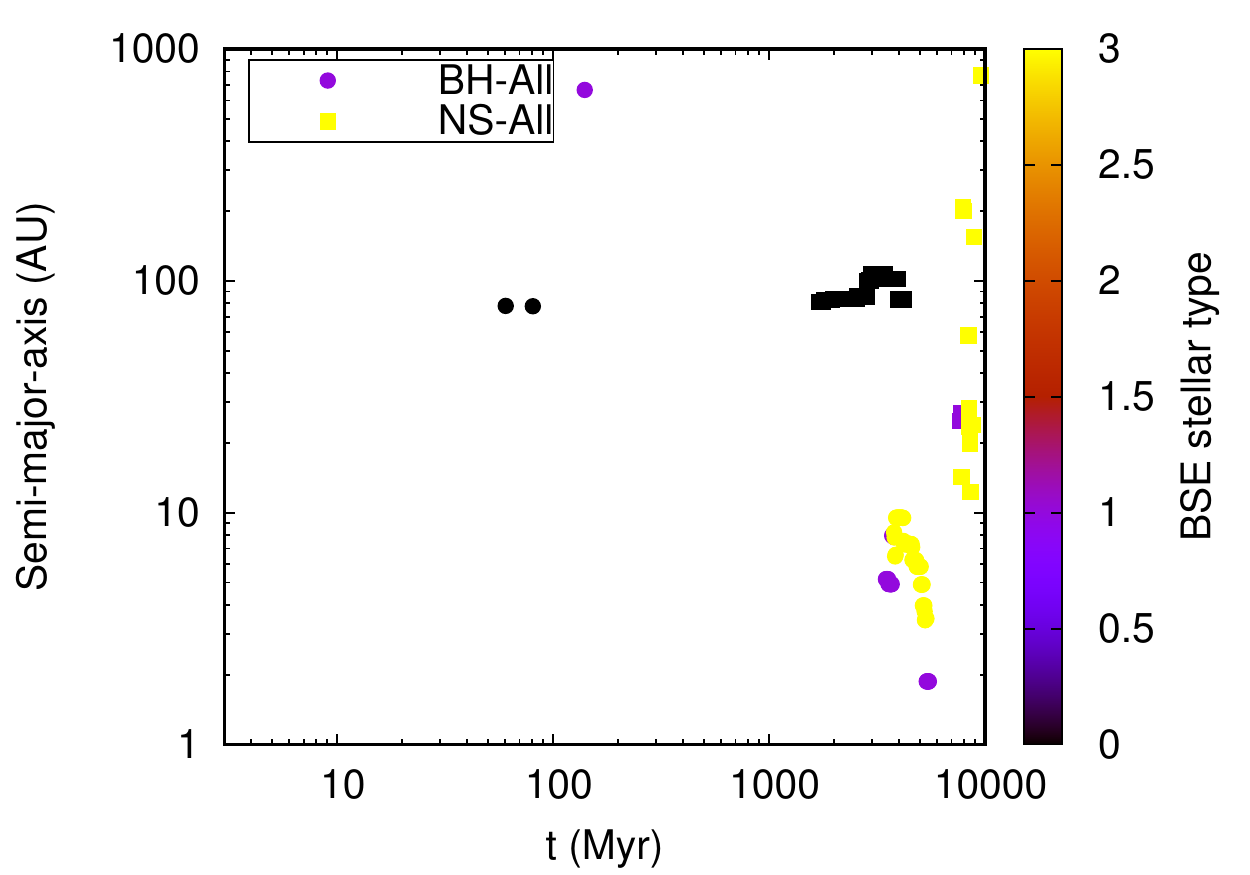}
\includegraphics[width=8.0cm,angle=0]{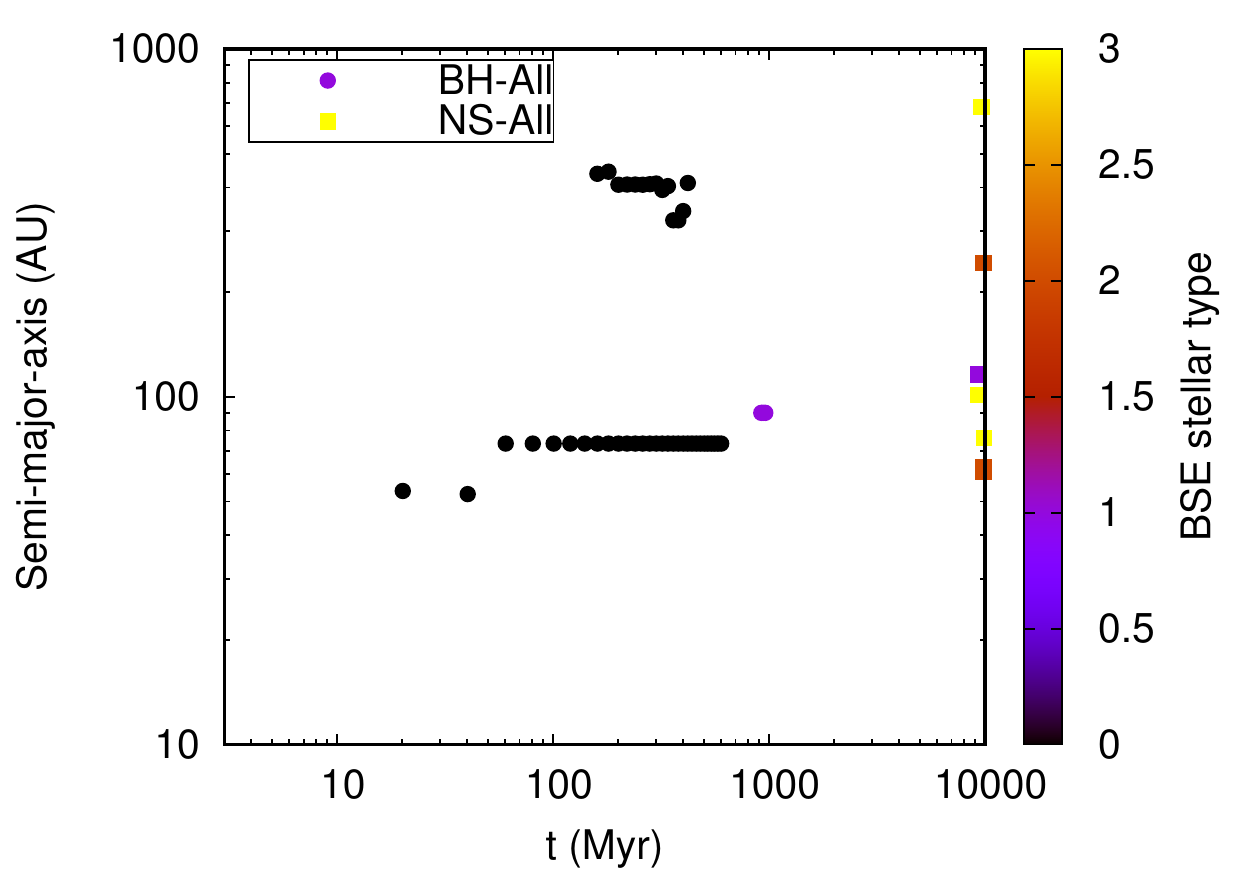}\\
\includegraphics[width=8.0cm,angle=0]{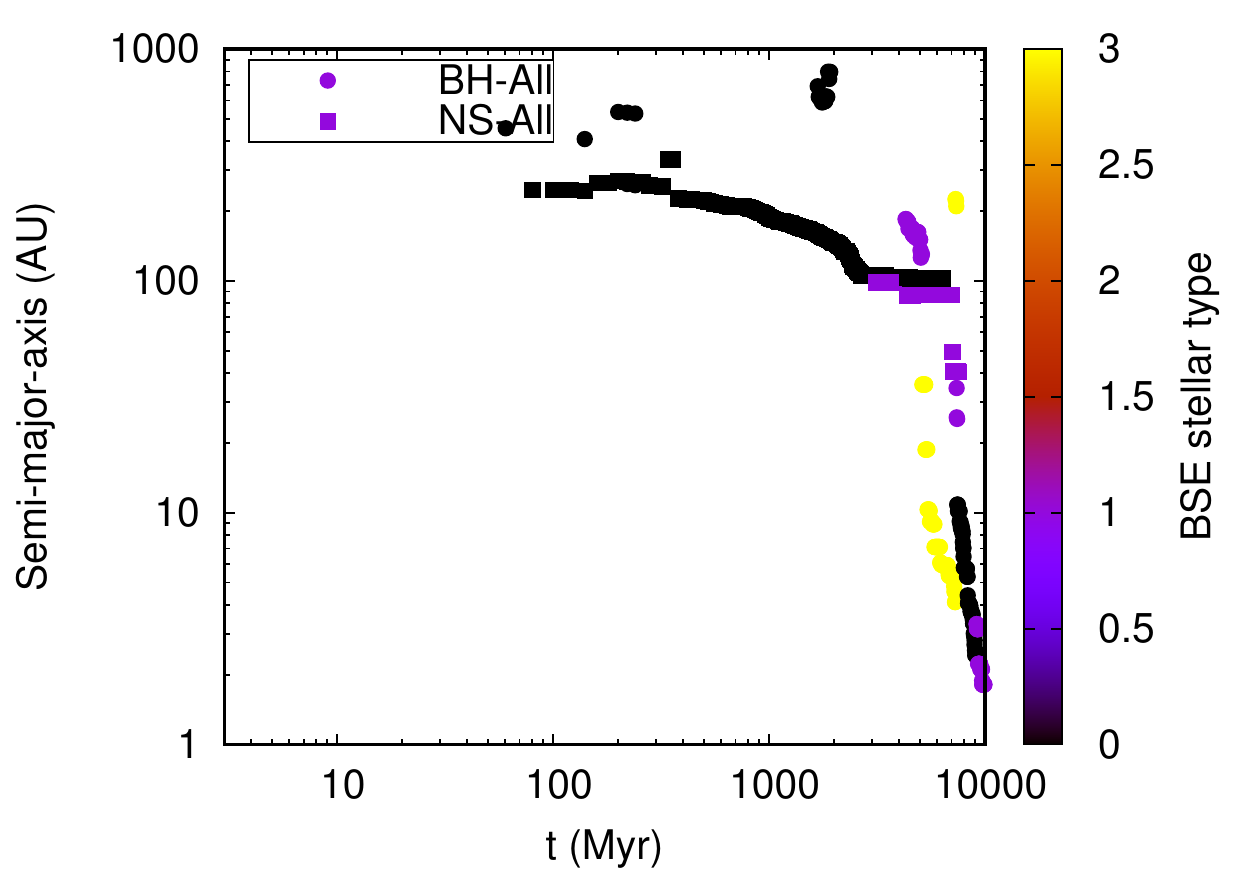}
\includegraphics[width=8.0cm,angle=0]{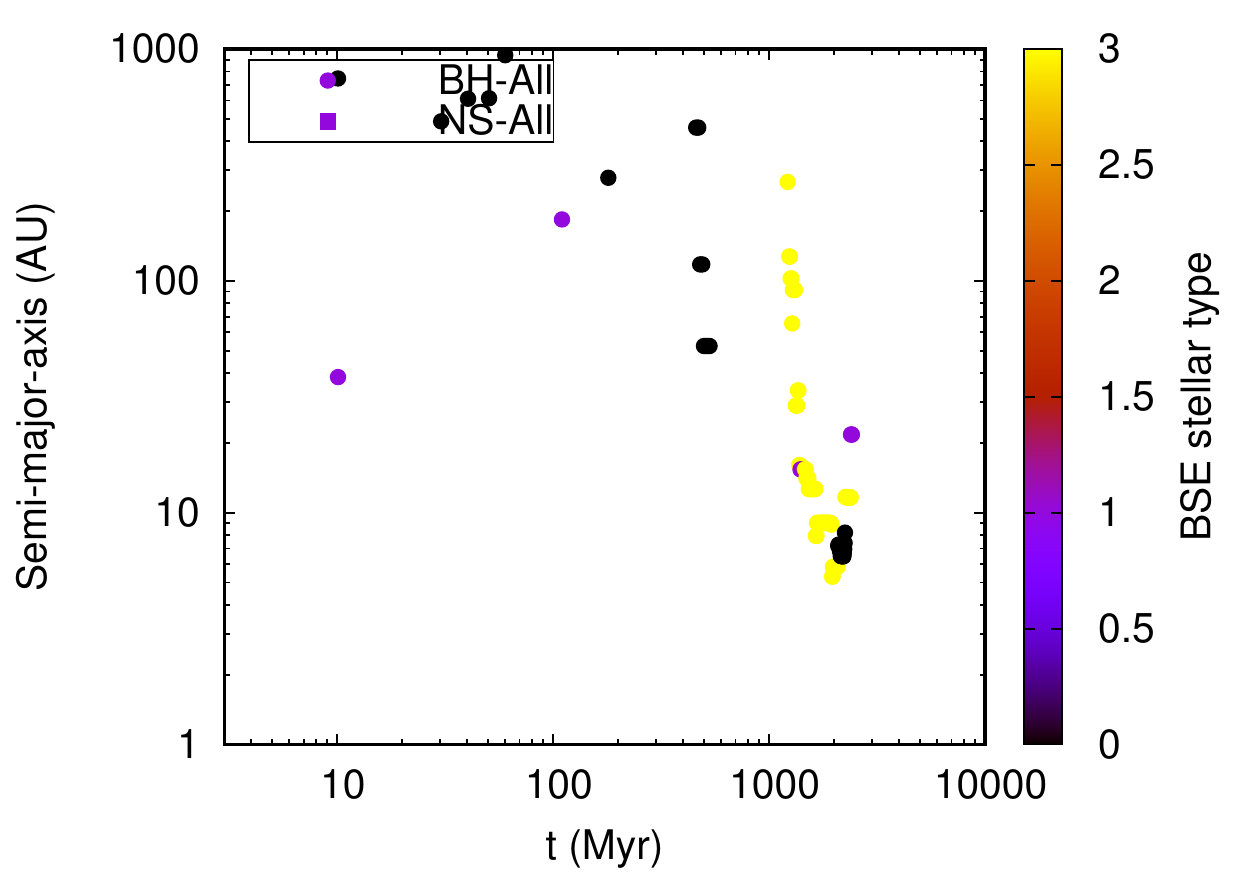}
	\caption{The computed models in Table~\ref{tab1} in which one or more BH/NS
	(filled circle/square)
	are found in binaries with a stellar companion that is evolved beyond
	the MS. In the order top left, top right, bottom left, bottom right
	with $\mcl(0)$, $\fbin(0)$, $Z$:
	$1.5\times10^4\Ms$, 0.05, 0.02; $1.5\times10^4\Ms$, 0.05, 0.005;
	$1.5\times10^4\Ms$, 0.30, 0.001; $7.5\times10^3\Ms$, 0.50, 0.02. In these panels,
	the points are colour-coded according to the corresponding stellar companion's
	\bse stellar type \citep{Hurley_2000}, indicating its evolutionary status ($0/1=$MS,
	$\geq2=$evolved).}
\label{fig:bhall_misc}
\end{figure*}

\begin{figure*}
\centering
\includegraphics[width=5.94cm,angle=0]{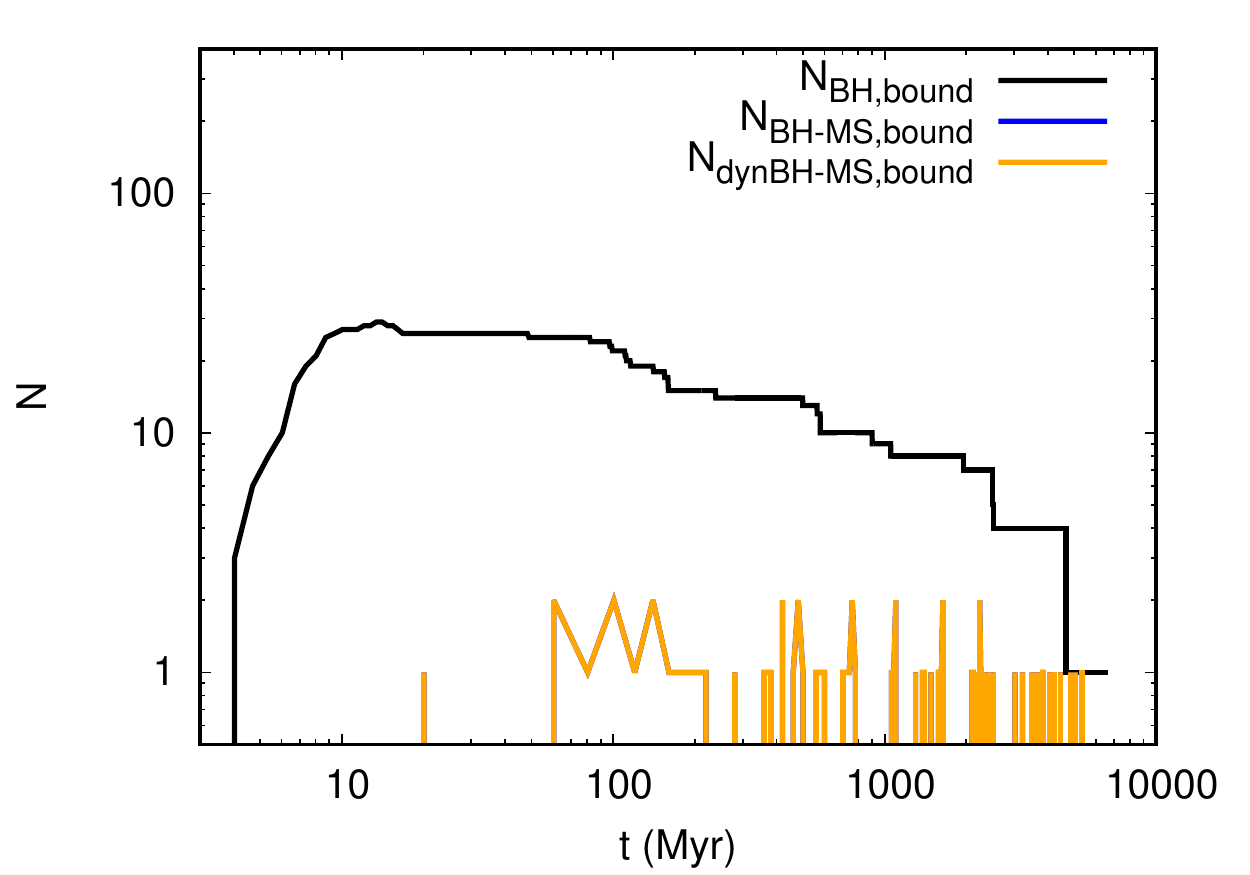}
\hspace{-0.25cm}
\includegraphics[width=5.94cm,angle=0]{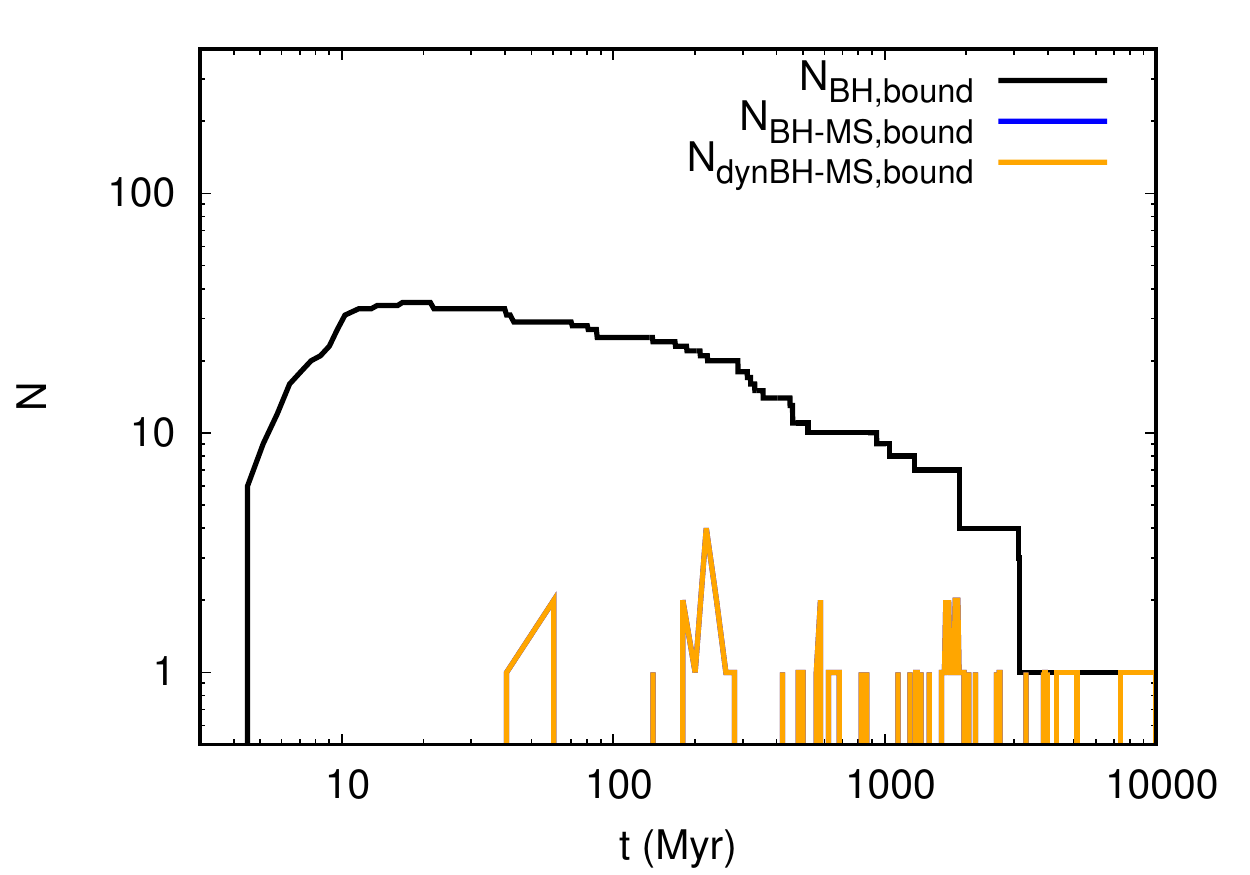}
\hspace{-0.25cm}
\includegraphics[width=5.94cm,angle=0]{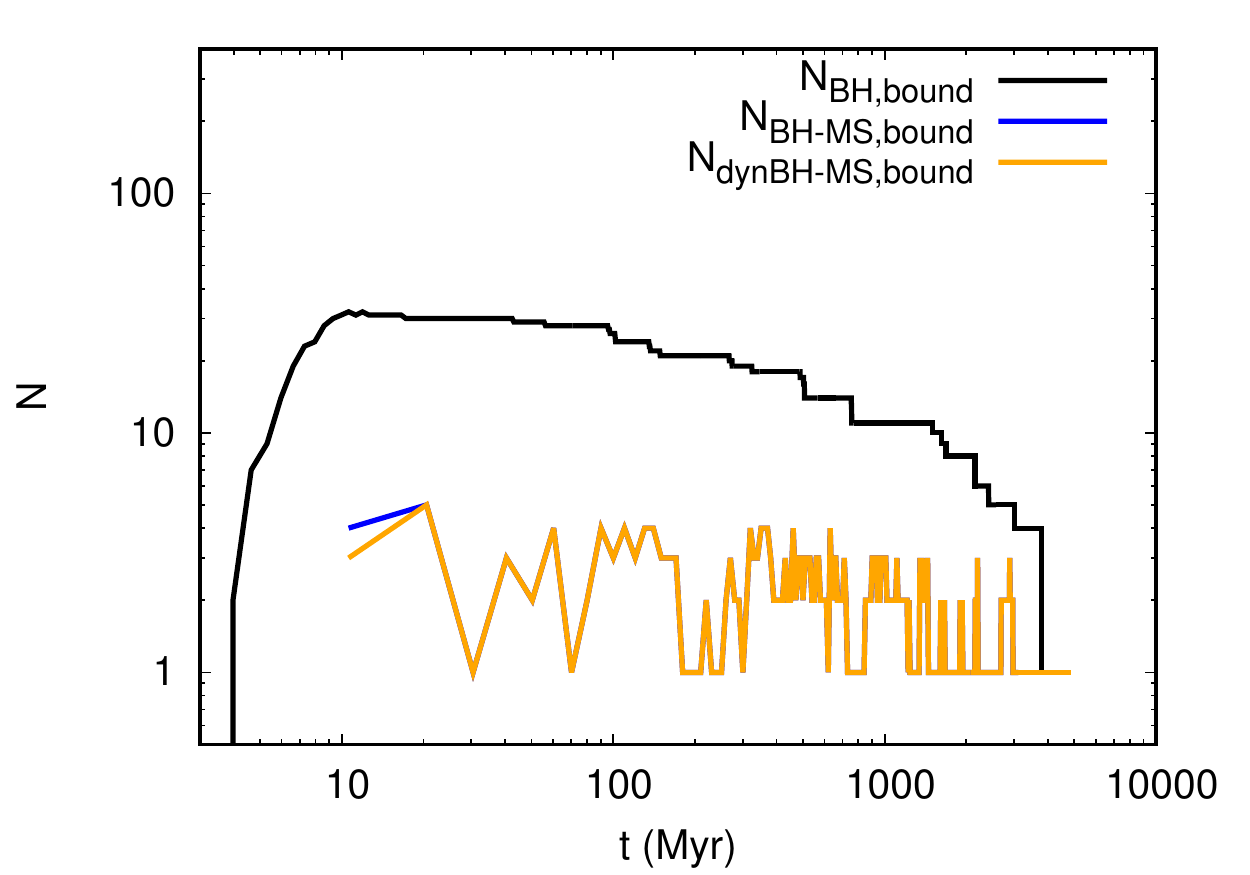}\\
\includegraphics[width=5.94cm,angle=0]{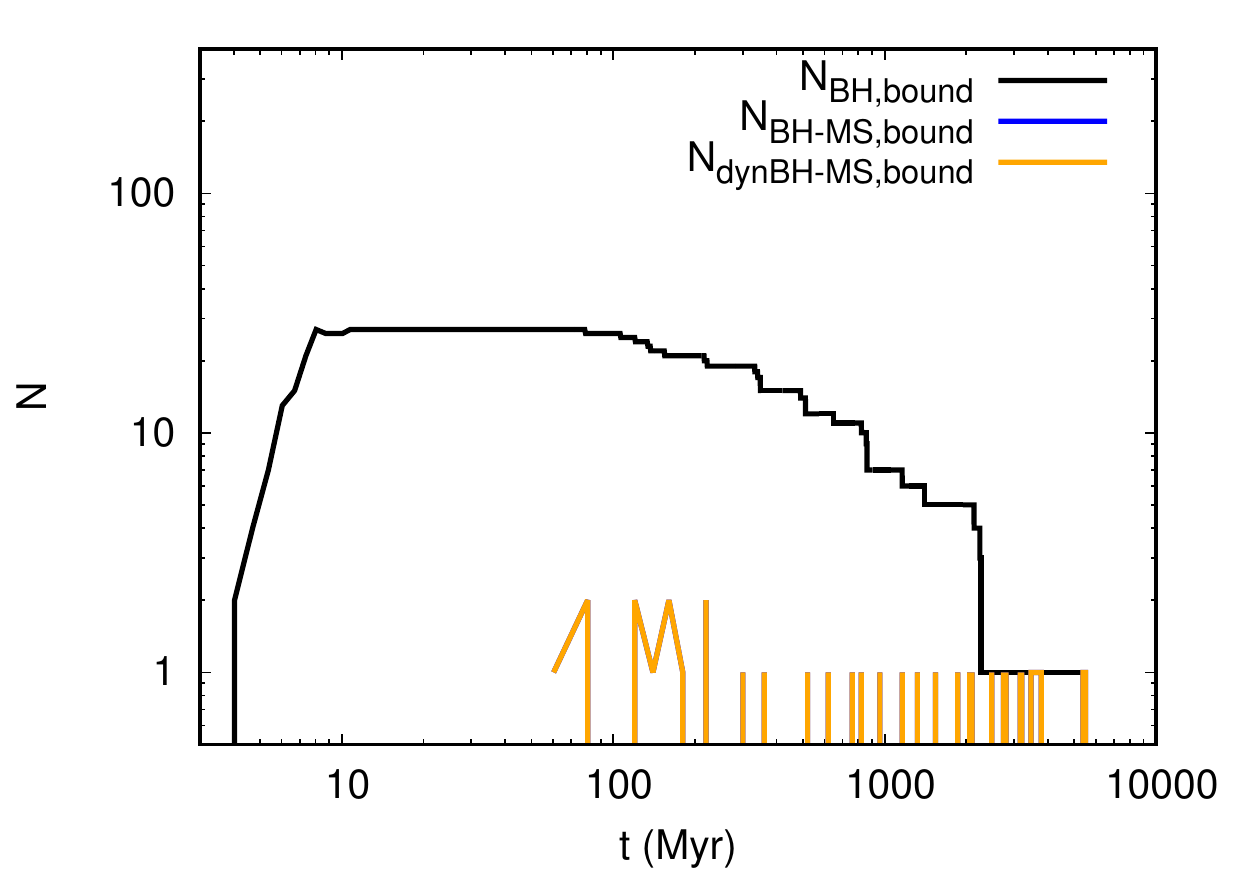}
\hspace{-0.25cm}
\includegraphics[width=5.94cm,angle=0]{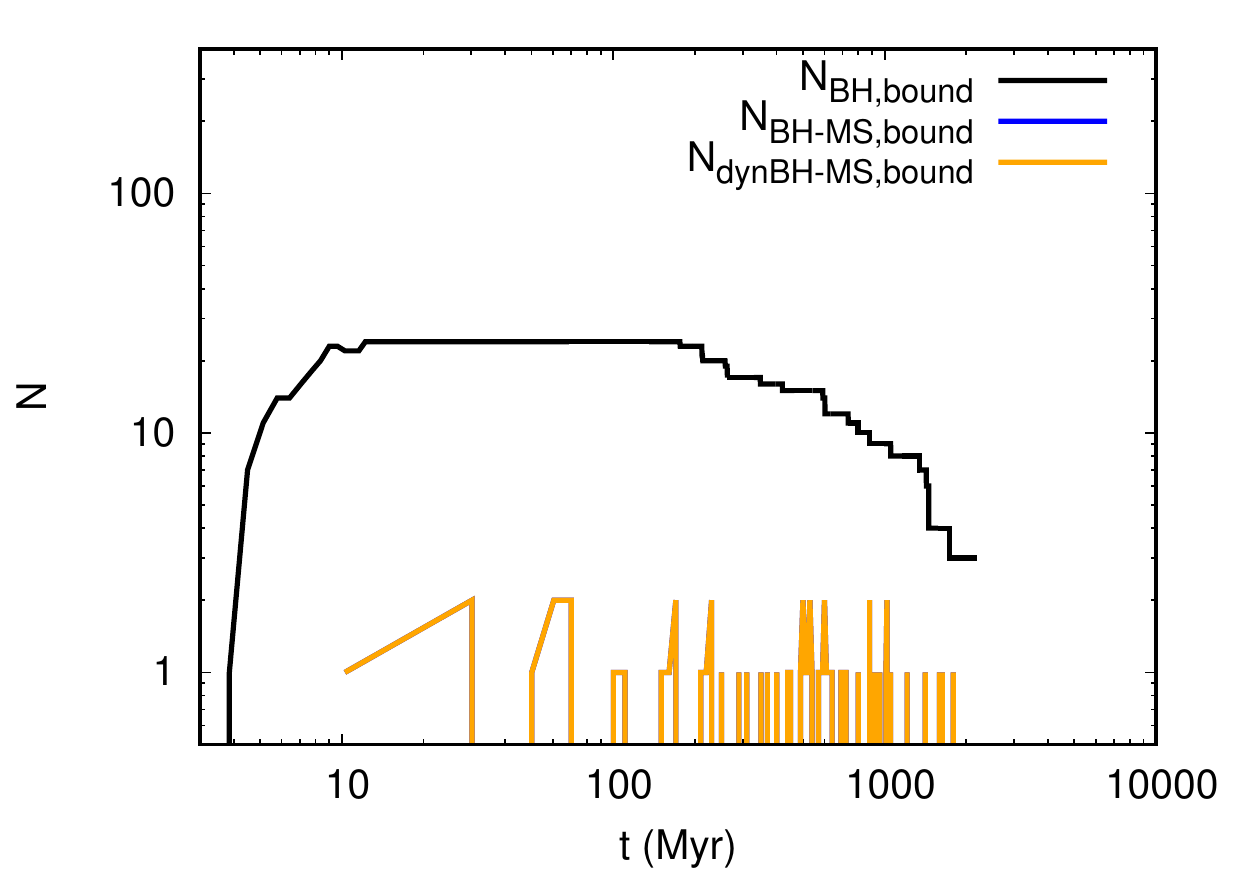}
\hspace{-0.25cm}
\includegraphics[width=5.94cm,angle=0]{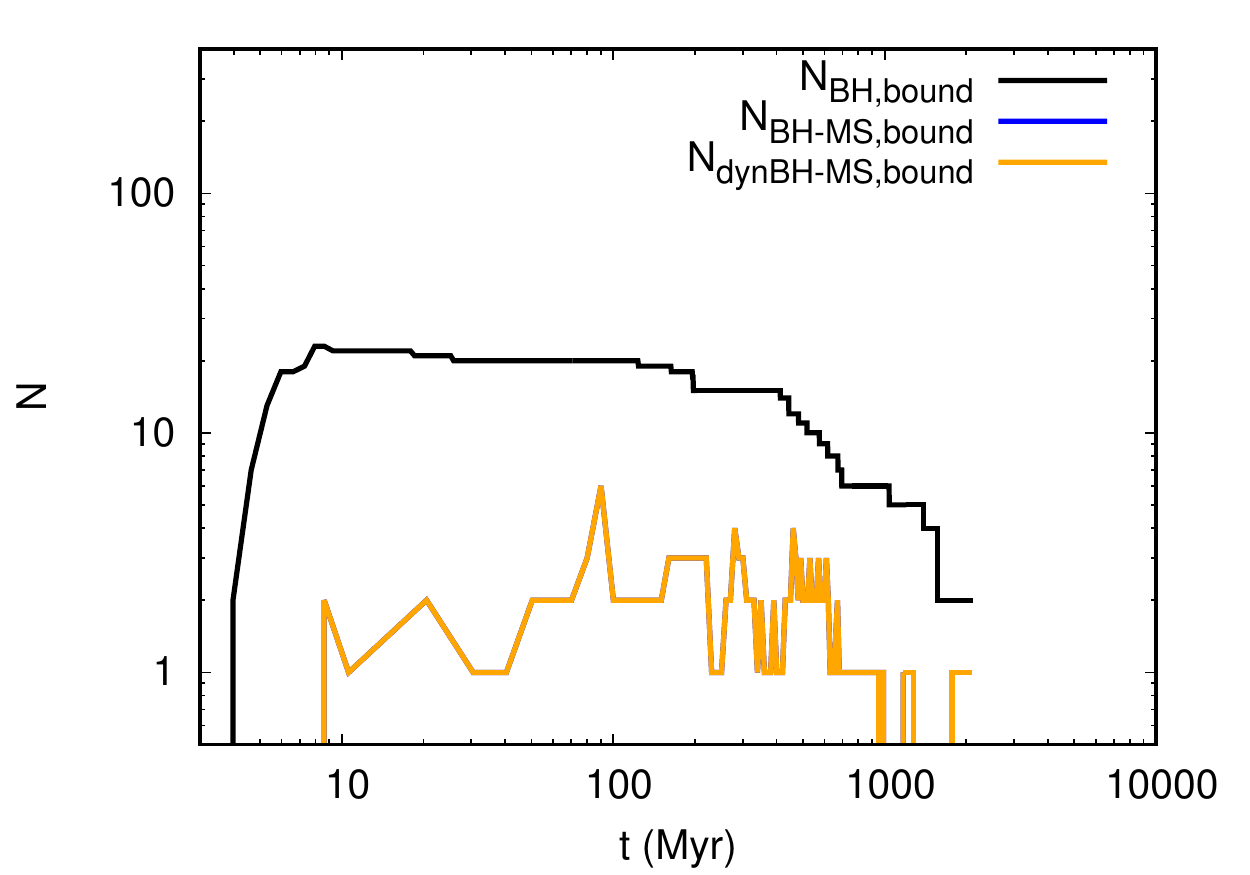}\\
\includegraphics[width=5.94cm,angle=0]{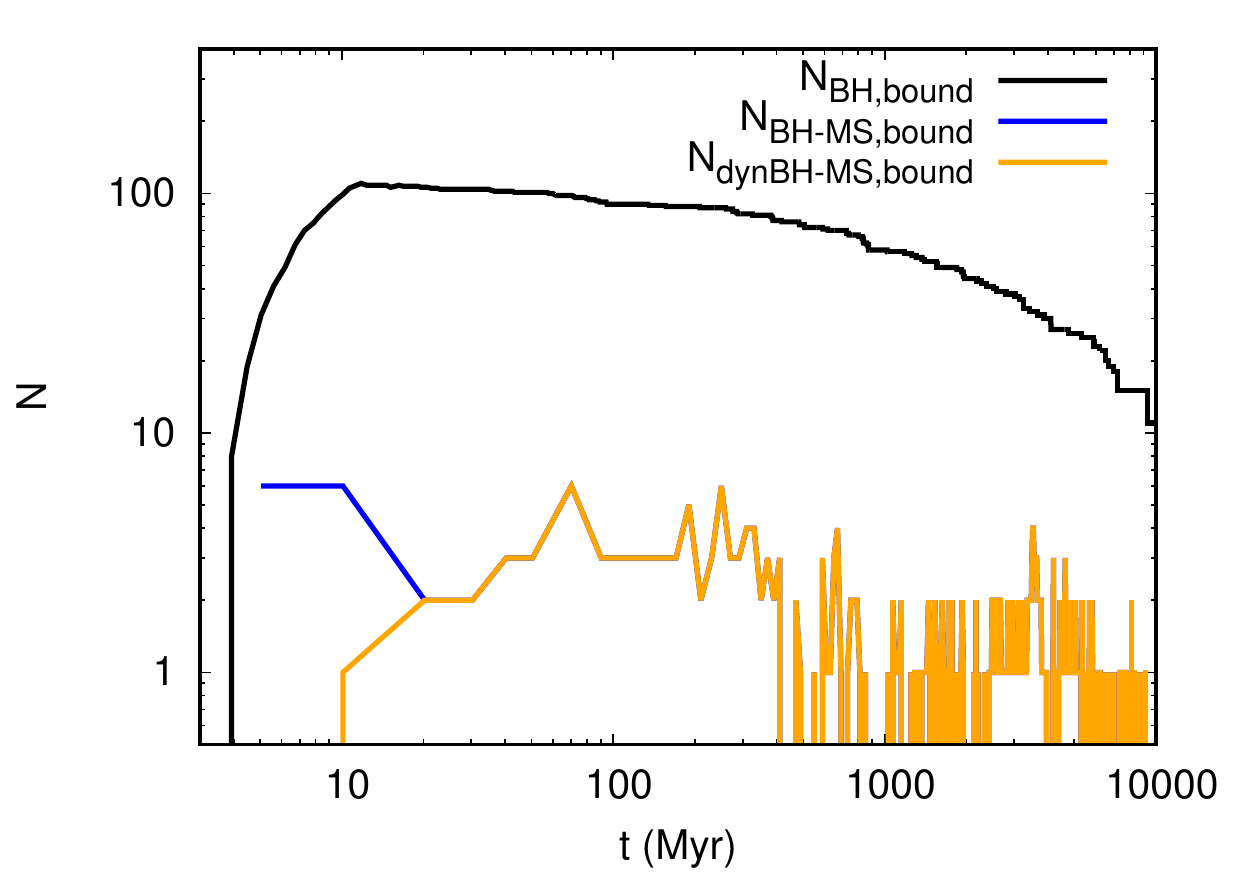}
\hspace{-0.25cm}
\includegraphics[width=5.94cm,angle=0]{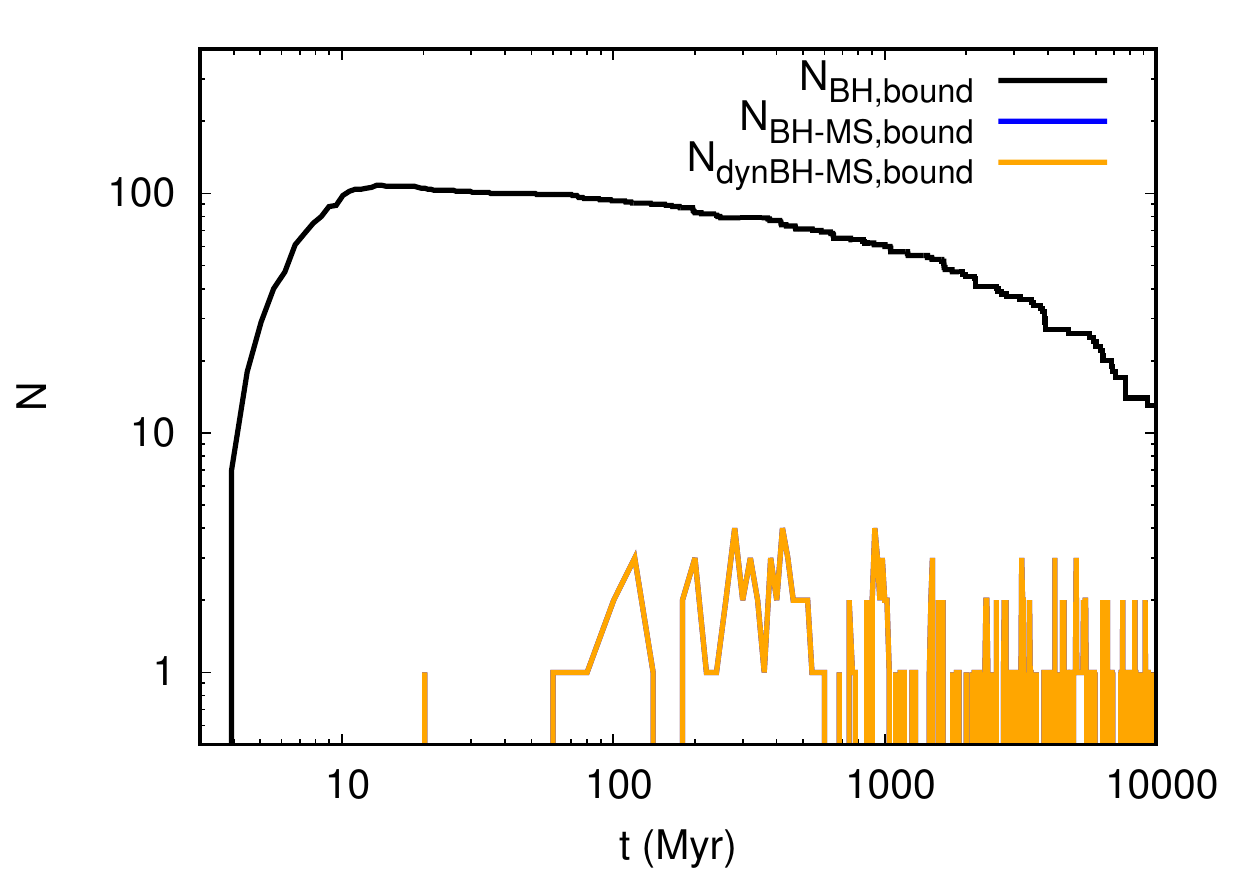}
\hspace{-0.25cm}
\includegraphics[width=5.94cm,angle=0]{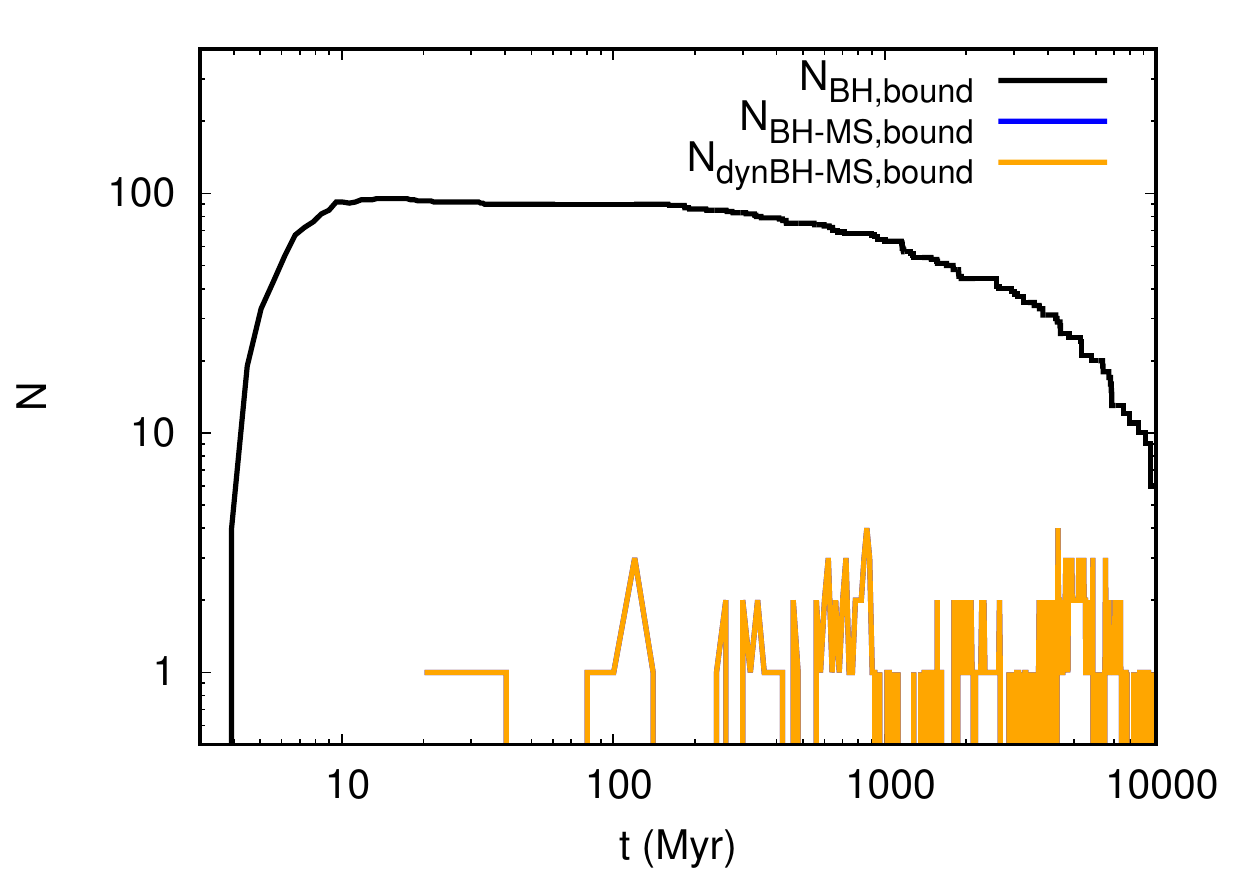}
\caption{Each of the panels represents the time evolutions of the total
number of BHs, $\nbhbound$ (black line), of the total number of BH-MS binaries, $\nbhmsbound$ (blue line),
and of the number of dynamically-formed BH-MS binaries, $\ndynbhmsbound$ (orange line),
that are bound to the cluster.
{\bf Top and middle:} for the same models as in Fig.~\ref{fig:bbh_15k} shown in the same order, \ie,
those with
$\mcl(0)\approx1.5\times10^4\Ms$, $\fbin(0)\approx0.05,0.30,0.50$ (left to right), and $Z=0.001$ (top),
$0.02$ (middle). {\bf Bottom:} for the models with $\mcl(0)\approx5.0\times10^4\Ms$, $\fbin(0)\approx0.05$,
$Z=0.001$, $0.005$, $0.02$ (left to right).}
\label{fig:bhmscnt}
\end{figure*}

\begin{figure*}
\centering
\includegraphics[width=5.94cm,angle=0]{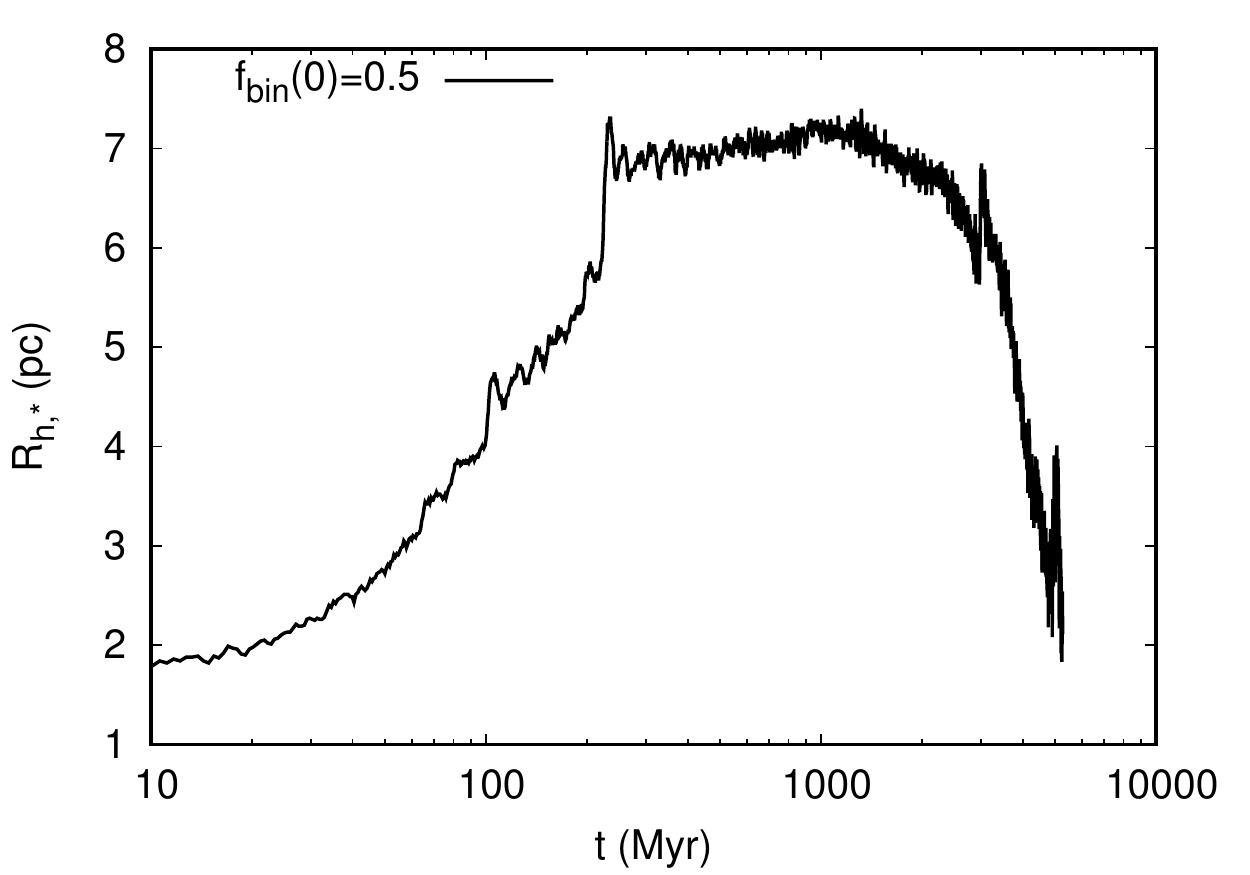}
\hspace{-0.25cm}
\hspace{5.94cm}
\hspace{-0.25cm}
\includegraphics[width=5.94cm,angle=0]{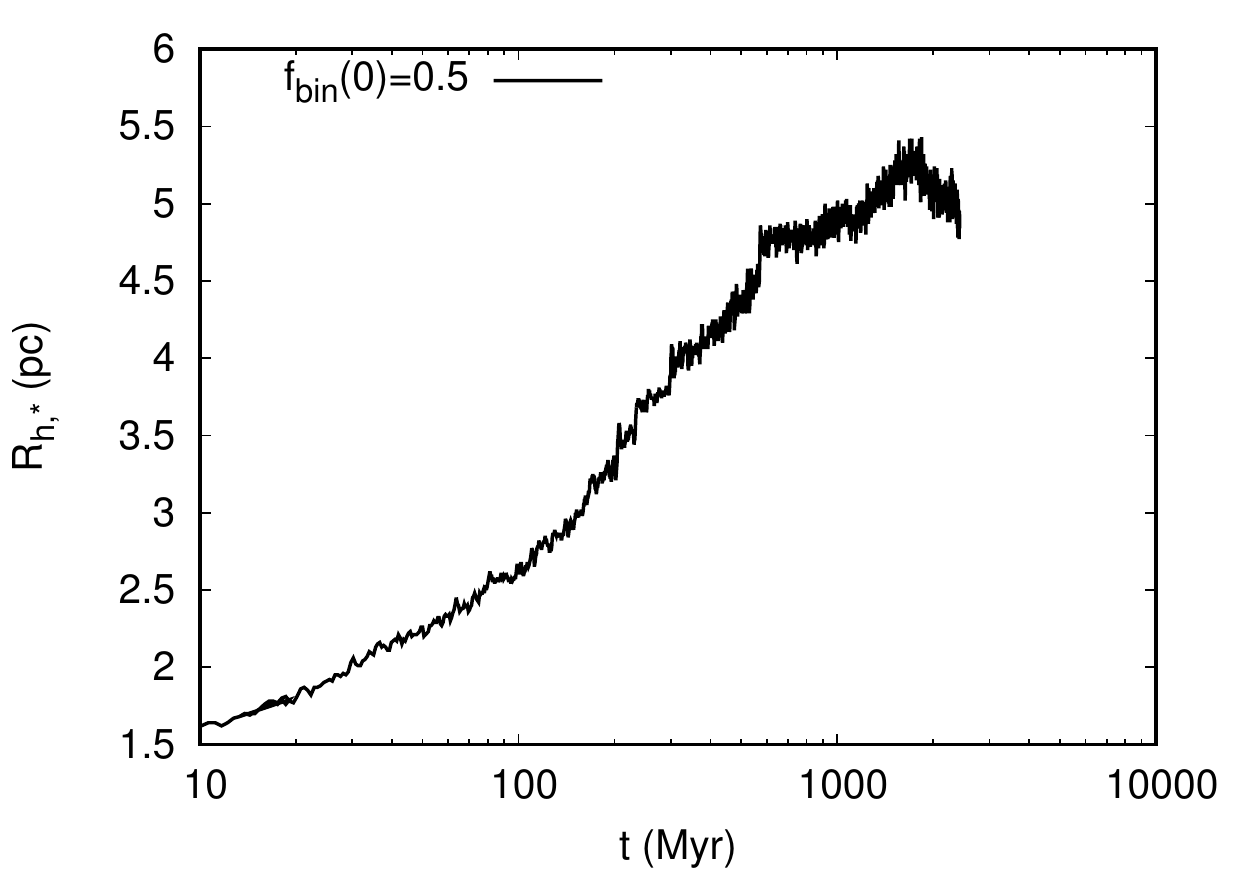}\\
\includegraphics[width=5.94cm,angle=0]{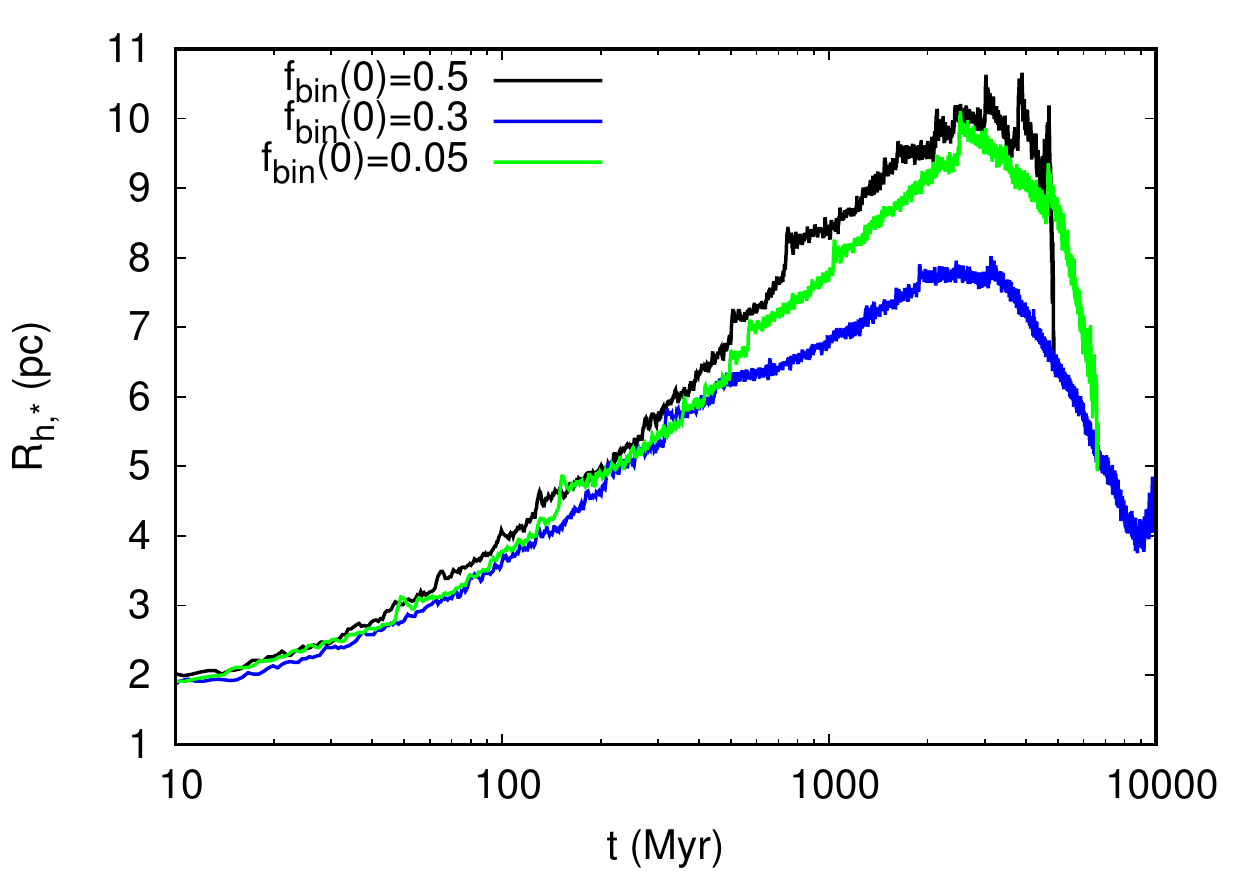}
\hspace{-0.25cm}
\includegraphics[width=5.94cm,angle=0]{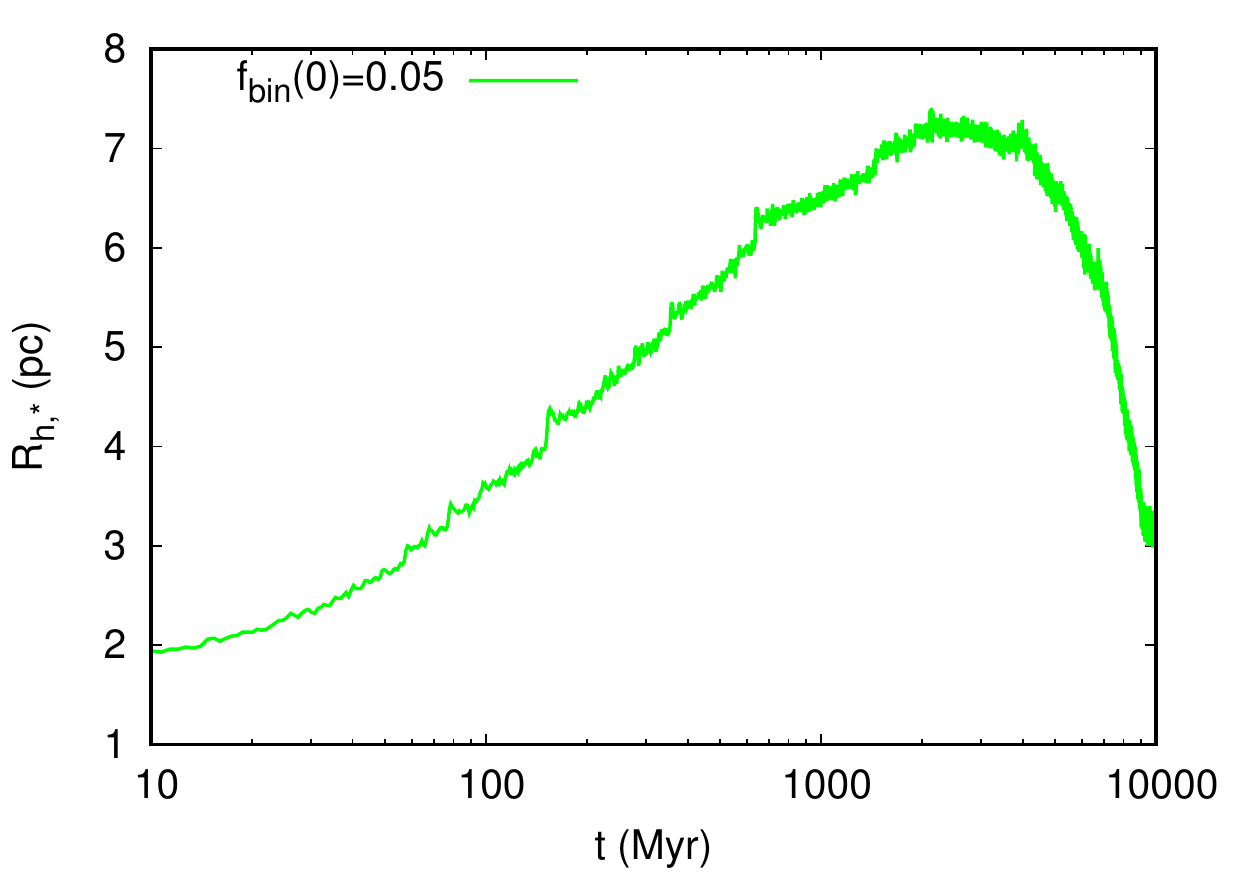}
\hspace{-0.25cm}
\includegraphics[width=5.94cm,angle=0]{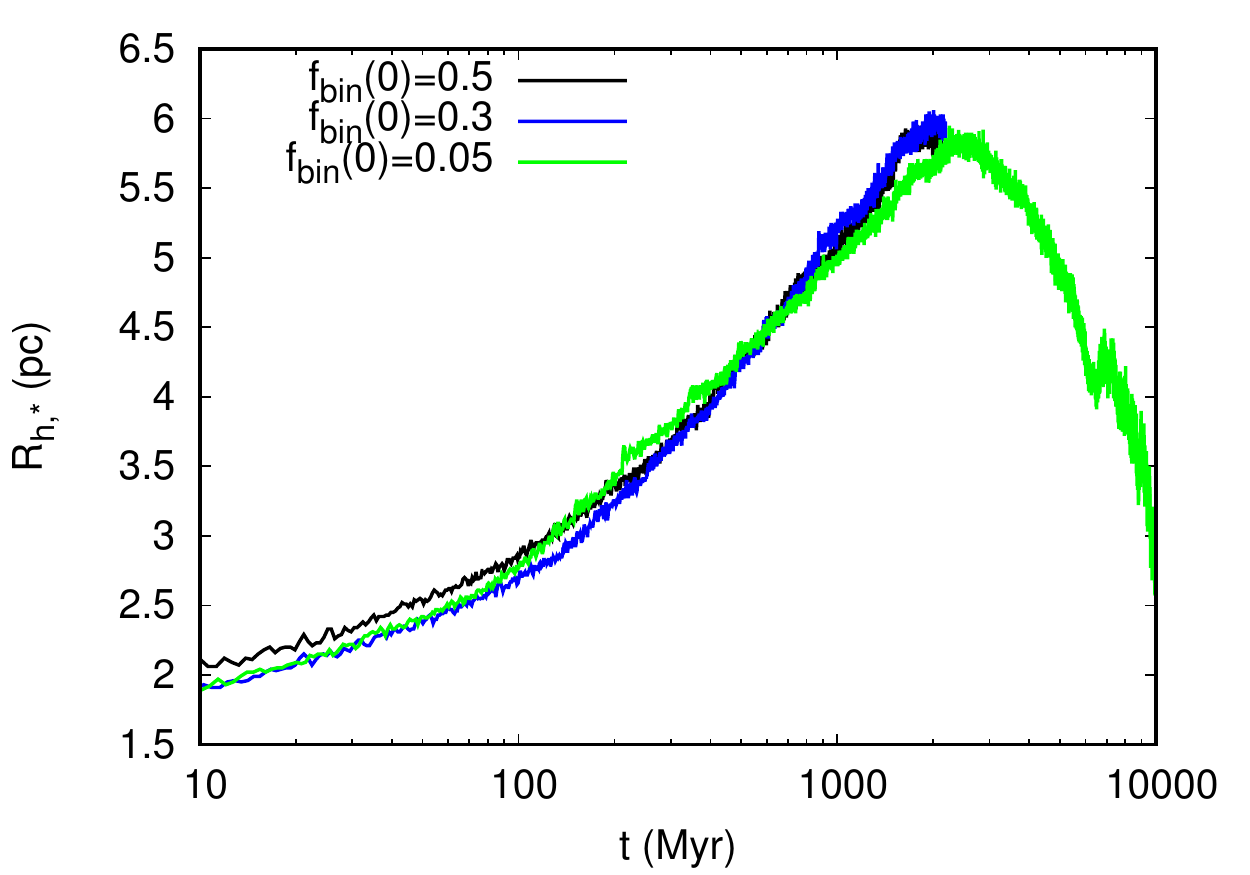}\\
\includegraphics[width=5.94cm,angle=0]{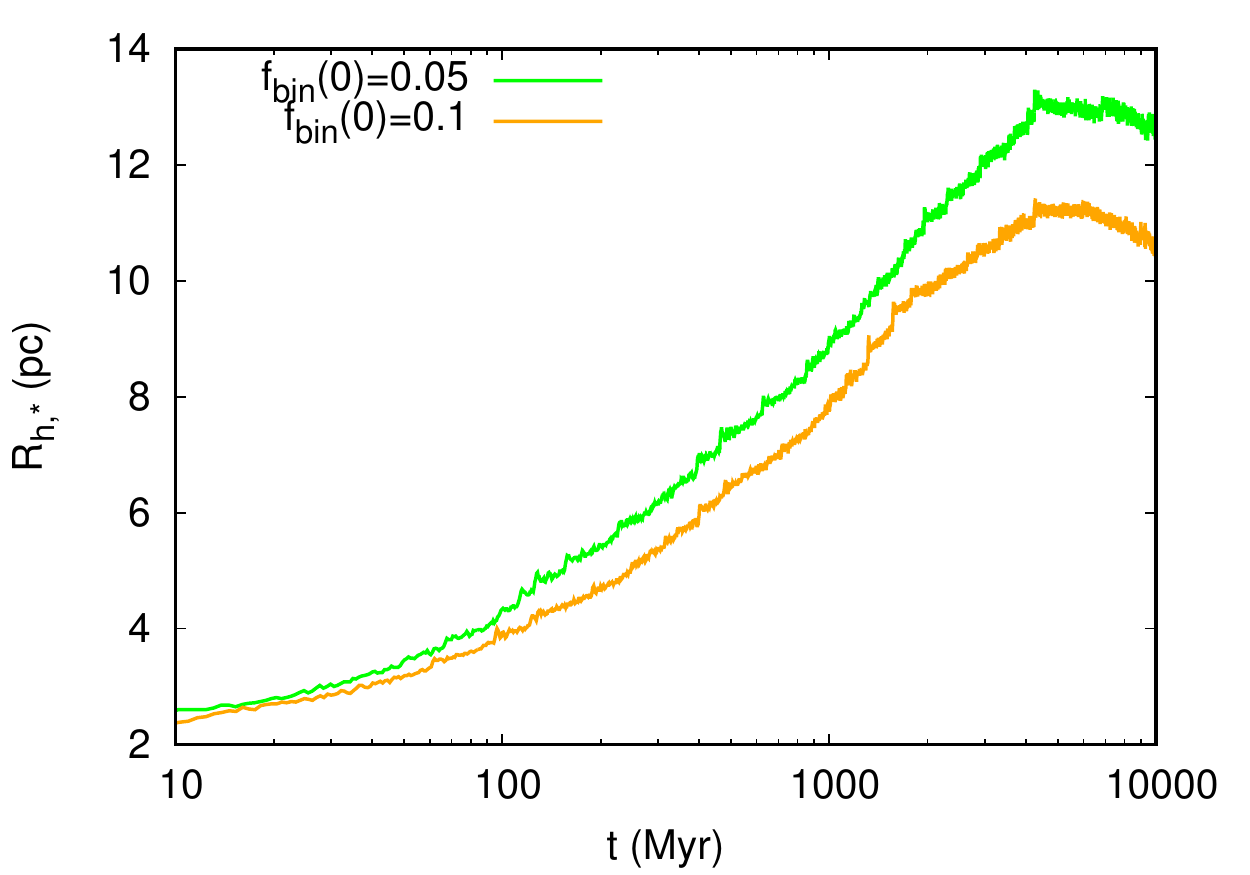}
\hspace{-0.25cm}
\includegraphics[width=5.94cm,angle=0]{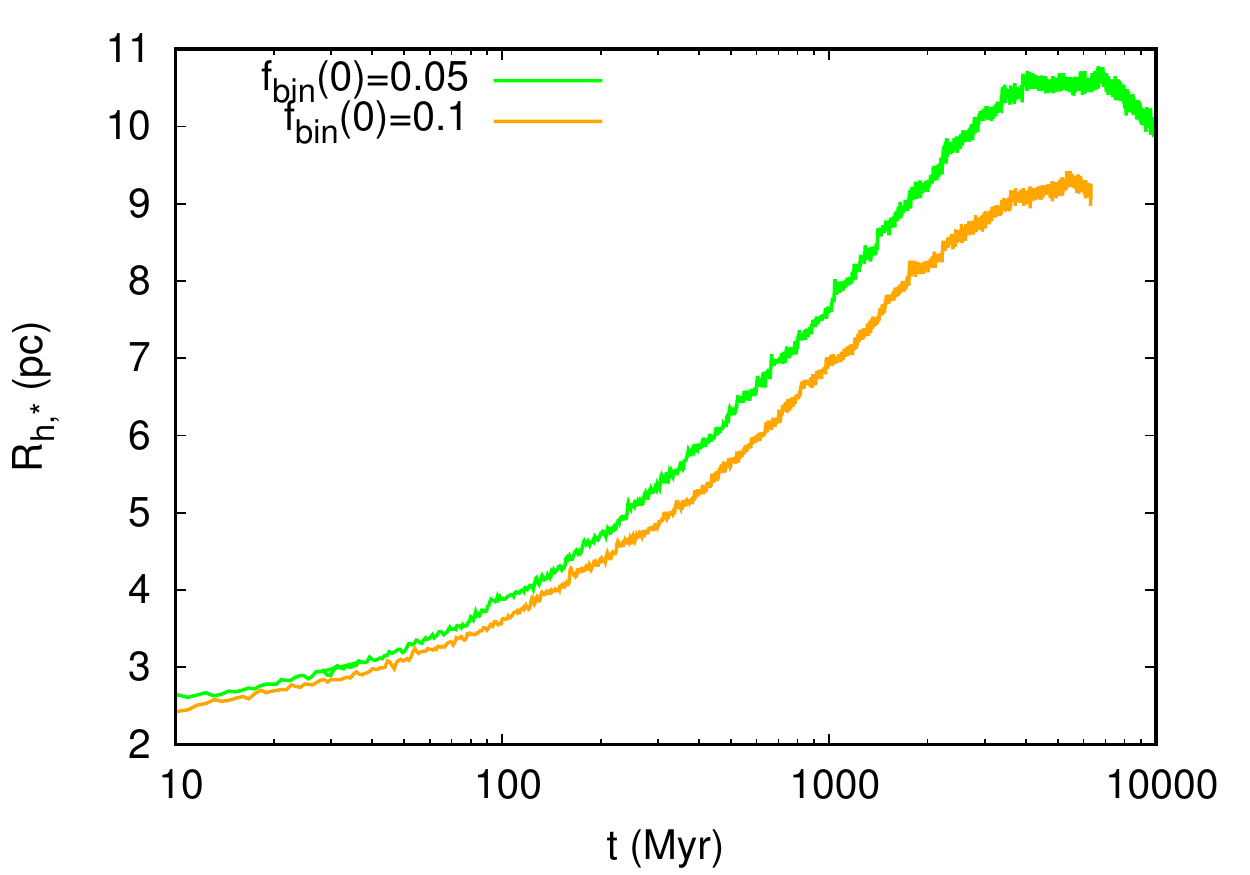}
\hspace{-0.25cm}
\includegraphics[width=5.94cm,angle=0]{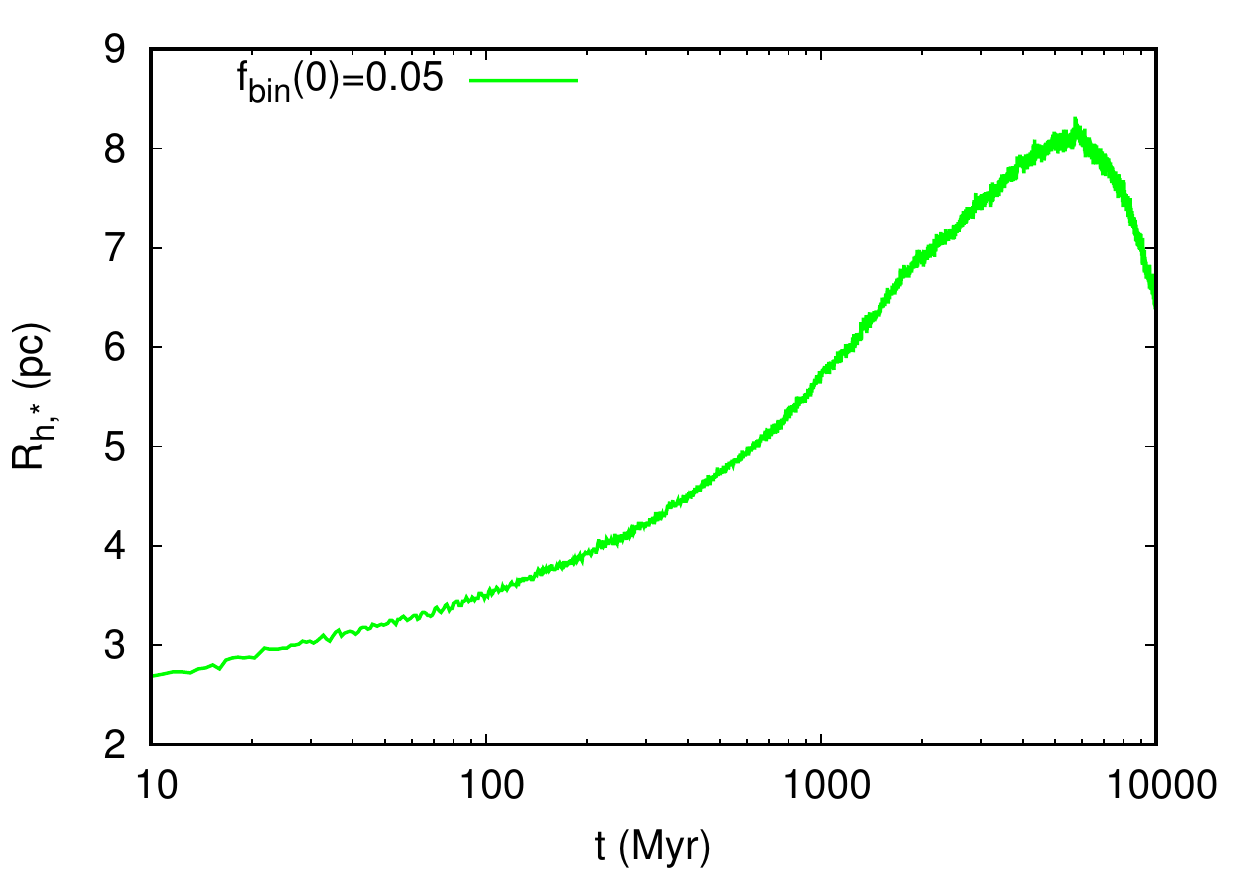}\\
\includegraphics[width=5.94cm,angle=0]{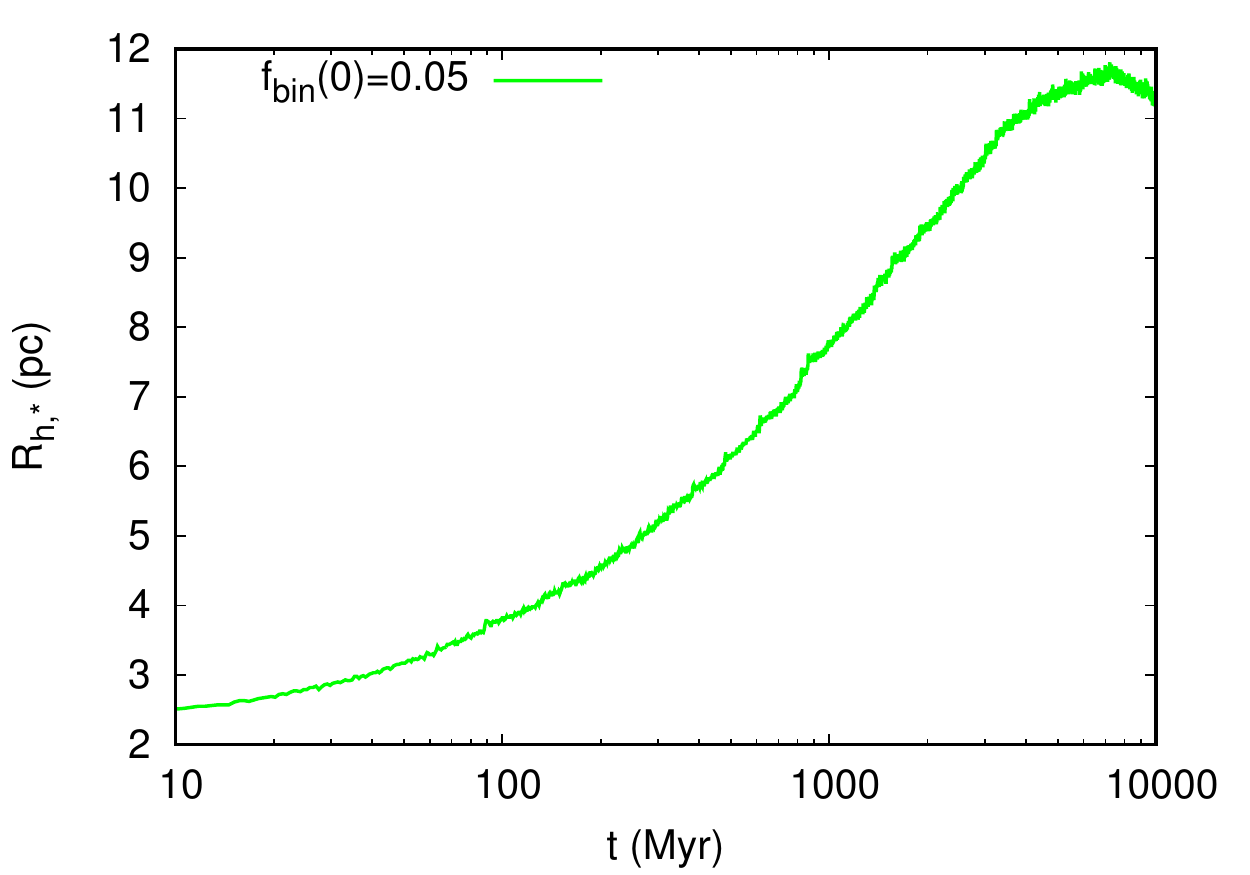}
\hspace{-0.25cm}
\includegraphics[width=5.94cm,angle=0]{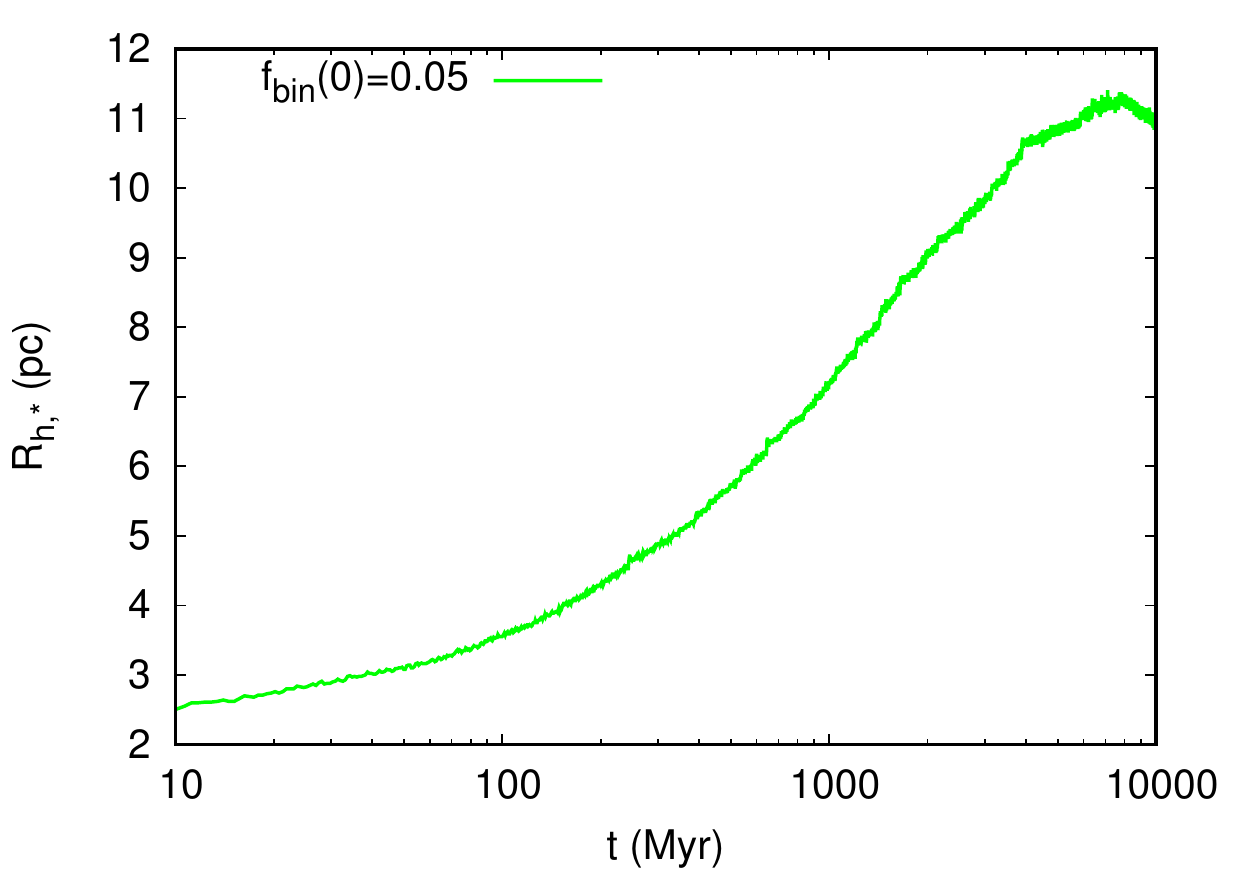}
\hspace{-0.25cm}
\includegraphics[width=5.94cm,angle=0]{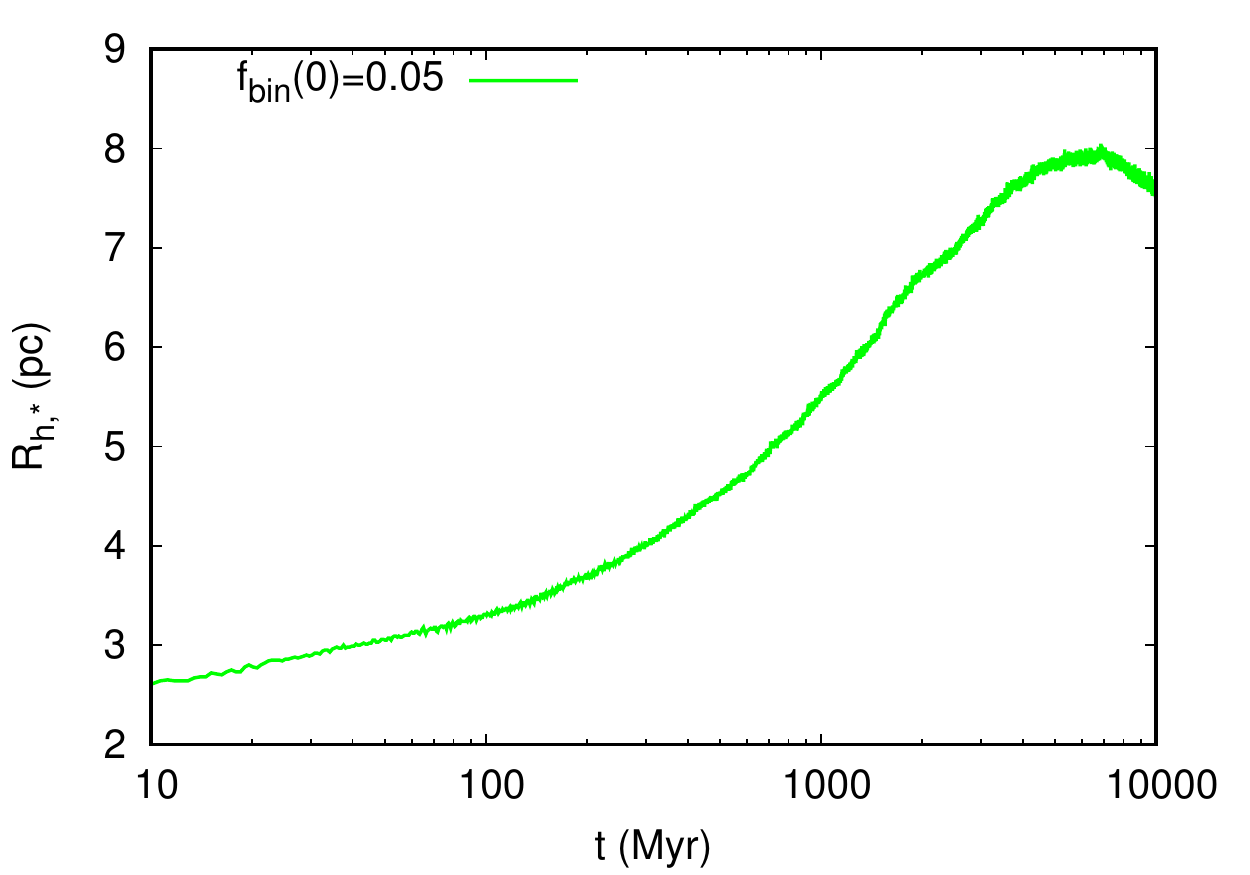}
\caption{The time evolution of the half-mass radii of the luminous component (all members except the NSs and
the BHs), $\rhstar$, in the computed models with $\mcl(0)\approx7.5\times10^3\Ms$, $1.5\times10^4\Ms$,
$3.0\times10^4\Ms$, and $5.0\times10^4\Ms$ (top to bottom) and $Z=0.001$, $0.005$, and $0.02$
(left to right). The primordial binary fractions are indicated in the legends.}
\label{fig:lagr}
\end{figure*}

\begin{figure*}
\centering
\includegraphics[width=5.94cm,angle=0]{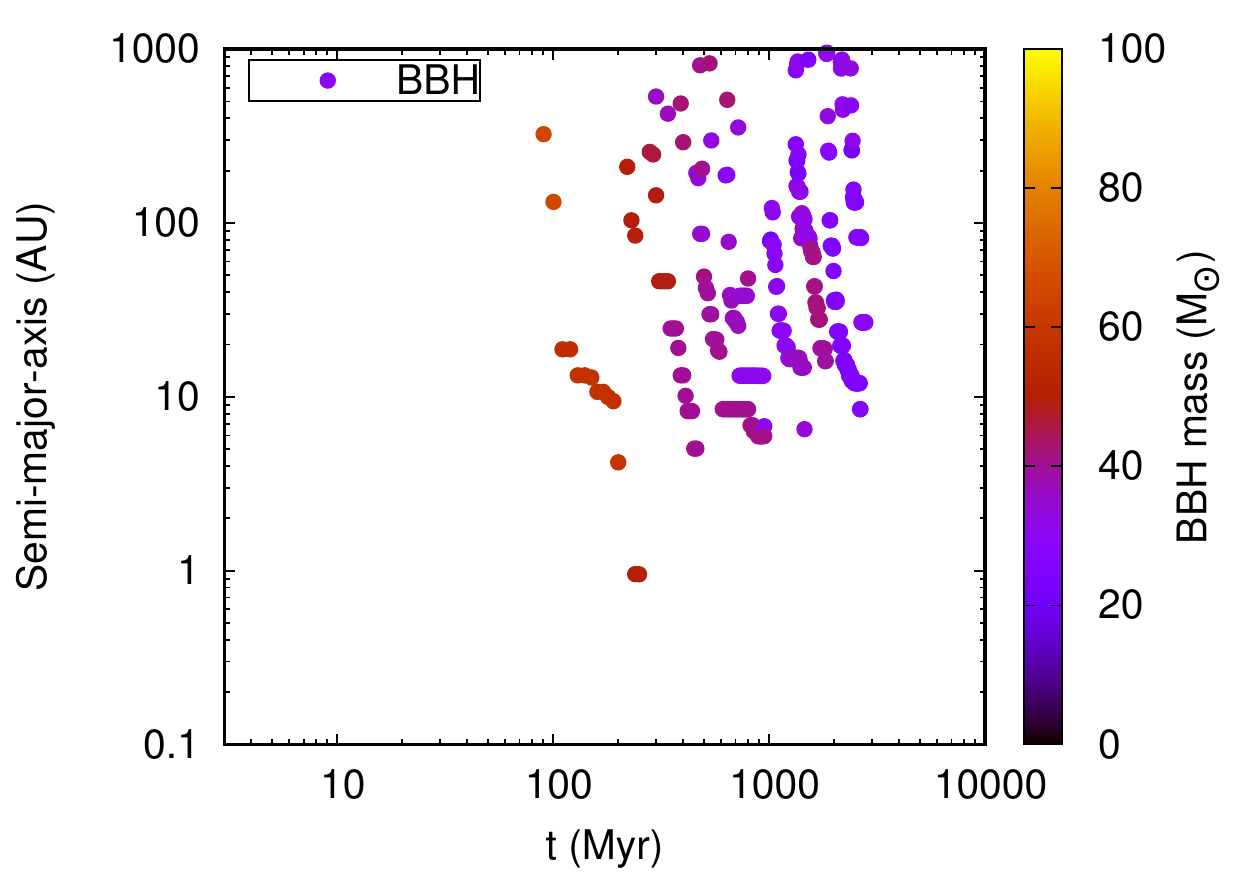}
\hspace{-0.25cm}
\includegraphics[width=5.94cm,angle=0]{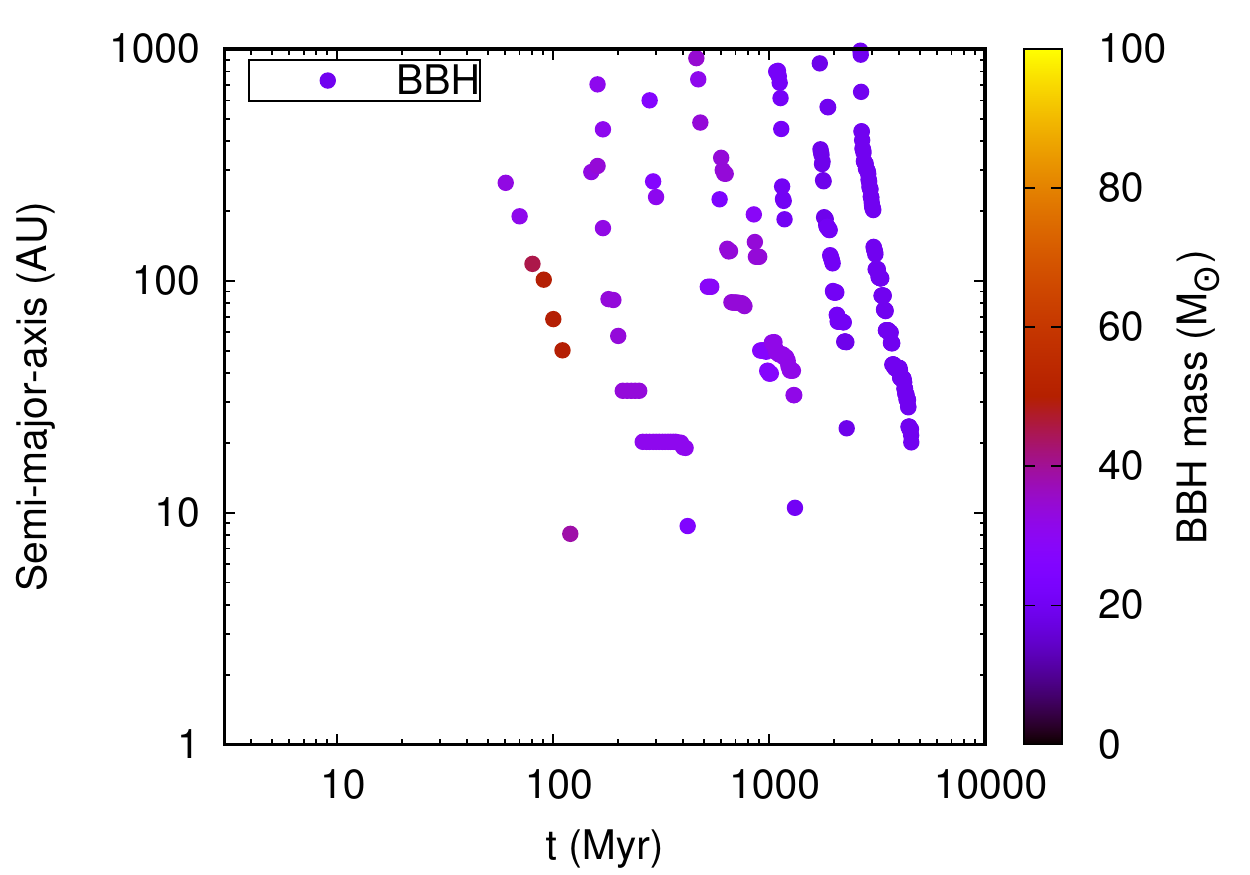}
\hspace{-0.25cm}
\includegraphics[width=5.94cm,angle=0]{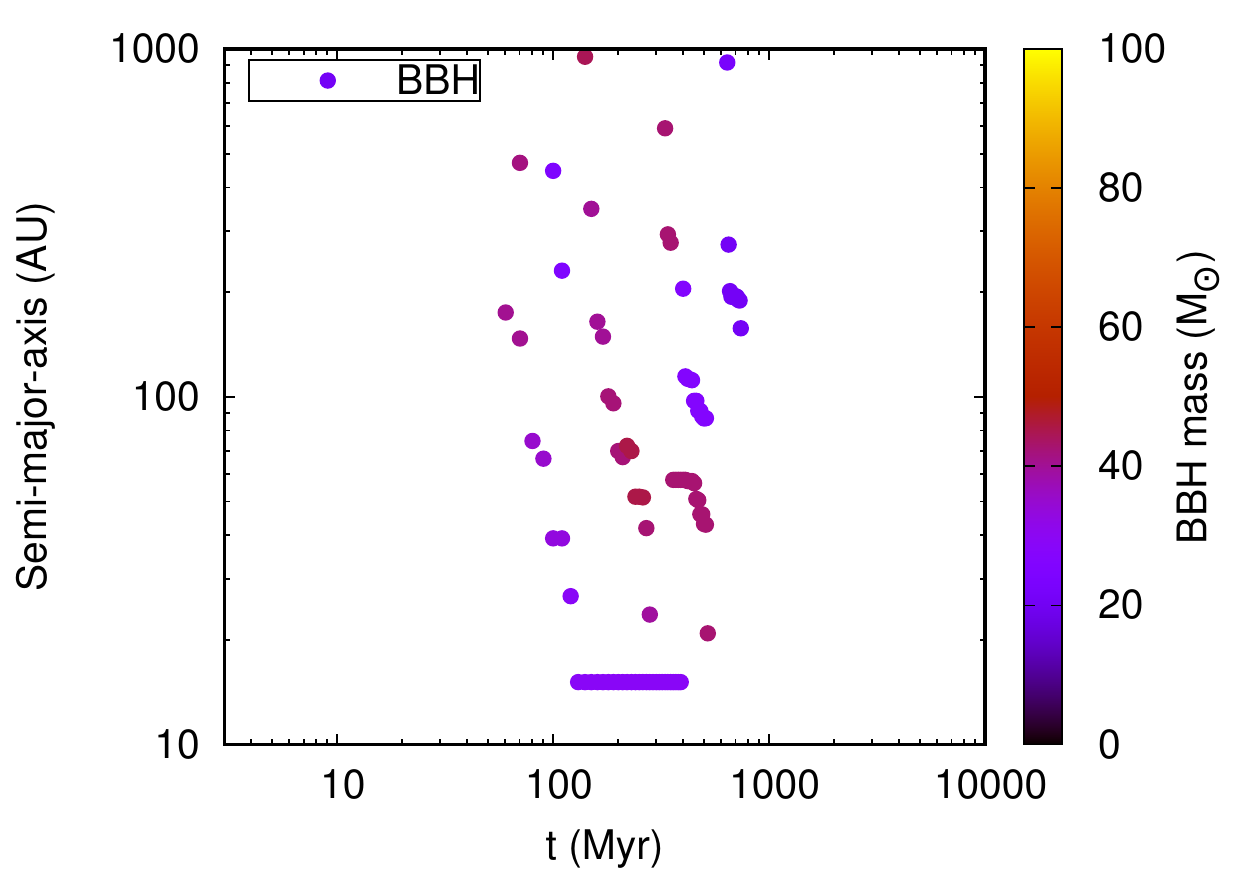}\\
\includegraphics[width=5.94cm,angle=0]{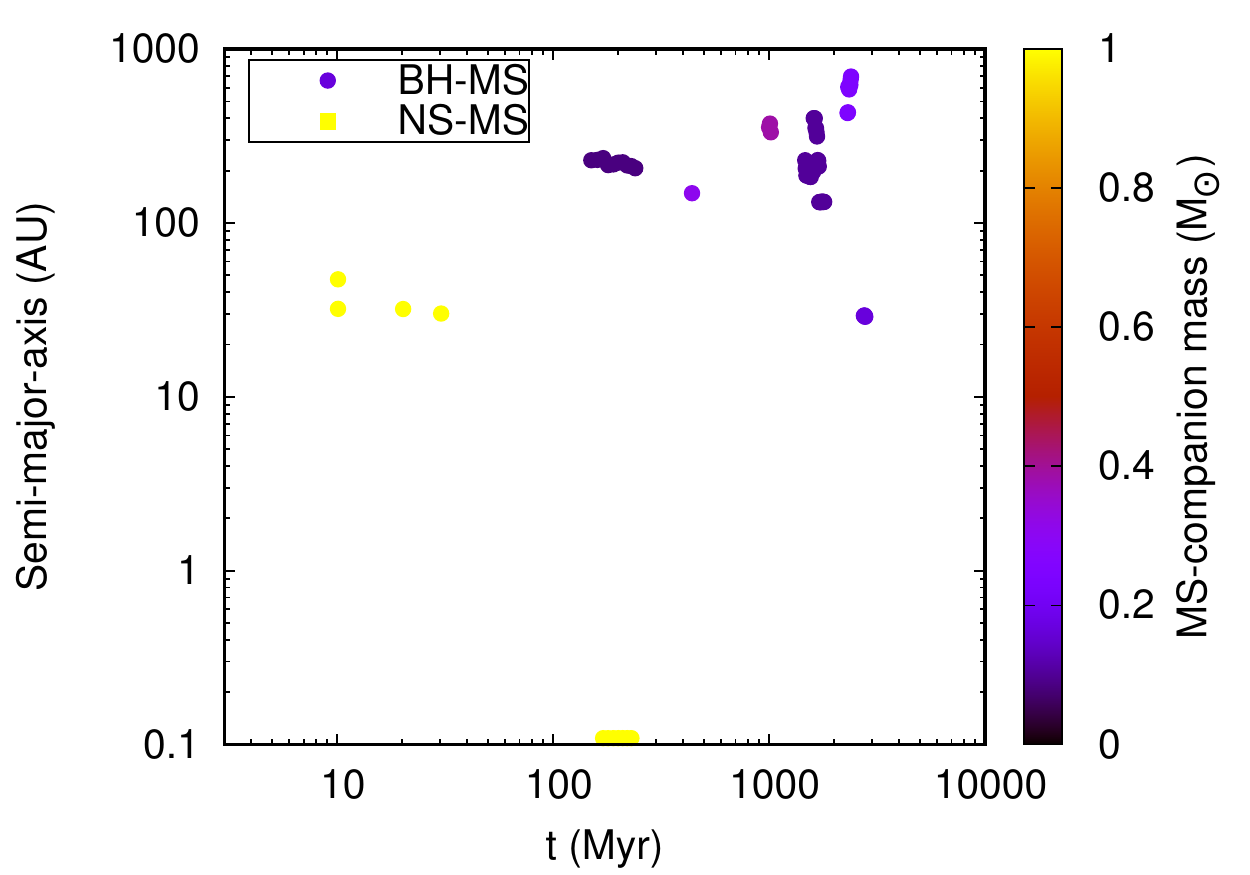}
\hspace{-0.25cm}
\includegraphics[width=5.94cm,angle=0]{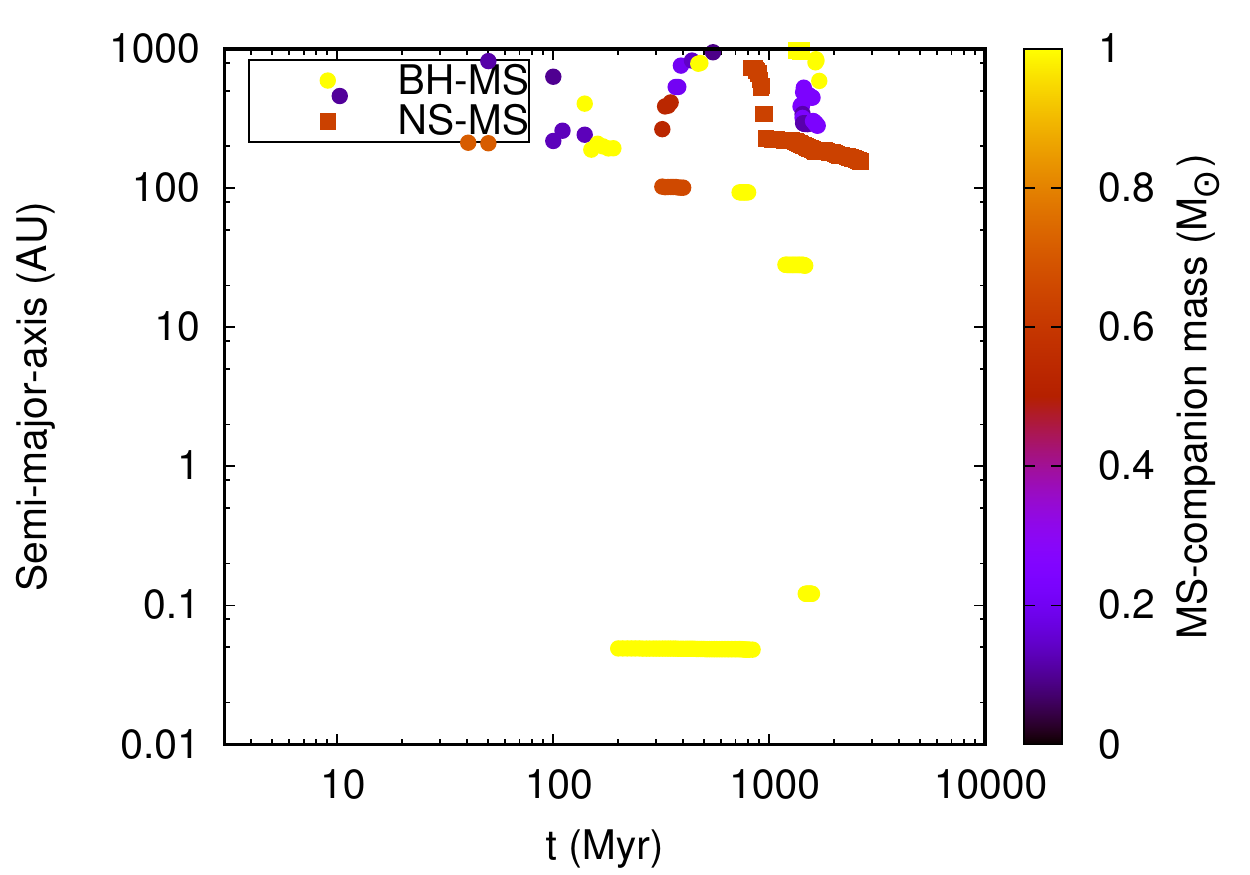}
\hspace{-0.25cm}
\includegraphics[width=5.94cm,angle=0]{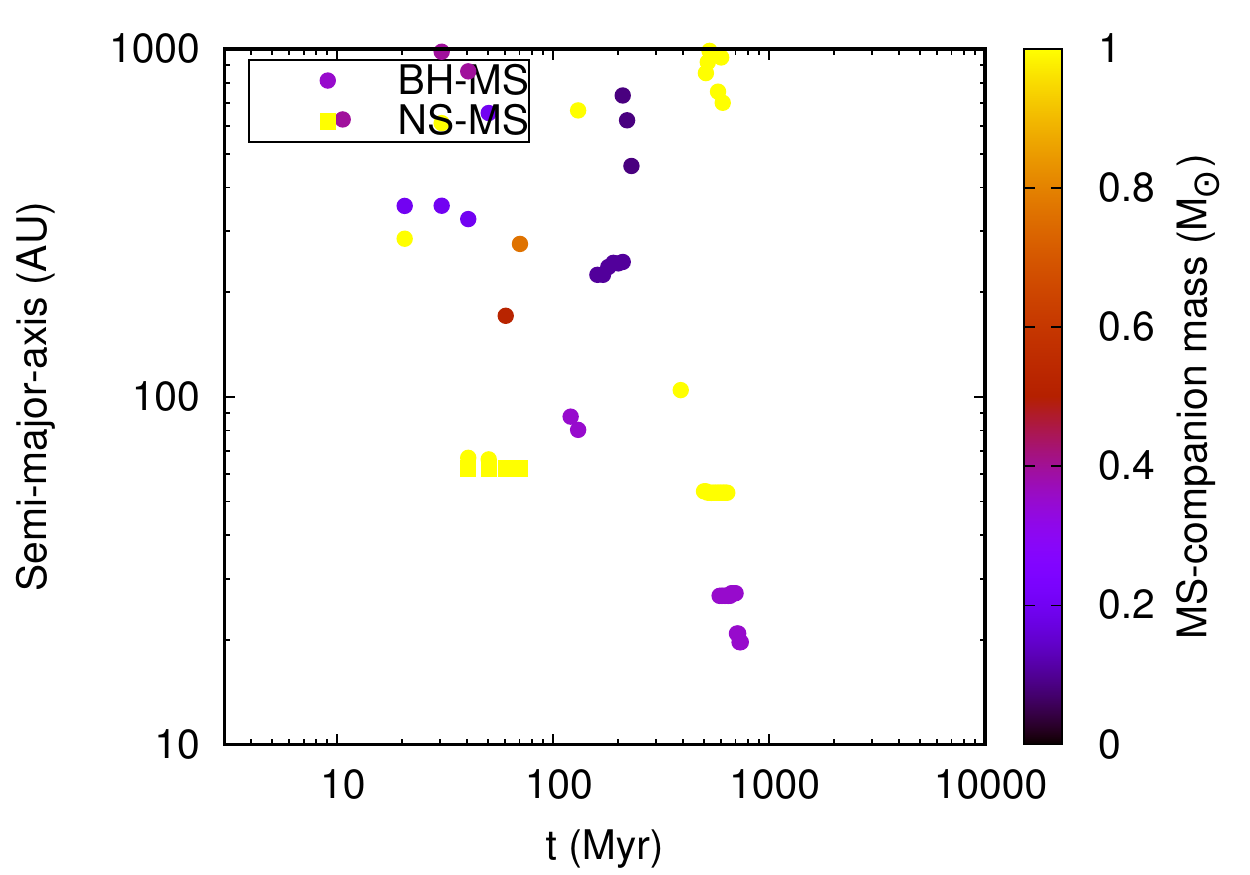}
\caption{{\bf Top:} The BBH content in the ``test'' evolutionary models that include
	explicit tidal cirularization within a binary that contain a non-compact member
	(see Table~\ref{tab1}; Sec.~\ref{stellrem}).
	The models with $\mcl(0)\approx5.0\times10^4\Ms$; $\rh(0)\approx2.0$ pc;
	$\fbin(0)\approx0.05$, $\mcl(0)\approx1.5\times10^4\Ms$; $\rh(0)\approx1.5$ pc; $\fbin(0)\approx0.30$,
	and $\mcl(0)\approx1.5\times10^4\Ms$; $\rh(0)\approx1.5$ pc; $\fbin(0)\approx0.50$ are
	shown in the panels from left to right respectively. All the three models have $Z=0.02$.
	{\bf Bottom:} The BH-MS/NS-MS binaries in these models displayed in the same
	order.}
\label{fig:bhbin_circ}
\end{figure*}

In terms of computed models' ingredients, a significant leap forward
in the present work from its precursors, namely Papers I \& II (Sec.~\ref{intro}), is the inclusion
of high primordial binary fractions, $\fbin(0)\approx30-50$\%, which
takes these models closer to realistic massive open clusters (Sec.~\ref{primbin}).
Since long-term direct N-body computations (Sec.~\ref{nbprog}) of massive clusters 
with such high binary fractions is challenging, such $\fbin(0)$ is used
for models with $\mcl(0)\lesssim1.5\times10^4\Ms$; for the rest of the models,
$\fbin(0)\lesssim0.1$ is used (Table~\ref{tab1}). Nevertheless, it is for
the first time that the dynamics of stellar-mass BHs is studied ab initio
in high-binary-fraction massive cluster models,
through detailed and completely self-consistent
relativistic direct N-body simulations. Another new aspect in this
work is to be able to live-track in direct N-body simulations (Sec.~\ref{getsys}),
the configurations of the few-body subsystems that drive the GR mergers inside the
massive clusters, for the first time. The key inferences based on the model computations
of open clusters in this study are as follows:

\begin{itemize}

\item GR inspiral and coalescences occur inside young massive and open clusters with
	birth masses $\mcl(0)\gtrsim7.5\times10^3\Ms$ (Table~\ref{tab1}),
	under conditions that are realistic for such clusters, as long as
	a population of compact stellar remnants (NSs and BHs) are retained in the
	clusters at birth.
	Although the majority of such mergers are of BBHs, it is possible that other types of
	stellar remnants also participate in GR coalescences inside
	the clusters, \eg, mergers of  BH-NS and BH-WD
	(Sec.~\ref{bbhmrg}). In addition to generating GW that is
	detectable by the LISA and LIGO-Virgo, the latter types of mergers
	would also generate electromagnetic imprints of the events.
	Young massive and open clusters would potentially
        contribute to present-day (dynamical) BBH mergers, visible by the LISA and the LIGO-Virgo,
        at a rate that is comparable to those from the GCs and the galactic nuclei
	(Sec.~\ref{remark}).

\item The GR inspirals take place inside clusters via intricate dynamical interactions, that involve
	dynamically-assembled compact subsystems which are most often triples (Sec.~\ref{bbhmrg}). Such triples
	typically range from being marginally hierarchical to hierarchical in configuration
	(Fig.~\ref{fig:stabplot}), although the onset of the GR inspiral of the
	inner binary and its merger would take place due to the chaotic interaction
	of the binary with its outer companion. Such GR-coalescence-hosting triples are typically composed
	of BHs although both the inner and the outer binary can, possibly, contain a non-BH member, \eg,
	an NS or a WD (Fig.~\ref{fig:stabplot_ratio}, Sec.~\ref{bbhmrg}). The subsystems leading
	to a GR merger can, possibly, be even more complex, \eg, a relatively long lasting
	triple undergoing multiple exchanges of its outer member. The mergers can also take place
	within an even higher-order subsystem, \eg, inside a quadruple (of both binary-binary and
	hierarchical configurations).

\item A close, fly-by encounter between a purely Keplerian compact binary and a single member can
	impart sufficient eccentricity (plus any hardening) onto the binary so that it becomes PN,
	which interactions are termed here as ``GR slingshot''s (Sec.~\ref{sling}). As in the
	case of triple-mediated GR inspirals (see above), the participants of such encounters
	are mainly BHs but NS, WD, and MS stars can also take part. Approximately half of such
	GR-slingshot binaries have their peak-power GW frequency in the LISA's characteristic
	detection band and a few of them also fall in the PTA's band (Fig.~\ref{fig:sling});
	that way a population of such binaries in the stellar clusters of the Milky Way and the
	other nearby galaxies would potentially contribute to the GW background noise for
	the LISA and the PTA.
	However, such fly-by interactions, which predominantly take place over the first
	$\sim$ Gyr of the host clusters' evolution, rarely lead to full GR inspiral and coalescence of
	the GR-slingshot binaries.

\item Intermediate-aged and old open clusters harbour detached BH-MS binaries that continue to form
	and destroy in (get ejected from) them
	via exchange (close) encounters (Sec.~\ref{bhmspop}; Figs.~\ref{fig:bhms_15k}, \ref{fig:bhms_30k},
	\ref{fig:bhms_f05}, \ref{fig:bhms_f50}, \ref{fig:bhbin_circ}).
	Such clusters also contain detached NS-MS binaries
	and also BH/NS paired with evolved stellar companions (Fig.~\ref{fig:bhall_misc}).
	In clusters that are relatively dense and contain high stellar binary fraction at late
	times, a few BH-MS binaries are formed tight, with $a<10$ AU, which are good candidates to
	be identified in the open clusters through radial-velocity measurements of the MS companion,
	as in the case of the GC NGC 3201. The BH---evolved-star companions, when formed, and, more rarely,
	the NS-MS binaries are also similarly tight. The BH/NS---MS/evolved-star binaries
	typically last for $\sim100$ Myr, before getting the MS companion exchanged
	by a BH/NS or getting dynamically ejected from the cluster,
	although they would last up to $\sim$ Gyr in lower-density environments. 
	Although the present models have produced detached BH-MS binaries with
	properties consistent to the one identified in NGC 3201 \citep{Giesers_2018}, open clusters would
	potentially harbour a wide range of remnant---normal-star binaries
	with prospects of discovery via forthcoming radial-velocity surveys.

\item Throughout their evolutionary times, young massive and open clusters harbour a population of
	BBHs that continue to form, get modified, get ejected or undergo GR coalescences via dynamical
	interactions (Sec.~\ref{bbhpop}; Figs.~\ref{fig:bbh_15k}, \ref{fig:bbh_30k},
	\ref{fig:bbh_f05}, \ref{fig:bbh_f50}, \ref{fig:bhbin_circ}). The BBHs which maintain
	the primordial pairings at their appearance are typically found in relatively low-density outer
	regions of the clusters (Fig.~\ref{fig:bbh_rad}) where they undergo gradual orbital decay via weak
	encounters (that would accelerate only after they sink, via dynamical friction,
	further into the cluster, where the stellar density is higher).
	In contrast, the dynamically-assembled BBHs, which are also
	typically more massive and initially wider, are assembled in the innermost region of the cluster
	where they can undergo dynamical hardening efficiently, owing to the much higher stellar density and
	the prominence of BHs. Irrespective of a cluster's parameters (\eg, its mass, binary fraction)
	and the BH population contained in it, the ambient number of BBHs and BH-MSs within it is
	always a only few (Secs.~\ref{bbhpop} \& \ref{bhmspop}; Figs.~\ref{fig:bbhcnt} \& \ref{fig:bhmscnt}).
	
\end{itemize}

\section*{Acknowledgements}
SB is thankful to the anonymous referee for constructive comments
which have helped to improve the presentation of this work.
SB is indebted to Sverre Aarseth of the Institute of Astronomy, Cambridge,
for his efforts in improving {\tt NBODY6/7}, without which this study would not
have been possible.
This work has been partly supported by the Deutsche Forschungsgemeinschaft
(DFG; German Research Foundation) through the individual research grant
``The dynamics of stellar-mass black holes in dense stellar systems and
their role in gravitational-wave generation'' (BA 4281/6-1; PI: Sambaran Banerjee).
An enhanced version of the {\tt Fortran} routine {\tt OrbitDat}, originally
written by Pavel Kroupa and Ladislav Subr, has been utilized in this work.
SB is thankful to the computing team of the Argelander-Institut f\"ur Astronomie,
University of Bonn, for their efficient maintenance of the workstations
on which all the computations have been performed.

\onecolumn
\begin{minipage}{\textwidth}
\begin{longtable}{rcclrccc}
	\caption[Summary of model calculations]
	{Summary of model evolutionary calculations.
	The columns from
	left to right respectively denote: (a) initial mass, $\mcl(0)$, of the model cluster,
	(b) initial half-mass radius, $\rh(0)$, (c) metallicity, $Z$,
	(d) overall primordial binary fraction, \fbin (see Sec.~\ref{primbin}), (e) total evolutionary time, \tevol,
	of the model, (f) the number of 
	(triple-mediated) binary black hole (BBH) coalescences, $\nmrgin$, that occurred within the clusters,
	(g) the number of BBH coalescences (in BBHs that are ejected from the clusters), $\nmrgout$, that occurred
	outside the clusters within the Hubble time, (h) the number of distinct ``GR slingshots'' within the
	clusters, $\nsling$, when available (Sec.~\ref{sling}).
	This table includes the new N-body computations (Sec.~\ref{runs})
	and as well those from \citet[Paper II]{Banerjee_2017b} that include a significant population
	of primordial binaries.}\label{tab1}\\

	\hline
	\hline
	\mcl(0)/\Ms     & \rh(0)/pc & $Z/\Zs$ & \fbin(0) & \tevol/Gyr & \nmrgin & \nmrgout & \nsling\\
	\hline
	\endfirsthead
        
	\multicolumn{7}{c}%
        {{\bfseries \tablename\ \thetable{} -- continued from previous page}} \\
        \hline
	\hline
	\mcl(0)/\Ms     & \rh(0)/pc & $Z/\Zs$ & \fbin & \tevol/Gyr & \nmrgin & \nmrgout & \nsling\\
	\hline
	\endhead

	\hline \multicolumn{7}{r}{{Continued on next page}} \\ \hline
        \endfoot

        \hline \hline
        \endlastfoot

	$7.5\times10^3$ & 1.0       & 0.05 & 0.50  & 5.2  &  1   &  0   &  1   \\
	\hline
	$7.5\times10^3$ & 1.0       & 1.00 & 0.50  & 2.4  &  0   &  0   &  0   \\
	\hline
	\hline
	$1.5\times10^4$ & 1.5       & 0.05 & 0.05  & 6.6  &  1   &  0   &  1   \\
	\hline
	\footnote{Speed of light c/100 assumed for this particular model.}
	$1.5\times10^4$ & 1.5       & 0.05 & 0.05  & 9.4  &  4   &  0   &  >10 \\
	\hline
	$1.5\times10^4$ & 1.5       & 0.25 & 0.05  & 9.9  &  1   &  0   &  4   \\
	\hline
	$1.5\times10^4$ & 1.5       & 1.00 & 0.05  & 11.0 &  0   &  0   &  3   \\
	\hline
	\hline
	$1.5\times10^4$ & 1.5       & 0.05 & 0.30  & 11.0 &  2   &  0   &  1   \\
	\hline
	$1.5\times10^4$ & 1.5       & 1.00 & 0.30  & 2.2  &  0   &  1   &  2   \\
	\hline
	\hline
	$1.5\times10^4$ & 1.5       & 0.05 & 0.50  & 4.9  &  0   &  0   &  1   \\
	\hline
	$1.5\times10^4$ & 1.5       & 1.00 & 0.50  & 2.1  &  1   &  0   &  1   \\
	\hline
	\hline
	\footnote{From Paper II.}
	$3.0\times10^4$ & 2.0       & 0.05 & 0.05 & 10.2  &  1   &  0   &  -   \\
	\hline
	\footnote{From Paper II.}
	$3.0\times10^4$ & 2.0       & 0.25 & 0.05 & 10.1  &  4   &  0   &  -   \\
	\hline
	$3.0\times10^4$ & 2.0       & 0.25 & 0.05 & 10.9  &  1   &  0   &  4   \\
	\hline
	$3.0\times10^4$ & 2.0       & 1.00 & 0.05 & 10.9  &  1   &  0   &  0   \\
	\hline
	\hline
	\footnote{From Paper II.}
	$3.0\times10^4$ & 2.0       & 0.05 & 0.10 & 10.2  &  3   &  0   &  -   \\
	\hline
	\footnote{From Paper II.}
	$3.0\times10^4$ & 2.0       & 0.25 & 0.10 & 6.3   &  3   &  0   &  -   \\
	\hline
	\hline
	$5.0\times10^4$ & 2.0       & 0.05 & 0.05 & 11.0  &  2   &  1   &  -   \\
	\hline
	$5.0\times10^4$ & 2.0       & 0.25 & 0.05 & 11.0  &  4   &  0   &  -   \\
	\hline
	$5.0\times10^4$ & 2.0       & 1.00 & 0.05 & 11.0  &  2   &  0   &  -   \\
	\hline
	\hline
	\footnote{Tidal circularization (\nbseven option $27=1$) applied.}
	$1.5\times10^4$ & 1.5       & 1.00 & 0.30 & 4.6   & 1    &  0   &   0   \\
	\hline
	\footnote{Tidal circularization (\nbseven option $27=1$) applied.}
	$1.5\times10^4$ & 1.5       & 1.00 & 0.50 & 0.75  & 0    &  0   &   2   \\
	\hline
	\footnote{Tidal circularization (\nbseven option $27=1$) applied.}
	$5.0\times10^4$ & 2.0       & 1.00 & 0.05 & 2.8  &  2    &  0    &  0   \\
	\hline

\end{longtable}
\end{minipage}
\twocolumn


\bibliographystyle{mnras}
\bibliography{bibliography/biblio.bib}

\label{lastpage}

\end{document}